\documentclass[twocolumn]{aa}
\usepackage{graphicx}
\usepackage{xcolor}
\usepackage{subcaption}
\usepackage{txfonts}
\usepackage{natbib}
\usepackage{supertabular}
\usepackage{bm}
\usepackage{placeins}

\bibpunct{(}{)}{;}{a}{}{,} 
\newcommand{\ha}  {H$\alpha$}
\newcommand{\ew}  {EW(H$\alpha$)}

\def\simless{\mathbin{\lower 3pt\hbox
     {$\rlap{\raise 5pt\hbox{$\char'074$}}\mathchar"7218$}}}   
\def\simmore{\mathbin{\lower 3pt\hbox
     {$\rlap{\raise 5pt\hbox{$\char'076$}}\mathchar"7218$}}}   

\begin{document}

   \title{Long-term optical variability of high-mass X-ray binaries. \\
   III. Polarimetry}

   \subtitle{}
  \author{
	P. Reig\inst{1,2}
	\and
  	D. Blinov\inst{1,2}
	\and
	A. Tzouvanou\inst{2,3}
          }

\authorrunning{Reig et al.}
\titlerunning{Polarimetric variability of BeXBs}

   \offprints{pau@physics.uoc.gr}

   \institute{Institute of Astrophysics, Foundation for Research and Technology-Hellas, 
   	71110, 	Heraklion, Greece 
	 \and Physics Department, University of Crete, 71003, 
   		Heraklion, Greece --
		\email{pau@physics.uoc.gr; blinov@physics.uoc.gr}
	 \and Max-Planck-Institut f\"ur Astronomie, K\"onigstuhl 17, 69117
	 Heidelberg, Germany --
	 \email{tzouvanou@mpia.de}
}

   \date{Received ; accepted}

\abstract
{Be/X-ray binaries are the most numerous group of high-mass X-ray binaries.
Their long-term optical and infrared variability reflects the
evolution of the circumstellar disk around the luminous companion. This
variability manifests photometrically as an excess of flux that increases
with wavelength and spectroscopically as line emission. The disk is
also expected to generate linear polarization. }
{We present a systematic study of the optical long-term polarimetric variability
of Be/X-ray binaries on data collected over  10 years. Our aim is to
characterize the polarimetric properties of these systems and to probe the
structure of their circumstellar disks.}
{We have been monitoring Be/X-ray binaries visible from the Northern hemisphere
with the RoboPol polarimeter. We performed a careful analysis of the interstellar
polarization in the direction of the sources to estimate their intrinsic
polarization. Stokes parameters for linear polarization were computed by means of
aperture photometry and corrected for instrumental polarization.}
{Optical polarimetric variability is a common trait in
Be/X-ray binaries. The variability can
be attributed to the Be star's circumstellar disk. Our polarization analysis
confirms previous claims based on spectroscopic data that the circumstellar disks in 
BeXBs are, on average, smaller and denser than Be stars in non-binary systems. 
Our data also confirms the presence of highly distorted disk prior to giant
X-ray outbursts, although this result is still affected by the lack of 
simultaneous and well-sampled observations during major X-ray outbursts.}
{}


   \maketitle

\section{Introduction}

Neutron star high-mass X-ray binaries (HMXB) are accretion-powered binary
systems where the neutron star orbits an early-type (O or B) companion. The
luminosity class of the optical companion subdivides HMXBs into Be/X-ray
binaries (BeXB), when the optical star is a dwarf, subgiant or giant OBe star
(luminosity class III, IV or V) and supergiant X-ray binaries (SGXBs), when they
contain an evolved star of luminosity class I-II. In SGXBs, the optical
star emits a  substantial  stellar  wind. A neutron star in a relatively  close
orbit will capture a  significant  fraction of this wind, sufficient to power a
bright X-ray source. In BeXB, the donor is a Be star, i.e. a rapidly rotating
early-type B star with a gaseous equatorial disk
\citep{reig11,rivinius13a}. 

The disk is formed when matter is ejected from the stellar photosphere with
sufficient angular momentum. Such disk is referred to as a
decretion disk. The ultimate mechanism that lifts material into the disk remains
an open question, but the most likely candidates are a combination of fast
rotation and non-radial pulsations \citep{baade92,rivinius13a} or magnetic
reconnection associated with localized magnetic fields generated by convection
\citep{balona03,balona21}. More recently, \citet{martin25} have proposed the
presence of a boundary layer that connects a geometrically thick disk to a
rotationally flattened star. The boundary layer would provide the require torque
and prevent the ejected material from falling back onto the star. In turn, the
decretion disk exerts a torque on the star that slows the Be star's rotation
rate while still allowing decretion to continue. 

The physical properties of the disk, such as density or physical extent, depend
on whether the mass-outflow mechanism is active. Disks in Be stars form and
dissipate contributing to their long-term optical/IR variability. Because the
disk is also the main reservoir of material available for accretion onto the
compact object, its evolution is likewise reflected in the X-ray variability of
Be/X-ray binaries. The largest-amplitude variations are typically observed on
timescales of years, corresponding to the characteristic timescale for disk
build-up and dissipation \citep{jones08,reig16,treiber25}.

Circumstellar disks in Be stars contribute to both the continuum and discrete
(e.g. line) emission and this contribution is observable in photometry, spectroscopy, and
polarimetry:

\begin{itemize}

\item Disk emission contributes to the overall brightness of the system and affects
the photometric magnitudes and colors. The excess continuum radiation is small
at short wavelengths (i.e $B$ band) and increases at longer wavelengths (i.e.
$H$ and $K$ bands). The main disk contribution to the $V$ band comes from a
relatively small area near the star, whereas the emission area increases for
longer wavelengths \citep{carciofi06}. As a result, the color indices increase
and the emission becomes redder, as the disk develops.

\item The most prominent feature in the optical spectra of Be stars is the
presence of emission lines, particularly those of the Balmer series. The hydrogen
lines are optically thick and are formed in a large part of the disk by
recombination. In particular, the \ha\ line stands out as the primary diagnostic
of the disk state \citep{quirrenbach97,tycner05,grundstrom06}. Large amplitude
variations in the strength and profile morphology are associated with structural changes
of the circumstellar decretion disk \citep{reig16}.

\item The continuum polarization is attributed to Thomson scattering of
initially unpolarized stellar radiation in the disk
\citep{coyne69,serkowski70,poeckert79}. Thomson scattering polarizes the
radiation perpendicular to the scattering plane. The amount of the polarization
gives information about the number of scatterers (i.e. density). Consequently,
changes in the polarization degree and position angle over time trace the
evolution of the disk's physical structure.

\end{itemize}

\begin{table*}
\caption{Target list and relevant information.}             
\label{tab:targets}      
\centering          
\begin{tabular}{l@{~~}l@{~~}c@{~~}c@{~~}c@{~~}c@{~~}c@{~~}l@{~~}}
\hline\hline
Source          	&Spectral   &$V$-band	&P$_{\rm spin}$ &P$_{\rm orb}$ &$e$   &Distance (kpc)   &References\\
name            	&type       &(mag)	&(s)    	&(days)	      &       &Gaia DR3    	&\\
\hline
4U\,0115+63     	&B0.2Ve     &15.4	&3.6    &24.3	&0.34   &$5.7^{+0.5}_{-0.4}$        &[1a], [1b] \\
IGR\,J01363+6610	&B1IV-Ve    &13.3	&-- 	&--	&--     &$5.6\pm0.4$	            &[2a], [2b]\\
RX\,J0146.9+6121	&B1Ve	    &11.3	&1400   &303?	&--     &$2.75\pm0.15$ 		    &[3a], [3b]      \\
IGR\,J01583+6713	&B2IVe	    &14.4	&--  	&--	&--     &$5.6^{+0.6}_{-0.4}$        &[4a], [4b] \\
RX\,J0240.4+6112	&B0Ve       &10.8	&--     &26.5	&0.54   &$2.50\pm0.07$	            &[5a], [5b]    \\
Swift\,J0243.6+6124	&O9.5Ve     &12.8	&9.9    &27.6	&0.10   &$5.2^{+0.3}_{-0.2}$        &[6a], [6b]      \\
V\,0332+53      	&O8--9Ve    &15.4	&4.4    &33.8	&0.37   &$5.9\pm0.4$   		    &[7a], [7b]    \\
RX\,J0440.9+4431	&B0.2Ve     &10.7	&202.5  &150	&--     &$2.44^{+0.08}_{-0.09}$     &[8a], [8b]    \\
1A\,0535+262    	&O9.7IIIe   &9.2	&105    &111	&0.47   &$1.77\pm0.06$		    &[9a], [9b]   \\
IGR\,J06074+2205	&B0.5Ve     &12.2	&373.2  &80?	&--     &$6.0^{+0.9}_{-0.6}$        &[10a], [10b]    \\
MXB\,0656--072  	&O9.5Ve     &12.3	&160.4  &101.2	&0.4?   &$5.7\pm0.5$	            &[11a], [11b]   \\
XTE\,J1946+274		&B0-1V-IVe  &16.6	&15.8	&172	&0.25	&$12.1^{+2.6}_{-2.2}$	    &[12a], [12b]  \\
KS\,1947+300    	&B0Ve       &14.5	&18.7   &40.4	&0.03   &$14^{+3}_{-2}$		    &[13a], [13b]    \\
EXO\,2030+375		&B0Ve	    &19.5	&42	&46.02	&0.41	&$2.4^{+0.5}_{-0.4}$	    &[14a], [14b]  \\
GRO\,J2058+42   	&O9.5-B0IV-Ve&14.9	&192	&110	&--	&$8.9^{+0.7}_{-0.8}$	    &[15a], [15b]   \\
SAX\,J2103.5+4545	&B0Ve       &13.9	&358	&12.7	&0.40   &$6.2^{+0.4}_{-0.5}$	    &[16a], [16b]  \\
IGR\,J21343+4738	&B1IVe      &14.1	&320	&--	&--	&$8.5^{+1.1}_{-0.8}$	    &[17]    \\
Cep\,X--4		&B1-2Ve     &14.3	&66.3   &--	&--     &$7.2^{+0.68}_{-0.6}$       &[18]    \\
4U\,2206+54     	&O9.5Ve	    &9.8	&5550   &19.2	&0.15   &$3.2^{+0.2}_{-0.1}$        &[19a], [19b]   \\
SAX\,J2239.3+6116	&B0Ve       &14.4	&1247   &263	&--     &$7.3^{+0.7}_{-0.5}$        &[20a], [20b]    \\
IGR\,J22534+6243	&B1Ve	    &15.3	&46.67	&--	&--	&$8.9^{+1.4}_{-0.9}$	    &[21]   \\
\hline\hline
\end{tabular}
\tablefoot{Distance from \citet{bailer-jones21}} 
\tablebib{
[1a] \citet{negueruela01a}; [1b] \citet{raichur10}; 
[2a] \citet{reig05a}; [2b] \citet{tomsick11}; 
[3a] \citet{reig97b}; [3b] \citet{sarty09}; 
[4a] \citet{wang10}; [4b] \citet{kaur08}; 
[5a] \citet{aragona09}; [5b] \citet{zamanov13}; 
[6a] \citet{wilson18}; [6b] \citep{reig20}; 
[7a] \citet{negueruela99}; [7b] \citet{doroshenko16}; 
[8a] \citet{reig05b}; [8b] \citet{ferrigno13};
[9a] \citet{haigh04}; [9b] \citet{grundstrom07b}; 
[10a] \citet{reig10b}; [10b] \citet{chhotaray24};
[11a] \citet{yan12a}; [11b] \citet{nespoli12};
[12a] \citet{verrecchia02}; [12b] \citet{marcu15};
[13a] \citet{galloway04};  [13b] \citet{liu25} ;
[14a] \citet{parmar89b}; [14b] \citet{wilson08};
[15a] \citet{wilson98}; [15b] \citet{kiziloglu07b};
[16a] \citet{baykal00};  [16b] \citet{reig04};
[17] \citet{reig14a}; 
[18] \citet{bonnet-bidaud98}; 
[19a] \citet{corbet07}; [19b] \citet{blay06} ;
[20a] \citet{intzand01}; [20b] \citet{reig17};
[21] \citet{esposito13}
}
\end{table*}

We have been monitoring the BeXBs visible from the Northern Hemisphere in the
optical band with the 1.3m telescope at Skinakas observatory since 1998,  with
the aim of investigating the evolution of their decretion disks. The monitoring
consists of $BVRI$ photometry and medium resolution spectroscopy around the \ha\
line.  The results from the photometric study were presented in \citet{reig15},
while those from the spectroscopic observations were reported in \citet{reig16}.
Since 2013, following the installation of the RoboPol polarimeter, we have also
obtained optical polarimetric measurements of the same sources. The present work
constitutes the third part of this series and reports the results of the
polarimetric observations.

\section{Data acquisition and analysis}

\subsection{Observations}

The data were obtained from the Skinakas Observatory (SKO),
located on the Ida mountain in central Crete (Greece) at an altitude of 1750 m. 
The observations were made with the RoboPol polarimeter attached to the 1.3m
modified Ritchey-Chr\'etien telescope, an Andor DW436 CCD with an array of
2048 $\times$ 2048 13.5 $\mu$m pixel size (corresponding to 0.435 arcsec/pixel
 on sky), hence providing a field of view of 13 square arcmin. The
camera was cooled to $-70^{\circ}$C ensuring negligible dark current. The
default monitoring set-up uses a Johnson-Cousins R-band filter. However,
observations at other bandpasses are possible thanks to a five-slot filter wheel.

RoboPol is a four-channel imaging polarimeter that uses two Wollaston prisms and
two non-rotating half-wave plates to produce simultaneous measurements of the
Stokes $Q$ and $U$ parameters \citep{ramaprakash19}.  RoboPol splits the
incoming light into two light beams one light beam horizontally and the other
one vertically \citep[see Fig. 1 in][]{king14}. Thus, every point in the sky
appears four times on the CCD. In each spot the photon counts, measured using
aperture photometry, are used to calculate the normalized Stokes parameters $q=Q/I$ and
$u=U/I$ of linear polarization, where $I$ is the total intensity of the source.
A mask is placed in the telescope focal plane to optimize the instrument
sensitivity. The absence of moving parts avoids possible errors caused by
variations in sky conditions between measurements.  The list of targets is given
in Table~\ref{tab:targets} and the polarization measurements of each individual
target are presented as supplementary material (Appendix D) and are archived
in the Zenodo repository: https://doi.org/10.5281/zenodo.18346735.

\subsection{Instrumental polarization}

When light passes through the various optical elements that constitute the
optical system (telescope plus polarimeter), a low level polarization is
introduced into the observed radiation. This instrumental polarization appears
as an additional polarization signal that must be corrected for. We accounted
for this effect by measuring zero-polarized standard stars observed over multiple
years. The average instrumental polarization in the R band is $0.4\pm0.2$\%.

The data must also be corrected to align the rotation of the instrumental $q-u$
plane with respect to the standard reference frame. This was done using highly
polarized standards that were monitored along with the unpolarized standard
candidates sample. More details on how these two corrections were performed can
be found in \citet{blinov23}.

\subsection{Field stars}

The light emitted by the star travels through the interstellar medium (ISM)
before is detected by the observer. ISM introduces extra polarization that
results from the non-sphericity of the dust grains that become oriented in a
preferred direction due to the galactic magnetic field
\citep{lazarian15,tram22}. To correct for this extra polarization, we observed a
number of field stars in the field of view  of each target. The field stars were
selected for its proximity in absolute distance with the targets. The
interstellar component is removed by vector subtraction. The polarization
measurements of the field stars are presented as supplementary material
(Appendix E) and are archived in the Zenodo repository:
https://doi.org/10.5281/zenodo.18346735.

\subsection{Data analysis}

The data were processed using the standard RoboPol pipeline, which is described
by \citet{king14}, with modifications presented by  
\citet{blinov21}. The pipeline reads the raw images from the telescope and
computes the Stokes parameters of the stars in the field of view by means of
aperture photometry. Basically, it performs five tasks: {\em i)} source
identification, {\em ii)} aperture photometry, {\em iii)} calibration, {\em iv)}
polarimetry, and {\em v)} relative photometry

The polarization degree ($p$) is defined as a function of the normalized Stokes
parameters as

\begin{equation}
p=\sqrt{q^2+u^2}
\end{equation}

\noindent and the corresponding uncertainty

\begin{equation}
\sigma_{p}=\sqrt{\frac{q^2\sigma_q^2+u^2\sigma_u^2}{q^2+u^2}}
\end{equation}

\noindent where $\sigma_q$ and $\sigma_u$ are the uncertainties of the
individual observations and include the contribution from the instrumental error uncertainty

\begin{equation}
\sigma_q = \sqrt{(\sigma_q^{\rm obs})^2 + (\sigma_q^{\rm inst})^2}
\end{equation}

\noindent and similarly for $u$. Unlike $u$ and $q$ which are unbiased
quantities and follow a normal distribution, the polarization degree is biased
toward higher values at low signal-to-noise ratios and follows a Rician
distribution. There are a variety of methods suggested for correction of this
bias \citep{simmons85,vaillancourt06,plaszczynski14}. In this work, we will use
the non debiased values of the polarization degree, unless stated otherwise.

The electric vector polarization angle (EVPA) is defined as

\begin{equation}
EVPA=\frac{1}{2} arctan (\frac{u}{q}) 
\end{equation}

Because EVPA does not follow a Gaussian distribution, the
uncertainty $\sigma_{\theta}$ was computed  numerically, by solving the following
integral

\begin{equation}
\int_{-1\sigma_{\theta}}^{+1\sigma_{\theta}} G(\theta;p_0) d\theta = 68.27\%
\end{equation}

\noindent where $G(\theta;\theta_0;p_0)$ is the probability density followed by EPVA
\citep{naghizadeh-khouei93}

\begin{equation}
G(\theta; \theta_0; p_0) =
\frac{e^{-\frac{p_0^2}{2}}}{\sqrt{\pi}}\left\{\frac{1}{\sqrt{\pi}}+\eta_0e^{\eta_0^2}[1+\rm{erf}(\eta_0)]\right\}
\end{equation}

\noindent where   $\eta_0 = P_0 2 \cos 2(\theta - \theta_0)$ and erf is the Gaussian error
function. We refer the reader to \citet{blinov23} for further details.

\begin{figure*}
\begin{center}
\includegraphics[width=16cm]{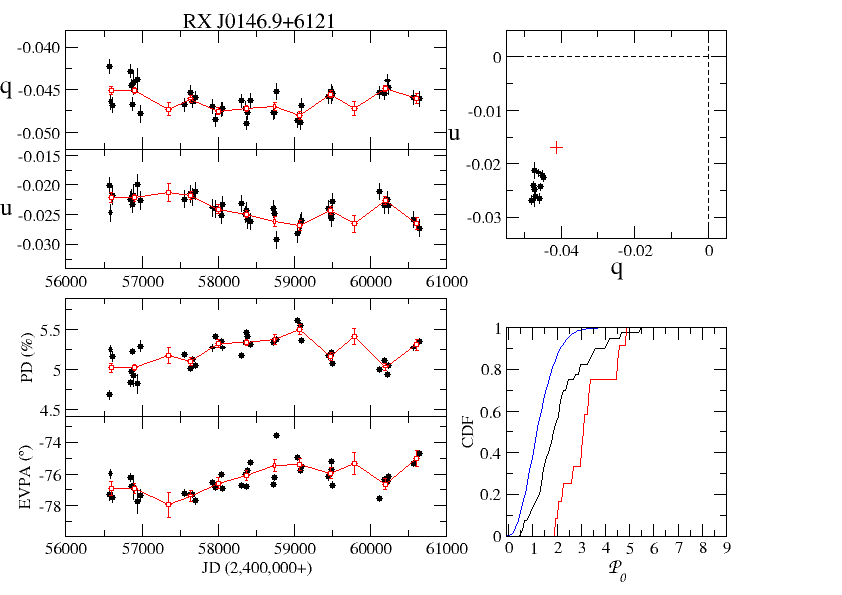} 
\caption[]{{\em Left}: Evolution of the Stokes parameters, polarization degree
and angle; {\em Top-right}: $q-u$ plane. Weighted mean of the source observations
(black circles) calculated yearly and of the field stars (red cross); {\em Bottom-right}:  
EDF of measured polarization using all data points (black line) or the weighted
averaged points (red line) compared with expected CDF of polarization
measurements (blue line). See Sect.~\ref{res:var} for the meaning of these
terms. The data shown correspond to the BeXBs RX J0146.9+6121.}
\label{var}
\end{center}
\end{figure*}

\begin{figure*}
\begin{center}
\begin{tabular}{cc}
\includegraphics[width=9.5cm]{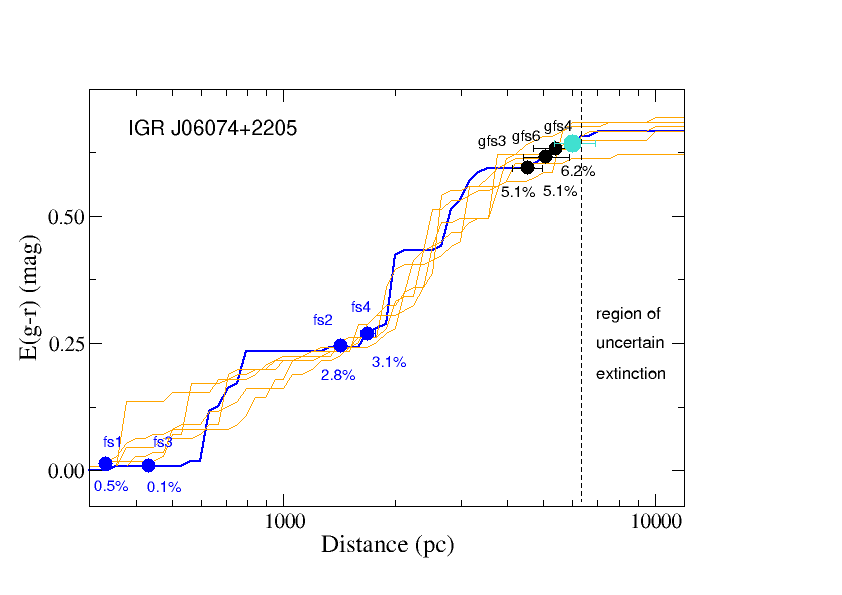} &
\includegraphics[width=9.5cm]{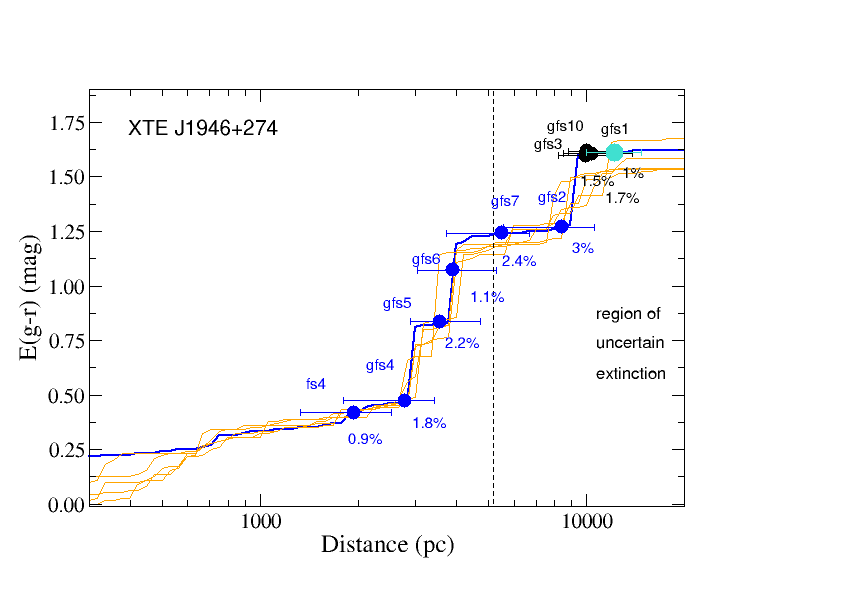} \\
\end{tabular}
\caption[]{Two representative examples of the extinction as a function of
distance  \citep[data from][]{green19}. The data points simply mark the assumed location
of the stars on the extinction curve, based on their distance but they do not
imply any extinction value.  Distances are from \citet{bailer-jones21}. The
turquoise circle represents the BeXB, the black circles are the field stars used
for the ISM correction, and the blue circles correspond to other field stars. We also
indicate the PD of the field stars. The vertical dashed line marks the region
of uncertain extinction. }
\label{extdist}
\end{center}
\end{figure*}

\section{Results}

In this section we present the results of our polarimetric analysis of the
targets (Table~\ref{tab:targets}). Our aim is to investigate if BeXBs are
intrinsically polarized and study the connection between the long-term
polarimetric variability and the X-ray activity.

To determine whether the source is intrinsically polarized, the ISM polarization
must be estimated and subtracted from the observed polarization. There are
various ways to assess whether the measured polarization can attributed entirely
to the ISM or whether some fraction is intrinsic to the source. These include:
{\em i)} multi-color polarimetry; {\em ii)} variability analysis; {\em iii)}
empirical relationships between extinction and polarization degree in the
Galaxy; {\em iv)}) measurements of the polarization of field stars around the
X-ray source.

\begin{itemize}

\item The wavelength dependence of interstellar polarization  is different from
that of a Be star. ISM polarization obeys an empirical relation given by \citet{serkowski75}

\begin{equation}
P(\lambda)/P_{\rm max}=\exp[-k \ln^2(\lambda_{\rm max}/\lambda)]
\label{serk}
\end{equation}

\noindent $k$ determines the width or sharpness of the curve. $\lambda_{\rm
max}$ is the wavelength at which the polarization is maximum and is directly
related to the size of the dust grains \citep{coyne74,serkowski75} and to the total
to selective extinction $R=A_V/E(B-V)$ \citep{whittet78}. The mean value of
$R=3.05$ corresponds to $\lambda_{\rm max}=0.545$ $\mu$m \citep{whittet78}. A
deviation of the Serkowsky's law would indicate a certain amount of intrinsic
polarization, although sometimes it can be caused by multiple dust clouds with
different properties along the line of sight \citep{mandarakas25}. On the other
hand, the polarization degree of unreddened Be stars peaks in the blue ($\lambda
\approx 0.45$ $\mu$m) and decreases with wavelength in the range 0.45-0.80
$\mu$m \citep{poeckert79,mcdavid01,halonen13a,haubois14}.

\item Because ISM polarization is expected to remain constant for a given direction,
if the polarization degree of a source is variable, then we can conclude that a
fraction of it is intrinsic to the source. However, quantifying the actual
amount of intrinsic polarization simply through a variability analysis will not
be possible unless the nature of the system is such that it allows to define a
zero-level intrinsic polarization. Be stars provide such a method. As mentioned
above, polarization in Be stars is due to electron scattering in the
circumstellar disk. If the disk is absent, then all the contribution to the
polarization parameters should come from the ISM. We will discuss disk-loss
events in Sect.~\ref{sect:disk-loss}.

\item As the accumulated mass of dust along any line of sight grows with
distance, it is expected that the degree of polarization associated with any
star will also depend on its distance, or equivalently with extinction. The
relationship between polarization and extinction has been studied by a number of
authors \citep{hiltner56,serkowski75,jones89,reiz98,fosalba02,ignace25}. These
studies confirms that the ISM polarization fraction increases as the extinction
(or distance) increases, albeit with a large scatter. However, these studies are
limited to relatively nearby stars with $E(B-V) \simless 1$ mag. The alignment
and uniformity of the magnetic field is expected to disappear as the line of
sight crosses more and more interstellar dust clouds. The resulting polarization
of the most distance objects becomes a function of the cloud parameters
(orientation and strength of their magnetic field). Thus, the more distant
objects are expected to suffer from depolarization.  For example,
\citet{fosalba02} found that the polarization degree as a function of distance
shows a maximum for stars at $2-4$ kpc and then decreases slightly for more
distant objects up to 6 kpc. Similarly, using $\sim$ 28000 stars from the catalog of
\citet{panopoulou25}, \citet{ignace25} reported that the median polarization
fraction increases up to $\sim$1 kpc and then flattens out for larger distances.

\item  The most reliable method for estimating the ISM polarization is to
observe a sample of field stars located at distances similar to that of the
source. The measured Stokes $u$ and $q$ of the field stars are then subtracted
from the measured polarization of our targets.

\end{itemize}

In the following sections, we address each one of these methods. We only show
the results for a selected number of targets, chosen as representative
cases. The results for all the sources are presented in the Appendices.

\subsection{Wavelength dependence}

Our polarimetric monitoring of the Galactic BeXBs visible from the northern
hemisphere was conducted in the $R$ band. However, we also observed several
sources in the $B$, $V$, and $I$ bands. This multi-color photometry enables, in
principle, an investigation of the wavelength dependence of the polarization. In
practice, however, due to the scarcity of the data (only four data points) and
limited wavelength ranged, the analysis did not yield conclusive results.
Therefore, we did not pursue this method.

\begin{table*}
\caption{Polarization mean and standard deviation. The $p$-value serves as an indicator of
variability. EVPA in the 0$^{\circ}$--180$^{\circ}$ interval.}
\label{varanal}
\centering
\begin{tabular}{l@{~~}c@{~~}c@{~~}c@{~~}c@{~~}c@{~~}c@{~~}c@{~~}c@{~~}c@{~~}c@{~~}}
\hline
Source		   &PD(\%) &$\sigma_{\rm PD}$(\%) & EVPA ($^{\circ}$) &$\sigma_{\rm EVPA}$ ($^{\circ}$) & q &$\sigma_q$ & u &$\sigma_u$ & p-value (all)  & p-value (mean) \\
\hline
4U 0115+63          &  3.66 &  0.39 &    110.6 &   3.1 & -0.0275 &  0.0041 & -0.0241 &  0.0038 &    3.2$\times 10^{-16}$ &	6.2$\times 10^{-08}$ \\ 
IGR J01363+6610     &  7.59 &  0.21 &    108.4 &   0.8 & -0.0607 &  0.0019 & -0.0456 &  0.0024 &    7.1$\times 10^{-01}$ &	1.3$\times 10^{-02}$ \\ 
RX J0146.9+6121     &  5.20 &  0.17 &    103.6 &   1.1 & -0.0462 &  0.0015 & -0.0238 &  0.0021 &    1.6$\times 10^{-05}$ &	2.6$\times 10^{-10}$ \\ 
IGR J0158+6713      & 10.31 &  0.27 &    110.4 &   0.7 & -0.0781 &  0.0025 & -0.0673 &  0.0028 &    4.2$\times 10^{-01}$ &	9.8$\times 10^{-01}$ \\ 
RX J0240.4+6112     &  1.22 &  0.17 &    138.0 &   4.1 &  0.0013 &  0.0018 & -0.0121 &  0.0017 &    4.8$\times 10^{-29}$ &	1.1$\times 10^{-14}$ \\ 
Swift J0243.6+6124  &  3.77 &  0.18 &    110.7 &   1.3 & -0.0282 &  0.0020 & -0.0249 &  0.0014 &    4.5$\times 10^{-02}$ &	9.6$\times 10^{-05}$ \\ 
V 0332+53           &  3.57 &  0.23 &    115.1 &   1.8 & -0.0229 &  0.0021 & -0.0274 &  0.0024 &    2.2$\times 10^{-08}$ &	9.8$\times 10^{-06}$ \\ 
RX J0441.0+4431     &  1.88 &  0.21 &    155.8 &   3.1 &  0.0125 &  0.0018 & -0.0141 &  0.0023 &    1.0$\times 10^{-12}$ &	7.3$\times 10^{-08}$ \\ 
1A 0535+26          &  0.79 &  0.19 &      0.3 &   5.1 &  0.0079 &  0.0019 &  0.0001 &  0.0014 &    1.1$\times 10^{-10}$ &	3.3$\times 10^{-10}$ \\ 
IGR J06074+2205     &  6.12 &  0.38 &    159.0 &   2.0 &  0.0454 &  0.0017 & -0.0410 &  0.0054 &    1.3$\times 10^{-10}$ &	2.0$\times 10^{-06}$ \\ 
MXB 0656-072        &  2.15 &  0.18 &    142.6 &   4.9 &  0.0056 &  0.0038 & -0.0207 &  0.0016 &    6.8$\times 10^{-08}$ &      1.5$\times 10^{-15}$ \\ 
XTE J1946+274       &  1.77 &  0.36 &     31.2 &   5.4 &  0.0082 &  0.0032 &  0.0157 &  0.0037 &    1.1$\times 10^{-17}$ &	1.1$\times 10^{-03}$ \\ 
KS 1947+300         &  2.62 &  0.23 &     23.0 &   2.5 &  0.0182 &  0.0021 &  0.0189 &  0.0025 &    4.8$\times 10^{-13}$ &	8.1$\times 10^{-03}$ \\ 
EXO 2030+375        & 19.06 &  1.38 &     41.2 &   1.7 &  0.0250 &  0.0116 &  0.1889 &  0.0139 &    8.3$\times 10^{-19}$ &	1.5$\times 10^{-10}$ \\ 
GRO J2058+42        &  4.69 &  0.33 &     67.3 &   2.0 & -0.0329 &  0.0031 &  0.0334 &  0.0035 &    1.6$\times 10^{-11}$ &	8.6$\times 10^{-05}$ \\ 
SAX J2103.5+4545    &  1.64 &  0.24 &      4.9 &   4.0 &  0.0161 &  0.0024 &  0.0028 &  0.0023 &    4.4$\times 10^{-21}$ &	4.6$\times 10^{-06}$ \\ 
IGR J21343+4738     &  1.61 &  0.28 &     41.6 &   4.1 &  0.0019 &  0.0023 &  0.0160 &  0.0028 &    1.8$\times 10^{-08}$ &	6.9$\times 10^{-09}$ \\ 
Cep X--4            &  2.16 &  0.20 &     17.6 &   3.0 &  0.0177 &  0.0018 &  0.0124 &  0.0025 &    2.1$\times 10^{-13}$ &	1.9$\times 10^{-02}$ \\ 
4U 2206+54          &  4.07 &  0.20 &     41.4 &   1.2 &  0.0051 &  0.0017 &  0.0404 &  0.0020 &    5.9$\times 10^{-15}$ &	6.0$\times 10^{-05}$ \\ 
SAX J2239.3+6116    &  7.38 &  0.22 &     62.8 &   0.9 & -0.0429 &  0.0026 &  0.0601 &  0.0020 &    9.4$\times 10^{-02}$ &	5.4$\times 10^{-02}$ \\ 
IGR J22534+6243     &  5.23 &  0.33 &     67.2 &   1.8 & -0.0366 &  0.0037 &  0.0374 &  0.0030 &    2.7$\times 10^{-05}$ &	3.9$\times 10^{-02}$ \\ 
\hline
\end{tabular}
\end{table*}

\subsection{Measuring variability: variability diagram}
\label{res:var}

In this section, we study the long-term polarimetric variability. Since the ISM
polarization is expected to remain constant for a given direction and distance
in the sky and in order to avoid introducing extra uncertainty inherent to the
field stars, the analysis was performed on the data after correcting for
instrumental polarization only, without removing the contribution from the ISM.

To study the variability, we follow the methodology explained in
\citet{blinov23}, which in turn, is based on an statistical test using the
cumulative distribution function (CDF) of the polarization data
\citep{clarke94,bastien07}. The method basically compares the theoretical and
empirical cumulative distribution functions (EDF) using a two-sided
Kolmogorov-Smirnov (KS) test. If the $p$-value of the KS test exceeds a given
threshold (e.g., $p=0.0027$ corresponds to 3 $\sigma$ confidence level), we
consider the star to be non-variable.

The theoretical CDF for a non-polarized source normalized by its errors
$\mathcal{P}=\sqrt{(q/\sigma_q)^2+(u/\sigma_u)^2}$ is given by

\begin{equation}
\label{cdf}
CDF(\mathcal{P})=1-e^{-\mathcal{P}^2/2}
\end{equation}

\noindent and the EDF by

\begin{equation}
EDF(\mathcal{P})=\frac{\rm number \,\,of\,\, observations < \mathcal{P}}{\rm total \,\,number \,\, of\,\, observations}
\end{equation}


\noindent where $q$ and $u$ are the Stokes parameters. For a polarized star,
eq.~(\ref{cdf}) does not apply. One way to circumvent this situation is to
subtract the weighted mean $\bar{q}$ and $\bar{u}$ to make the source appear as unpolarized
\citep{clarke94,blinov23}

\begin{equation}
\mathcal{P}_0 = \sqrt{\left(\frac{q-\bar{q}}{\sigma_q}\right)^2+\left(\frac{u-\bar{u}}{\sigma_u}\right)^2}
\end{equation}

The plots showing the long-term evolution of the PD and EVPA of all the targets
are given in Appendix~\ref{app:longterm}. Figure~\ref{var} shows the case for
the BeXB RXJ0146.9+6121. The left panels in this figure show the evolution of
the Stokes parameters $u$ and $q$ and the polarization degree $PD$ and
polarization angle $EVPA$.  The upper-right panel in Fig.~\ref{var} displays the
$q-u$ plane. In this panel, the black circles correspond to the weighted mean of
the source computed yearly, while the red cross represents the weighted mean of
the field stars used later to derive the intrinsic polarization. The
bottom-right panel shows the comparison of the theoretical CDF (blue line) and
the empirical EDFs for the entire set of observations (i.e. the black circles in
the evolution plots, black line) and the seasonal weighted mean (red circles,
red line). Instead of assuming a given threshold to assess the variability of
the sources, we provide the $p$-value derived from the KS test\footnote{As a
reference, the corresponding $p$-value for a 2$\sigma$ confidence level is 
$p=0.0455$, for a 3$\sigma$ is $p=0.0027$, and for 5$\sigma$ is $P=5.733\times
10^{-7}$. If the computed probability is smaller than $p$-value, then the source
is variable at the given confidence level.}. Table~\ref{varanal} gives the normal
mean and standard deviation of the entire data set for each BeXB. Columns 10 and
11 in this table show the computed $p$-value of the KS test. We conclude that 18
out of 22 sources are variable at $>3\sigma$ level on timescales of years.

\subsection{Extinction}
\label{res:ext}

The relationship between polarization degree and extinction shows large
dispersion. For many years, the theoretical upper limit for optimal alignment
efficiency of dust grains with external  magnetic fields was $P(\%) = 9E(B - V)$
\citep{serkowski75,reiz98}.  However, cases where the observed starlight
polarization exceeds the classical upper limit and reach $P_V/E(B-V)=14$\% have
been reported \citep{panopoulou19, bartlett25}. Nevertheless, the observed mean
correlation between polarization degree and extinction for data averaged in
extinction bins is much smaller and deviates slightly from the simple linear
correlation, $P(\%) = 3.5E(B - V)^{0.8}$ \citep{fosalba02}.

Figure~\ref{extdist} shows the extinction curve for IGR J06074+2205 and XTE
J1946+274 as examples of a well-behaved case where the PD increases with
distance and a more complex case associated with a larger and more uncertain
distance, respectively. We marked  the location of field stars and the source
according to their distances. The extinction data were taken from the dust maps
by \citet{green19}. These figure provides a simple way to estimate the number of
molecular clouds to the source, although it should be noted that the range where
the data are reliable is limited to a certain distance range due to the lack of
sufficient number of stars with reddening estimates. This range varies from
source to source and it is indicated by a vertical dashed line in the figures.
The individual extinction curves for each source are presented in
Appendix~\ref{app:extinction}.

\subsection{Field stars}
\label{sect:fs}

The most common method to account for the ISM polarization is to use field
stars. We observed several stars in the vicinity of each target (within $\sim 7$
arcmin) and at comparable distances. Distances were taken from Gaia DR3
\citep{bailer-jones21}. The polarization measurements of the field stars are
given in Appendix E (Supplementary material in Zenodo
repository: https://doi.org/10.5281/zenodo.18346735). The weighted means of the
Stokes parameters $u$ and $q$ are also shown in the variability plots
(Fig.~\ref{var}). However, this method is not free from uncertainty  because 
the dust distribution (e.g. orientation of the dust grains) and number of
intervening molecular clouds along the line of sight may be complex (see the
right panel in Fig.~\ref{extdist}). In addition, some field stars may exhibit
non-zero intrinsic polarization, contrary to the assumption of this method that
they are intrinsically unpolarized. It is therefore advisable to employ multiple
field stars. Owing to this uncertainty, the results of the observations given in
the Appendix correspond to the the measured polarization values, without
correction for ISM contribution. This approach allows readers to apply their own
selection of field stars for ISM subtraction. Appendix~\ref{app:charts} presents
the sky charts with the identification of the target and the field stars.

\section{Discussion}

One of the objectives of this work is to study the evolution of decretion disks
in BeXBS through polarization observations. Changes in the polarization angle
and polarization degree trace evolutionary changes in the geometry and physical
extend of the disk. Electron scattering of unpolarized light generates polarized
radiation perpendicular to the scattering plane, which in Be stars coincides
with the decretion disk. This is because multiple scattering occurs
predominantly in the optically thicker equatorial regions. This biases the
orientation of the scattering planes of the multiply scattered photons toward
the equatorial plane \citep{wood96}. Therefore variability in the polarization
angle may reveal the presence of warped disks \citep{reig18b}. Likewise, since
the degree of the polarization gives information about the number of
scatterers, it can be linked to the density of the disk.

\begin{table}
\caption{Minimum and maximum values of the \ha\ equivalent width: $N$ is the number of
measurements. The period analyzed is MJD 56445--60645. Typical errors in \ew\ are
$\simless5$\%.}
\label{disk-loss}
\centering
\begin{tabular}{l@{~~}c@{~~}c@{~~}c@{~~}}
\hline
Source			&Min.		&Max. 		&N	\\
			& \ew\		& \ew\		&	\\
\hline
4U\,0115+63     	&+0.3		&--12.1		&51	 \\
IGR\,J01363+6610	&--47.0		&--74.7		&33	 \\
RX\,J0146.9+6121	&--6.3		&--15.4		&64	 \\
IGR\,J01583+6713	&--59.0		&--73.9		&23	 \\
RX\,J0240.4+6112	&--9.3		&--14.5		&22	 \\
SWIFT\,J0243.6+6124	&--4.6		&--11.1		&39	 \\
V\,0332+53      	&--4.0		&--10.5		&33	 \\
RX\,J0440.9+4431	&--1.6		&--12.8		&40	 \\
1A\,0535+262    	&--8.3		&--22.1 	&22	 \\
IGR\,J06074+2205	&+2.6		&--11.4 	&31	 \\
MXB\,0656--072  	&--3.2		&--19.0 	&21	 \\
XTE\,J1946+274		&--33.6		&--44.6 	&30	 \\
KS\,1947+300    	&--7.1		&--16.5 	&40	 \\
EXO\,2030+375		&--		&--		&--	 \\
GRO\,J2058+42   	&--0.7		&--13.9 	&60	 \\
SAX\,J2103.5+4545	&+3.1		&--5.6  	&53	 \\
IGR\,J21343+4738	&+3.0		&--8.9		&51	 \\
Cep\,X--4		&--47.7		&--55.0		&29	 \\
SAX\,J2239.3+6116	&--0.8		&--21.9		&58	 \\
IGR\,J22534+6243	&--2.8		&--39.9		&20	 \\
\hline
\end{tabular}
\end{table}

\begin{table}
\caption{Comparison of the polarization degree during low-optical states
including disk-loss events and that obtained from the analysis of field stars. }
\label{pdcomp}
\centering
\begin{tabular}{l@{~~}c@{~~}c@{~~}c@{~~}}
\hline
Source			&PD (\%) 		&PD (\%) 		&Low state 	 \\
			&low states		&field stars		&(MJD) 		 \\
\hline
4U\,0115+63     	&$3.6\pm0.1$		&$2.6\pm0.1$		&59000-59500	 \\
RX\,J0440.9+4431	&$2.3\pm0.1$		&$3.0\pm0.1$		&56300-56700	 \\	    
IGR\,J06074+2205	&$6.5\pm0.1$		&$5.5\pm0.1$		&57950-58050	 \\	    
MXB\,0656--072  	&$2.2\pm0.1$		&$2.3\pm0.1$		&59000-60000	 \\	    
GRO\,J2058+42   	&$4.4\pm0.1$		&$2.5\pm0.1$		&59700-60500	 \\
SAX\,J2103.5+4545	&$1.6\pm0.1$		&$1.6\pm0.1$		&59700-59830	 \\
IGR\,J21343+4738	&$1.3\pm0.1$		&$1.4\pm0.1$		&56500-57300	 \\
SAX\,J2239.3+6116	&$7.2\pm0.1$		&$5.8\pm0.1$		&57000-57800	 \\
\hline
\end{tabular}
\end{table}

\subsection{Long-term variability}

In Sect.~\ref{res:var}, we showed that most BeXBs are polarimetric variable
sources on timescales of years. Only three systems, SAX J2239.3+6116, IGR
J01363+6610 and IGR J0158+6713 display stable long-term polarization
($<2\sigma$). But even in these cases, the standard deviation computed over the
entire data set (Table~\ref{varanal}) is significantly larger than the accuracy
of the instrument, estimated to be not worse than 0.015\% based on measurements
of standard stars. These timescales are associated with the viscous time over
which the Be star's disks evolve. Two natural consequences of the long-term
variability associated disk evolution are the loss of the disk and the
occurrence of X-ray outbursts. 

\subsubsection{Disk-loss episodes}
\label{sect:disk-loss}

Circumstellar disks in Be stars, and therefore also in BeXBs, are not permanent
features. They form, grow and dissipate on timescales of years. Disk-loss
episodes are of paramount importance because they enable the separation of
intrinsic polarization from that due to the ISM.  In Be stars, linear
polarization originates in the circumstellar disk. Thus, in the absence of the
disk, the measured polarization must be entirely interstellar in origin. The
H$\alpha$ line is the most reliable indicator of disk dissipation: it changes
from emission to a pure absorption profile once the disk has been lost.
Consequently, its equivalent width (EW) serves as a robust proxy for disk
presence or absence. By convention, EW is negative for emission lines and
positive for absorption lines.

Table~\ref{disk-loss} lists the maximum and minimum values of the \ew\ for the
studied targets during the period 2013-2024 that corresponds to our polarimetric
campaigns. The spectroscopic data were obtained with the 1.3m telescope at
Skinakas observatory and are part of a long-term monitoring project, as explained
in the introduction of this paper\footnote{The spectroscopic observations will be
presented in a forthcoming paper. We refer the reader to \citet{reig16} for
details on the monitoring program.}. Three BeXBs went through disk-loss
episodes:  IGR\,J06074+2205, SAX\,J2103.5+4545, and IGR\,J21343+4738, while
other five showed evidence for highly debilitated disks: 4U\,0115+63,
RX\,J0440.9+4431, MXB\,0656--072, GRO\,J2058+42, and SAX\,J2239.3+6116.
Table~\ref{pdcomp} compares the PD during low optical states, including
disk-loss events with the PD from the analysis of the field stars. In general,
the agreement is good, especially for the cases for which the absence of the disk
is more evident. For the sources with \ew$>0$, a small residual disk is expected
to be present. For RX\,J0440.9+4431, the adequacy of the field stars may be
questioned, as they are located at longer distance than the target.


\begin{table*}
\caption{BeXBs that displayed type II outbursts during the period covered by the
observations. The flux and luminosity correspond to the peak of the outburst.}
\label{tab:typeII}
\centering
\begin{tabular}{lcccccc}
\hline
X-ray source 	& Start  & End  & Peak 	 & BAT flux$^{(2)}$  & Flux$^{(2),(3)}$ & $L_X^{(1),(2),(4)}$  \\
		&(MJD)	&(MJD)	&(MJD)	&(ct/cm$^2$/s)		&(mCrab)	&(erg s$^{-1}$) \\
\hline
XTE J1946+274 	& 58275 & 58325 & 58283 & 0.035 & 159 & 3.56$\times$10$^{37}$ \\
KS 1947+300 	& 56548 & 56691 & 56610 & 0.071 & 323 & 8.48$\times$10$^{37}$ \\
EXO 2030+375 	& 59406 & 59547 & 59465 & 0.125 & 568 & 3.73$\times$10$^{37}$ \\
GRO J2058+42 	& 58545 & 58614 & 58569 & 0.050 & 227 & 2.86$\times$10$^{37}$ \\
4U 0115+63 	& 57306 & 57342 & 57318 & 0.100 & 455 & 2.30$\times$10$^{37}$ \\
4U 0115+63 	& 57958 & 57997 & 57971 & 0.064 & 291 & 1.47$\times$10$^{37}$ \\
4U 0115+63 	& 60028 & 60070 & 60042 & 0.150 & 682 & 3.45$\times$10$^{37}$ \\
Swift J0243.6+6124 & 60092 & 60204 & 60136 & 0.260 & 1182 & 4.97$\times$10$^{37}$ \\
V 0332+53 	& 57185 & 57311 & 57234 & 0.220 & 1000 & 5.41$\times$10$^{37}$ \\
RX J0441.0+4431 & 59935 & 60080 & 59977 & 0.500 & 2273 & 2.04$\times$10$^{37}$ \\
1A 0535+26 	& 59149 & 59213 & 59172 & 2.420 & 11000 & 5.54$\times$10$^{37}$ \\
\hline
\end{tabular}
\tablefoot{
\tablefoottext{1}Assumed distances from Table~\ref{tab:targets}; 
\tablefoottext{2}Energy range 15--50 keV;
\tablefoottext{3}$1\, mCrab = 0.000220$ ct cm$^{-2}$ s$^{-1}$; 
\tablefoottext{4}$1\, mCrab=1.3\times10^{-11}$ erg s$^{-1}$ cm$^{-2}$} 
\end{table*}

\begin{figure*}
\begin{center}
\includegraphics[width=16cm]{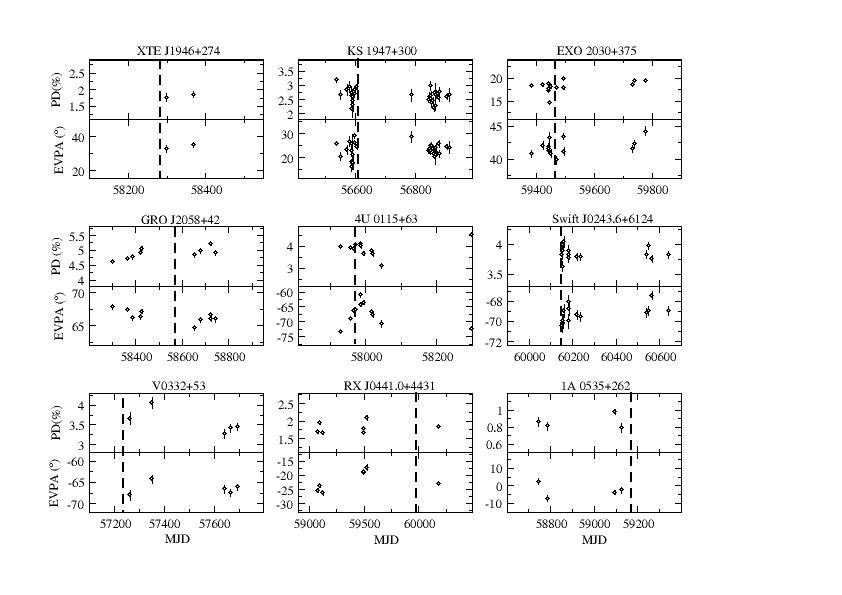} 
\caption[]{Polarization degree and angle around the time of a major (type II) X-ray outburst.
The vertical dashed lines mark the peak of the X-ray outburst.}
\label{fig:typeII}
\end{center}
\end{figure*}

\subsubsection{Giant X-ray outbursts}

It is generally accepted that X-ray outbursts in Be/X-ray binaries originate
when mass from the circumstellar disk of the Be star is transferred to the
neutron star. However, the exact mechanism of how this happens is still not well
understood.
The disks in Be/X-ray binaries are often truncated due to the gravitational
effects of the neutron star \citep{reig97a,negueruela01a,okazaki01,reig16}. This truncation
raises the question of how substantial amounts of matter can be transferred to
the neutron star to trigger the X-ray outbursts. If the disk is truncated, then
accretion of large amounts of matter is only feasible if the disk becomes
sufficiently asymmetric and dense to exceed the truncation radius.  The current
theory suggests that giant outbursts occur when the neutron star captures a
significant amount of gas from a warped, highly misaligned, and eccentric Be
disk \citep{martin11,okazaki13,martin14a}. Models indicate that such highly
distorted disks can result in enhanced mass accretion when the neutron star
passes through the warped region. This interaction increases the density and
volume of gas available for accretion onto the neutron star, thereby triggering
intense X-ray emission.

Observational evidence of warped disks comes from the spectral lines. Low
inclination systems give rise to single peak profiles. Double-peak profiles are
typical of intermediate inclination systems, while systems seen at high
inclination generate shell profiles \citep{rivinius13a}. Therefore, if all these
different profiles are seen in the same source and since the spin axis of the Be
star is not expected to change on timescales of days, the appearance of
different profiles in the same star can only imply that the disk axis is
changing direction. In other words, that there is a warped precessing disk.
Changes in the emission line profile in Be and BeXBs have been interpreted as
observational evidence for misaligned disks that become warped in their outer
parts \citep{hummel98,negueruela01b,reig07b,moritani11,moritani13}.

Further evidence for warped disks should come from the polarization angle
because light is polarized perpendicular to the scattering plane in Thompson
scattering. Therefore, if the orientation of the disk changes, the polarization
angle should change as well. Table~\ref{tab:typeII} lists the sources that
exhibited a giant (type II) outburst during the period covered by the
polarimetric observations.  Data were computed from the {\it Swift}/BAT light
curves by applying the average conversion factors shown in the table footnote.
We searched for changes in the polarization parameters during type II outbursts.
Figure~\ref{fig:typeII} shows the polarization degree and angle at the time of
type II outbursts for the sources that displayed those events during the course
of the observations. The vertical dashed lines mark the peak of the X-ray
outburst. Unfortunately, the lack of data during many of the
outbursts prevents us from providing strong evidence for warped precessing
disks. Nevertheless, we found some promising cases. In addition to 4U 0115+63,
which was already reported in \citet{reig18b}, KS 1947+300, EXO 2030+375, and
Swift J0243.6+6124 showed changes in the EVPA in coincidence with the X-ray
outburst. 

The outer parts of the disk are expected to be the most affected by warping
effects \citep{martin21}, while the inner regions, where most of the
polarization is produced, would be less affected. Linear polarization
signatures arise primarily in the inner parts of the disk, where the density of
scatterers is highest \citep{carciofi11,halonen13b}. This fact may explain
the relatively small observed EVPA variation ($\lesssim 10^{\circ}$), which
would be associated with the precession angle.

An interesting result that can be extracted from Fig.~\ref{fig:typeII} is that
the distorted disk appears to precede the X-ray event. Except for Swift
J0243.6+6124, for which no data are available prior to the outburst, the
polarization parameters had already begun to vary before the onset of the
outburst in the other three cases. In other words, this suggests that the
presence of a distorted disk seems to be a prerequisite for triggering an X-ray outburst,
consistent with theoretical predictions \citep{martin11,okazaki13,martin14a}.

\subsection{Short-term variability}

In addition to long-term variations, Fig.~\ref{var} (and
Appendix~\ref{app:longterm}) also shows significant  scatter with variations of
10\%--30\% of the polarization parameters within timescales of a few weeks to
months. This behavior has been seen in isolated Be stars
\citep{vince95,draper14}. The origin of this generally non-periodic variability
is not clear and most likely stems from a diversity of phenomena that lead to 
changes in the distribution and density of the light scattering material in the
disk. Longer-term phenomena (months) associated with this apparent stochastic jumps in the
polarization could be a rotating or precessing tilted disk \citep{marr18} or
one-armed density waves \citep{halonen13b,draper14}; on much shorter timescales
($\simless 1$ day), polarimetric variability may be associated with mass
ejection episodes from the Be star into the disk and with stochastic changes in
the mass decretion rate  \citep{carciofi07,wisniewski10,carciofi25}. On
intermediate-term timescales (weeks), orbital modulation could also contribute
to the scatter \citep{kravtsov20}.

\begin{figure}
\begin{center}
\includegraphics[width=9cm]{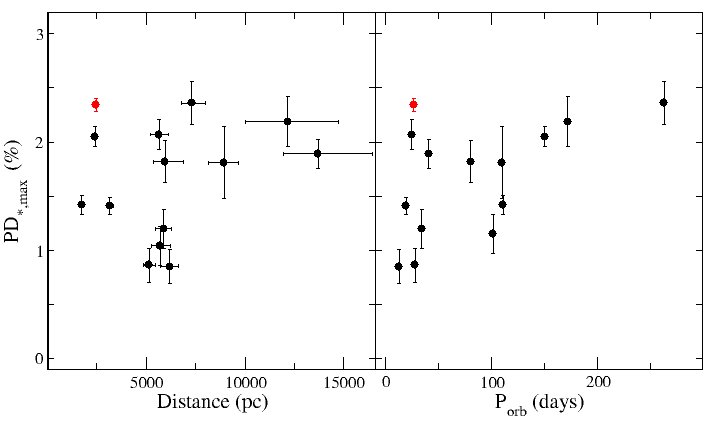} 
\caption[]{Maximum intrinsic polarization degree as a function of distance
(left) and orbital period (right), excluding the outlier EXO 2030+375. The red point
corresponds to the Be/$\gamma$-ray binary RX J0240.4+6112/LS I +61303}
\label{PDmax-Porb}
\end{center}
\end{figure}

\begin{figure}
\begin{center}
\includegraphics[width=9cm]{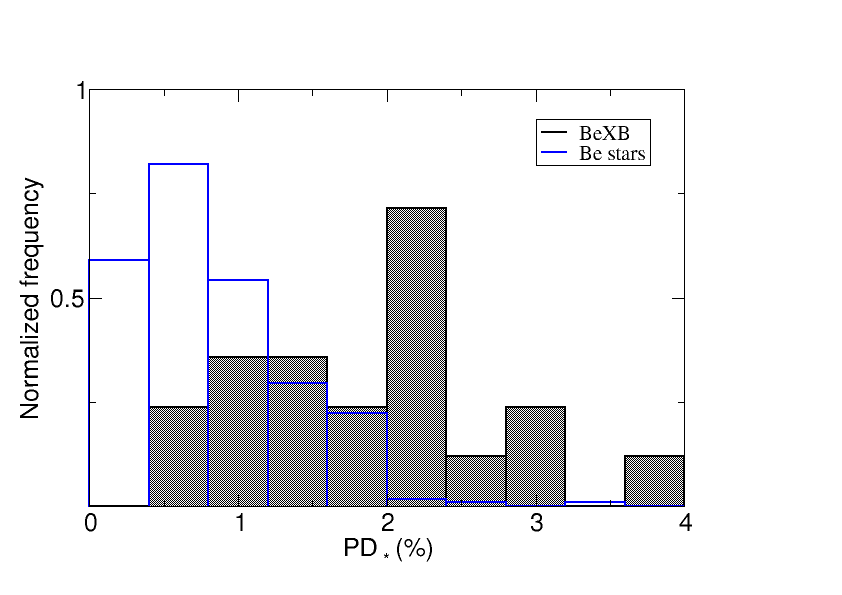} 
\caption[]{Normalize histogram (the $Y$-axis is computed so that the total area under 
the histogram equals 1) of the maximum intrinsic polarization (corrected for the ISM) of our 
sample of BeXBs (excluding the outlier EXO 2030+375),  compared to the
polarization distribution of classical Be stars of similar spectral type
\citep[data from][]{yudin01}.}
\label{PDmax-hist}
\end{center}
\end{figure}

\subsection{Comparison with Be stars}
\label{compBe}

The circumstellar disks in BeXBs are, on average, smaller and denser than those
of Be stars in non-binary systems. Evidence for this result comes from both
observations \citep{reig97a,reig00,reig16,zamanov01} and theory
\citep{okazaki02,panoglou16,cyr17}.  
This is explained by the interaction between the disk and the neutron star, and
more precisely, by disk truncation. The closer the two components of the binary,
the stronger the tidal torque exerted by the neutron star on the disk. Systems
with short orbital periods ($P_{\rm orb} \simless 50$ d) exhibit faster
variability (i.e., rapid, irregular changes) than systems with longer orbital
periods \citep[see Fig. 1 in][]{reig16}. Clearly, the disk in narrow-orbit
systems is exposed to larger torques from the neutron star, preventing it from
reaching a stable configuration for a long period of time. Another example is
the correlation between the maximum \ew\ and the orbital period in BeXBs
\citep{reig97a,reig11}. Because \ew\ provides a measure of the extent of the
disk \citep{tycner05,grundstrom06}, the correlation implies that wider-orbit
systems are able to develop stable and larger disks.

The question that we investigate here is whether polarization data agree with
these results. For this analysis to be meaningful, the intrinsic polarization
must be used. The measured polarization was corrected for the contribution of
the ISM as indicated in Sect.~\ref{sect:fs}. The right panel in
Fig.~\ref{PDmax-Porb} shows the maximum intrinsic polarization degree, $P_*$, as
a function of the orbital period, $P_{\rm orb}$. There is a trend of larger
intrinsic PD for wider-orbit systems in BeXBs. The correlation is significant if
the outlier EXO 2030+375 and the Be/$\gamma$-ray binary RX J0240.4+6112/LS I
+61303 are excluded\footnote{Be/$\gamma$-ray binaries differ from typical BeXBs
in the observational as well as emission models for the high-energy radiation.
See e.g. \citet{dubus13}.}: the correlation coefficient and the probability that
the null hypothesis (that the two variables are not correlated) is true are
$r=0.66$ and $p$-value=0.015, respectively. There are a number of factors that
could contribute to the dispersion in this plot. {\em i)} uncertainty in the ISM
polarization estimation, {\em ii)}  the dependence of polarization fraction on
inclination angle \citep{wood96,halonen13a} and/or {\em iii)} the fact that
polarization takes place closer to the Be star compared to the formation region
of the  \citep{carciofi11} \ha\ line. In this respect, \ew\ represents a better
proxy for the extent of the disk and the effects of the tidal interaction with
the neutron star manifest in a stronger way.  The intrinsic polarization
fraction and angle strongly depends on the polarization of the field stars.
However, the fact that there is no clear trend ($r=0.26$, $p$-value=0.34)
between the maximum intrinsic polarization and distance (left panel in
Fig.~\ref{PDmax-Porb}) implies that, overall, the selection of the field stars
was adequate. Nevertheless, a bad representation of the ISM for a few individual
systems cannot be ruled out.  The case of EXO 2030+375 is unique. It exhibits
the highest polarization degree ever measured in either a Be star or a BeXB,
well above the typical values observed in these systems (see discussion below).
One possible explanation for this apparent extreme behavior is the high and
spatially complex extinction toward the source. As shown in Fig.~\ref{charts}
and Appendix E (Supplementary material in Zenodo repository:
https://doi.org/10.5281/zenodo.18346735), the polarization degree measured in
the vicinity of the X-ray source varies substantially. For instance, clear
differences are observed when comparing the spatial locations and polarization
degrees of the field stars gfs1 and gfs3, as well as gfs9 and gfs10. In this
scenario, the high polarization observed in EXO 2030+375 would not be intrinsic,
but rather the result of an inaccurate correction for ISM polarization. An
alternative physical explanation for the unusually large PD in EXO 2030+375 was
proposed by \citet{reig14d}, who speculated that the high polarization may arise
from the alignment of non-spherical ferromagnetic grains in the Be-star disk
under the influence of the neutron star's strong magnetic field. 

The scatter is larger among the short-period systems ($P_{\rm orb}\simless 50$
d). This is also observed in the \ha\ equivalent width versus orbital period
diagram \citep{reig16} and also in the X-ray band \citep{reig07a}. A simple
explanation is that wider orbit systems are able to develop stable disks during
longer periods of time. In contrast, the systems with short orbital periods feel
the tidal truncation exerted by the neutron star more strongly and more
frequently, so that the disk does not easily achieve a stable configuration on
shorter timescales. 

\citet{yudin01} performed a statistical analysis of the intrinsic polarization
of 497 isolated Be stars covering the entire spectral range. He found that for
the early-type group (B0-B3), which is the relevant group for comparison with
BeXBs, the values show larger scatter but  on average, Be stars in this spectral
range exhibit larger values of polarization. The maximum values of intrinsic
polarization for B0-B3 stars are $PD_*(\%)\simless 2$\%. In \citet{yudin01}
sample,  only 5 stars exhibited polarization degree above 2\% (all in the B0-B2
spectral range). The average intrinsic polarization for different spectral
subtypes in Yudin's sample are: $0.84\pm0.48$\% for O-B1.5 and $0.75\pm0.48$\%
for B2-B2.5. 

We find that about half of our sample of BeXBs display intrinsic maximum
polarization degree above 2\%.  Figure~\ref{PDmax-hist} shows the normalized
histogram of the maximum measured intrinsic PD in BeXBs compared to classical Be
stars of similar spectral type range \citep[data from][]{yudin01}. Since the
polarization degree reflects the number of scatterers, models predict that as
the density increases the polarization level increases because the number of
electrons available for Thomson scattering increases  \citep{halonen13c}.
Therefore, our findings support the idea that disks in BeXBs are denser than in
isolated Be stars.

\section{Conclusion}

We have performed a systematic study of the polarization properties of BeXBs
visible from the northern hemisphere to characterized their long-term optical
variability and to probe the dynamics and structure of their circumstellar
disks. Our findings confirm that BeXBs are intrinsically variable in
polarization, with changes in polarization degree and angle directly tracing the
evolution of the disk size and geometry.
A key result of this work is the confirmation that circumstellar disks in BeXBs
are, on average, smaller and significantly denser than those found in isolated
Be stars. This is evidenced by the higher intrinsic polarization values observed
in our sample, with approximately half of the targets exceeding the typical
maximum polarization of isolated Be stars. This finding strongly supports
theoretical models of disk truncation, where the gravitational influence of the
neutron star tidally truncates and confines the Be star's disk.
We also find further evidence linking giant X-ray outbursts to the presence
of warped and precessing disks, where changes in the polarization angle,
indicating a distorted disk, appear to be a necessary precursor to those
high-energy events. 

In conclusion, this study highlights the importance of polarimetry as a
diagnostic tool, together with spectroscopy and photometry, to investigate the
physical mechanisms governing the behavior of these complex binary systems.

\section{Data availability}

The polarization measurements, including $q$, $u$, $PD$, and $EVPA$ with the
corresponding uncertainties, are available on the Zenodo repository
https://doi.org/10.5281/zenodo.18346735.

\begin{acknowledgements}

Skinakas Observatory is a collaborative project of the University of Crete and
the Foundation for Research and Technology-Hellas.  D.B. acknowledges support
from the European Research Council (ERC) under the Horizon ERC Grants 2021
programme under grant agreement No. 101040021. This work made use of NASA's
Astrophysics Data System Bibliographic Services and of the SIMBAD database,
operated at the CDS, Strasbourg, France. This work has made use of data from the
European Space Agency (ESA) mission {\it Gaia}
(\url{https://www.cosmos.esa.int/gaia})

\end{acknowledgements}

\bibliographystyle{aa}
\bibliography{./artBex_bib}

@ARTICLE{aragona09,
   author = {{Aragona}, C. and {McSwain}, M.~V. and {Grundstrom}, E.~D. and 
	{Marsh}, A.~N. and {Roettenbacher}, R.~M. and {Hessler}, K.~M. and 
	{Boyajian}, T.~S. and {Ray}, P.~S.},
    title = "{The Orbits of the {$\gamma$}-Ray Binaries LS I +61 303 and LS 5039}",
  journal = {\apj},
archivePrefix = "arXiv",
   eprint = {0902.4015},
 primaryClass = "astro-ph.HE",
     year = 2009,
    month = jun,
   volume = 698,
    pages = {514-518},
      doi = {10.1088/0004-637X/698/1/514},
   adsurl = {http://cdsads.u-strasbg.fr/abs/2009ApJ...698..514A}
}

@ARTICLE{bailer-jones21,
       author = {{Bailer-Jones}, C.~A.~L. and {Rybizki}, J. and {Fouesneau}, M. and {Demleitner}, M. and {Andrae}, R.},
        title = "{Estimating Distances from Parallaxes. V. Geometric and Photogeometric Distances to 1.47 Billion Stars in Gaia Early Data Release 3}",
      journal = {\aj},
         year = 2021,
        month = mar,
       volume = {161},
       number = {3},
          eid = {147},
        pages = {147},
          doi = {10.3847/1538-3881/abd806},
archivePrefix = {arXiv},
       eprint = {2012.05220},
 primaryClass = {astro-ph.SR},
       adsurl = {https://ui.adsabs.harvard.edu/abs/2021AJ....161..147B}
}

@ARTICLE{balona03,
       author = {{Balona}, Luis A.},
        title = "{Mode Identification from Line Profiles using the Direct Fitting Technique}",
      journal = {\apss},
         year = 2003,
        month = jan,
       volume = {284},
       number = {1},
        pages = {121-124},
       adsurl = {https://ui.adsabs.harvard.edu/abs/2003Ap&SS.284..121B}
}

@ARTICLE{balona21,
       author = {{Balona}, Luis A. and {Ozuyar}, Dogus},
        title = "{TESS Observations of Be Stars: General Characteristics and the Impulsive Magnetic Rotator Model}",
      journal = {\apj},
         year = 2021,
        month = nov,
       volume = {921},
       number = {1},
          eid = {5},
        pages = {5},
          doi = {10.3847/1538-4357/ac1a77},
archivePrefix = {arXiv},
       eprint = {2008.06288},
 primaryClass = {astro-ph.SR},
       adsurl = {https://ui.adsabs.harvard.edu/abs/2021ApJ...921....5B}
}

@ARTICLE{bartlett25,
       author = {{Bartlett}, Jordan A. and {Kobulnicky}, Henry A.},
        title = "{Extreme Starlight Polarization Efficiency toward {\ensuremath{\zeta}} Ophiuchi: A Case for Line-of-sight Foreground Subtraction}",
      journal = {\aj},
         year = 2025,
        month = nov,
       volume = {170},
       number = {5},
          eid = {265},
        pages = {265},
          doi = {10.3847/1538-3881/ae01a2},
archivePrefix = {arXiv},
       eprint = {2509.04427},
 primaryClass = {astro-ph.GA},
       adsurl = {https://ui.adsabs.harvard.edu/abs/2025AJ....170..265B}
}

@INPROCEEDINGS{bastien07,
       author = {{Bastien}, P. and {Vernet}, E. and {Drissen}, L. and {M{\'e}nard}, F. and {Moffat}, A.~F.~J. and {Robert}, C. and {St-Louis}, N.},
        title = "{The Variability of Polarized Standard Stars}",
    booktitle = {The Future of Photometric, Spectrophotometric and Polarimetric Standardization},
         year = 2007,
       editor = {{Sterken}, C.},
       series = {Astronomical Society of the Pacific Conference Series},
       volume = {364},
        month = apr,
        pages = {529},
       adsurl = {https://ui.adsabs.harvard.edu/abs/2007ASPC..364..529B}
}

@ARTICLE{baykal00,
   author = {{Baykal}, A. and {Stark}, M.~J. and {Swank}, J.},
    title = "{Discovery of the Orbit of the Transient X-Ray Pulsar SAX J2103.5+4545}",
  journal = {\apjl},
   eprint = {astro-ph/0009481},
     year = 2000,
    month = dec,
   volume = 544,
    pages = {L129-L132},
      doi = {10.1086/317320},
   adsurl = {http://cdsads.u-strasbg.fr/abs/2000ApJ...544L.129B}
}

@ARTICLE{blay06,
   author = {{Blay}, P. and {Negueruela}, I. and {Reig}, P. and {Coe}, M.~J. and 
	{Corbet}, R.~H.~D. and {Fabregat}, J. and {Tarasov}, A.~E.},
    title = "{Multiwavelength monitoring of <ASTROBJ>BD +53{\deg}2790</ASTROBJ>, the optical counterpart to <ASTROBJ>4U 2206+54</ASTROBJ>}",
  journal = {\aap},
   eprint = {astro-ph/0510400},
     year = 2006,
    month = feb,
   volume = 446,
    pages = {1095-1105},
      doi = {10.1051/0004-6361:20053951},
   adsurl = {http://cdsads.u-strasbg.fr/abs/2006A%26A...446.1095B}
}

@ARTICLE{blinov21,
       author = {{Blinov}, D. and {Kiehlmann}, S. and {Pavlidou}, V. and {Panopoulou}, G.~V. and {Skalidis}, R. and {Angelakis}, E. and {Casadio}, C. and {Einoder}, E.~N. and {Hovatta}, T. and {Kokolakis}, K. and {Kougentakis}, A. and {Kus}, A. and {Kylafis}, N. and {Kyritsis}, E. and {Lalakos}, A. and {Liodakis}, I. and {Maharana}, S. and {Makrydopoulou}, E. and {Mandarakas}, N. and {Maragkakis}, G.~M. and {Myserlis}, I. and {Papadakis}, I. and {Paterakis}, G. and {Pearson}, T.~J. and {Ramaprakash}, A.~N. and {Readhead}, A.~C.~S. and {Reig}, P. and {S{\l}owikowska}, A. and {Tassis}, K. and {Xexakis}, K. and {{\.Z}ejmo}, M. and {Zensus}, J.~A.},
        title = "{RoboPol: AGN polarimetric monitoring data}",
      journal = {\mnras},
         year = 2021,
        month = mar,
       volume = {501},
       number = {3},
        pages = {3715-3726},
          doi = {10.1093/mnras/staa3777},
archivePrefix = {arXiv},
       eprint = {2012.00008},
 primaryClass = {astro-ph.HE},
       adsurl = {https://ui.adsabs.harvard.edu/abs/2021MNRAS.501.3715B}
}

@ARTICLE{blinov23,
       author = {{Blinov}, D. and {Maharana}, S. and {Bouzelou}, F. and {Casadio}, C. and {Gjerl{\o}w}, E. and {Jormanainen}, J. and {Kiehlmann}, S. and {Kypriotakis}, J.~A. and {Liodakis}, I. and {Mandarakas}, N. and {Markopoulioti}, L. and {Panopoulou}, G.~V. and {Pelgrims}, V. and {Pouliasi}, A. and {Romanopoulos}, S. and {Skalidis}, R. and {Anche}, R.~M. and {Angelakis}, E. and {Antoniadis}, J. and {Medhi}, B.~J. and {Hovatta}, T. and {Kus}, A. and {Kylafis}, N. and {Mahabal}, A. and {Myserlis}, I. and {Paleologou}, E. and {Papadakis}, I. and {Pavlidou}, V. and {Papamastorakis}, I. and {Pearson}, T.~J. and {Potter}, S.~B. and {Ramaprakash}, A.~N. and {Readhead}, A.~C.~S. and {Reig}, P. and {S{\l}owikowska}, A. and {Tassis}, K. and {Zensus}, J.~A.},
        title = "{The RoboPol sample of optical polarimetric standards}",
      journal = {\aap},
         year = 2023,
        month = sep,
       volume = {677},
          eid = {A144},
        pages = {A144},
          doi = {10.1051/0004-6361/202346778},
archivePrefix = {arXiv},
       eprint = {2307.06151},
 primaryClass = {astro-ph.IM},
       adsurl = {https://ui.adsabs.harvard.edu/abs/2023A&A...677A.144B}
}

@ARTICLE{bonnet-bidaud98,
   author = {{Bonnet-Bidaud}, J.~M. and {Mouchet}, M.},
    title = "{The identification of the transient X-ray pulsar Cepheus X-4 with a Be/X-ray binary}",
  journal = {\aap},
   eprint = {astro-ph/9801215},
     year = 1998,
    month = apr,
   volume = 332,
    pages = {L9-L12},
   adsurl = {http://cdsads.u-strasbg.fr/abs/1998A%26A...332L...9B}
}

@ARTICLE{carciofi06,
   author = {{Carciofi}, A.~C. and {Bjorkman}, J.~E.},
    title = "{Non-LTE Monte Carlo Radiative Transfer. I. The Thermal Properties of Keplerian Disks around Classical Be Stars}",
  journal = {\apj},
   eprint = {astro-ph/0511228},
     year = 2006,
    month = mar,
   volume = 639,
    pages = {1081-1094},
      doi = {10.1086/499483},
   adsurl = {http://cdsads.u-strasbg.fr/abs/2006ApJ...639.1081C}
}

@ARTICLE{carciofi07,
       author = {{Carciofi}, A.~C. and {Magalh{\~a}es}, A.~M. and {Leister}, N.~V. and {Bjorkman}, J.~E. and {Levenhagen}, R.~S.},
        title = "{Achernar: Rapid Polarization Variability as Evidence of Photospheric and Circumstellar Activity}",
      journal = {\apjl},
         year = 2007,
        month = dec,
       volume = {671},
       number = {1},
        pages = {L49-L52},
          doi = {10.1086/524772},
archivePrefix = {arXiv},
       eprint = {0710.4163},
 primaryClass = {astro-ph},
       adsurl = {https://ui.adsabs.harvard.edu/abs/2007ApJ...671L..49C}
}

@INPROCEEDINGS{carciofi11,
   author = {{Carciofi}, A.~C.},
    title = "{The circumstellar discs of Be stars}",
booktitle = {IAU Symposium},
     year = 2011,
   series = {IAU Symposium},
   volume = 272,
archivePrefix = "arXiv",
   eprint = {1009.3969},
 primaryClass = "astro-ph.SR",
   editor = {{Neiner}, C. and {Wade}, G. and {Meynet}, G. and {Peters}, G.
	},
    month = jul,
    pages = {325-336},
      doi = {10.1017/S1743921311010738},
   adsurl = {http://cdsads.u-strasbg.fr/abs/2011IAUS..272..325C}
}

@ARTICLE{carciofi25,
       author = {{Carciofi}, Alex C. and {Bolzan}, Guilherme P.~P. and {Querido}, P{\^a}mela R. and {Rubio}, Amanda C. and {Labadie-Bartz}, Jonathan and {de Amorim}, Tajan H. and {Fonseca Silva}, Ariane C. and {Schiavolim}, Vitt{\'o}ria L.},
        title = "{Mass Loss in Be Stars: News from Two Fronts}",
      journal = {Galaxies},
         year = 2025,
        month = jul,
       volume = {13},
       number = {4},
          eid = {77},
        pages = {77},
          doi = {10.3390/galaxies13040077},
       adsurl = {https://ui.adsabs.harvard.edu/abs/2025Galax..13...77C}
}

@ARTICLE{chhotaray24,
       author = {{Chhotaray}, Birendra and {Naik}, Sachindra and {Jaisawal}, Gaurava K. and {Ahuja}, Goldy},
        title = "{Optical and X-ray studies of the Be/X-ray binary IGR J06074+2205}",
      journal = {\mnras},
         year = 2024,
        month = nov,
       volume = {534},
       number = {3},
        pages = {2830-2847},
          doi = {10.1093/mnras/stae2282},
archivePrefix = {arXiv},
       eprint = {2410.00747},
 primaryClass = {astro-ph.HE},
       adsurl = {https://ui.adsabs.harvard.edu/abs/2024MNRAS.534.2830C}
}

@ARTICLE{clarke94,
       author = {{Clarke}, D. and {Naghizadeh-Khouei}, J.},
        title = "{A Reassessment of Some Polarization Standard Stars}",
      journal = {\aj},
         year = 1994,
        month = aug,
       volume = {108},
        pages = {687},
          doi = {10.1086/117105},
       adsurl = {https://ui.adsabs.harvard.edu/abs/1994AJ....108..687C}
}

@ARTICLE{corbet07,
       author = {{Corbet}, R.~H.~D. and {Markwardt}, C.~B. and {Tueller}, J.},
        title = "{Swift BAT and RXTE Observations of the Peculiar X-Ray Binary 4U 2206+54: Disappearance of the 9.6 Day Modulation}",
      journal = {\apj},
         year = 2007,
        month = jan,
       volume = {655},
       number = {1},
        pages = {458-465},
          doi = {10.1086/509319},
archivePrefix = {arXiv},
       eprint = {astro-ph/0609328},
 primaryClass = {astro-ph},
       adsurl = {https://ui.adsabs.harvard.edu/abs/2007ApJ...655..458C}
}

@ARTICLE{coyne69,
       author = {{Coyne}, G.~V. and {Kruszewski}, A.},
        title = "{Wavelength dependence of polarization. XVII. Be-type stars.}",
      journal = {\aj},
         year = 1969,
        month = may,
       volume = {74},
        pages = {528-532},
          doi = {10.1086/110830},
       adsurl = {https://ui.adsabs.harvard.edu/abs/1969AJ.....74..528C}
}

@ARTICLE{coyne74,
   author = {{Coyne}, G.~V. and {Gehrels}, T. and {Serkowski}, K.},
    title = "{Wavelength dependence of polarization. XXVI. The wavelength of maximum polarization as a characteristic parameter of interstellar grains.}",
  journal = {\aj},
     year = 1974,
    month = may,
   volume = 79,
    pages = {581-589},
      doi = {10.1086/111578},
   adsurl = {http://cdsads.u-strasbg.fr/abs/1974AJ.....79..581C}
}

@ARTICLE{cyr17,
       author = {{Cyr}, I.~H. and {Jones}, C.~E. and {Panoglou}, D. and {Carciofi}, A.~C. and {Okazaki}, A.~T.},
        title = "{Be discs in binary systems - II. Misaligned orbits}",
      journal = {\mnras},
         year = 2017,
        month = oct,
       volume = {471},
       number = {1},
        pages = {596-605},
          doi = {10.1093/mnras/stx1427},
archivePrefix = {arXiv},
       eprint = {1706.07029},
 primaryClass = {astro-ph.SR},
       adsurl = {https://ui.adsabs.harvard.edu/abs/2017MNRAS.471..596C}
}

@ARTICLE{doroshenko16,
       author = {{Doroshenko}, V. and {Tsygankov}, S. and {Santangelo}, A.},
        title = "{Orbital parameters of V 0332+53 from 2015 giant outburst data}",
      journal = {\aap},
         year = 2016,
        month = may,
       volume = {589},
          eid = {A72},
        pages = {A72},
          doi = {10.1051/0004-6361/201527756},
archivePrefix = {arXiv},
       eprint = {1509.04490},
 primaryClass = {astro-ph.HE},
       adsurl = {https://ui.adsabs.harvard.edu/abs/2016A&A...589A..72D}
}

@ARTICLE{draper14,
       author = {{Draper}, Zachary H. and {Wisniewski}, John P. and {Bjorkman}, Karen S. and {Meade}, Marilyn R. and {Haubois}, Xavier and {Mota}, Bruno C. and {Carciofi}, Alex C. and {Bjorkman}, Jon E.},
        title = "{Disk-loss and Disk-renewal Phases in Classical Be Stars. II. Contrasting with Stable and Variable Disks}",
      journal = {\apj},
         year = 2014,
        month = may,
       volume = {786},
       number = {2},
          eid = {120},
        pages = {120},
          doi = {10.1088/0004-637X/786/2/120},
archivePrefix = {arXiv},
       eprint = {1402.5240},
 primaryClass = {astro-ph.SR},
       adsurl = {https://ui.adsabs.harvard.edu/abs/2014ApJ...786..120D}
}

@ARTICLE{dubus13,
   author = {{Dubus}, G.},
    title = "{Gamma-ray binaries and related systems}",
  journal = {\aapr},
archivePrefix = "arXiv",
   eprint = {1307.7083},
 primaryClass = "astro-ph.HE",
     year = 2013,
    month = aug,
   volume = 21,
      eid = {64},
    pages = {64},
      doi = {10.1007/s00159-013-0064-5},
   adsurl = {http://cdsads.u-strasbg.fr/abs/2013A%26ARv..21...64D}
}

@ARTICLE{esposito13,
   author = {{Esposito}, P. and {Israel}, G.~L. and {Sidoli}, L. and {Mason}, E. and 
	{Rodr{\'{\i}}guez Castillo}, G.~A. and {Halpern}, J.~P. and 
	{Moretti}, A. and {G{\"o}tz}, D.},
    title = "{Discovery of 47-s pulsations in the X-ray source 1RXS J225352.8+624354}",
  journal = {\mnras},
archivePrefix = "arXiv",
   eprint = {1305.3439},
 primaryClass = "astro-ph.HE",
     year = 2013,
    month = aug,
   volume = 433,
    pages = {2028-2035},
      doi = {10.1093/mnras/stt870},
   adsurl = {http://cdsads.u-strasbg.fr/abs/2013MNRAS.433.2028E}
}

@ARTICLE{ferrigno13,
   author = {{Ferrigno}, C. and {Farinelli}, R. and {Bozzo}, E. and {Pottschmidt}, K. and 
	{Klochkov}, D. and {Kretschmar}, P.},
    title = "{RX J0440.9 + 4431: a persistent Be/X-ray binary in outburst}",
  journal = {\aap},
archivePrefix = "arXiv",
   eprint = {1303.7087},
 primaryClass = "astro-ph.HE",
     year = 2013,
    month = may,
   volume = 553,
      eid = {A103},
    pages = {A103},
      doi = {10.1051/0004-6361/201321053},
   adsurl = {http://cdsads.u-strasbg.fr/abs/2013A%26A...553A.103F}
}

@ARTICLE{fosalba02,
   author = {{Fosalba}, P. and {Lazarian}, A. and {Prunet}, S. and {Tauber}, J.~A.
	},
    title = "{Statistical Properties of Galactic Starlight Polarization}",
  journal = {\apj},
   eprint = {astro-ph/0105023},
     year = 2002,
    month = jan,
   volume = 564,
    pages = {762-772},
      doi = {10.1086/324297},
   adsurl = {http://cdsads.u-strasbg.fr/abs/2002ApJ...564..762F}
}

@ARTICLE{galloway04,
   author = {{Galloway}, D.~K. and {Morgan}, E.~H. and {Levine}, A.~M.},
    title = "{A Frequency Glitch in an Accreting Pulsar}",
  journal = {\apj},
   eprint = {astro-ph/0401476},
     year = 2004,
    month = oct,
   volume = 613,
    pages = {1164-1172},
      doi = {10.1086/423265},
   adsurl = {http://cdsads.u-strasbg.fr/abs/2004ApJ...613.1164G}
}

@ARTICLE{green19,
       author = {{Green}, Gregory M. and {Schlafly}, Edward and {Zucker}, Catherine and {Speagle}, Joshua S. and {Finkbeiner}, Douglas},
        title = "{A 3D Dust Map Based on Gaia, Pan-STARRS 1, and 2MASS}",
      journal = {\apj},
         year = 2019,
        month = dec,
       volume = {887},
       number = {1},
          eid = {93},
        pages = {93},
          doi = {10.3847/1538-4357/ab5362},
archivePrefix = {arXiv},
       eprint = {1905.02734},
 primaryClass = {astro-ph.GA},
       adsurl = {https://ui.adsabs.harvard.edu/abs/2019ApJ...887...93G}
}

@ARTICLE{grundstrom06,
   author = {{Grundstrom}, E.~D. and {Gies}, D.~R.},
    title = "{Estimating Be Star Disk Radii using H{$\alpha$} Emission Equivalent Widths}",
  journal = {\apjl},
   eprint = {arXiv:astro-ph/0609602},
     year = 2006,
    month = nov,
   volume = 651,
    pages = {L53-L56},
      doi = {10.1086/509635},
   adsurl = {http://cdsads.u-strasbg.fr/abs/2006ApJ...651L..53G}
}

@ARTICLE{grundstrom07b,
   author = {{Grundstrom}, E.~D. and {Boyajian}, T.~S. and {Finch}, C. and 
	{Gies}, D.~R. and {Huang}, W. and {McSwain}, M.~V. and {O'Brien}, D.~P. and 
	{Riddle}, R.~L. and {Trippe}, M.~L. and {Williams}, S.~J. and 
	{Wingert}, D.~W. and {Zaballa}, R.~A.},
    title = "{Joint H{$\alpha$} and X-Ray Observations of Massive X-Ray Binaries. III. The Be X-Ray Binaries HDE 245770 = A0535+26 and X Persei}",
  journal = {\apj},
   eprint = {arXiv:astro-ph/0702283},
     year = 2007,
    month = may,
   volume = 660,
    pages = {1398-1408},
      doi = {10.1086/514325},
   adsurl = {http://cdsads.u-strasbg.fr/abs/2007ApJ...660.1398G}
}

@ARTICLE{halonen13a,
   author = {{Halonen}, R.~J. and {Mackay}, F.~E. and {Jones}, C.~E.},
    title = "{Computing the Continuum Polarization from Thomson Scattering in Gaseous Circumstellar Disks}",
  journal = {\apjs},
     year = 2013,
    month = jan,
   volume = 204,
      eid = {11},
    pages = {11},
      doi = {10.1088/0067-0049/204/1/11},
   adsurl = {http://cdsads.u-strasbg.fr/abs/2013ApJS..204...11H}
}

@ARTICLE{halonen13b,
   author = {{Halonen}, R.~J. and {Jones}, C.~E.},
    title = "{On the Intrinsic Continuum Linear Polarization of Classical Be Stars: The Effects of Metallicity and One-armed Density Perturbations}",
  journal = {\apjs},
archivePrefix = "arXiv",
   eprint = {1307.6220},
 primaryClass = "astro-ph.SR",
     year = 2013,
    month = sep,
   volume = 208,
      eid = {3},
    pages = {3},
      doi = {10.1088/0067-0049/208/1/3},
   adsurl = {http://cdsads.u-strasbg.fr/abs/2013ApJS..208....3H}
}

@ARTICLE{halonen13c,
   author = {{Halonen}, R.~J. and {Jones}, C.~E.},
    title = "{On the Intrinsic Continuum Linear Polarization of Classical Be Stars during Disk Growth and Dissipation}",
  journal = {\apj},
archivePrefix = "arXiv",
   eprint = {1307.6207},
 primaryClass = "astro-ph.SR",
     year = 2013,
    month = mar,
   volume = 765,
      eid = {17},
    pages = {17},
      doi = {10.1088/0004-637X/765/1/17},
   adsurl = {http://cdsads.u-strasbg.fr/abs/2013ApJ...765...17H}
}

@ARTICLE{haubois14,
   author = {{Haubois}, X. and {Mota}, B.~C. and {Carciofi}, A.~C. and {Draper}, Z.~H. and 
	{Wisniewski}, J.~P. and {Bednarski}, D. and {Rivinius}, T.},
    title = "{Dynamical Evolution of Viscous Disks around Be Stars. II. Polarimetry}",
  journal = {\apj},
archivePrefix = "arXiv",
   eprint = {1402.1968},
 primaryClass = "astro-ph.SR",
     year = 2014,
    month = apr,
   volume = 785,
      eid = {12},
    pages = {12},
      doi = {10.1088/0004-637X/785/1/12},
   adsurl = {http://cdsads.u-strasbg.fr/abs/2014ApJ...785...12H}
}

@ARTICLE{haigh04,
   author = {{Haigh}, N.~J. and {Coe}, M.~J. and {Fabregat}, J.},
    title = "{Cyclical behaviour and disc truncation in the Be/X-ray binary A0535+26}",
  journal = {\mnras},
   eprint = {arXiv:astro-ph/0305194},
     year = 2004,
    month = jun,
   volume = 350,
    pages = {1457-1466},
      doi = {10.1111/j.1365-2966.2004.07743.x},
   adsurl = {http://cdsads.u-strasbg.fr/abs/2004MNRAS.350.1457H}
}

@ARTICLE{hiltner56,
   author = {{Hiltner}, W.~A.},
    title = "{Photometric, Polarization, and Spectrographic Observations of O and B Stars.}",
  journal = {\apjs},
     year = 1956,
    month = oct,
   volume = 2,
    pages = {389},
      doi = {10.1086/190029},
   adsurl = {http://cdsads.u-strasbg.fr/abs/1956ApJS....2..389H}
}

@ARTICLE{hummel98,
       author = {{Hummel}, W.},
        title = "{On the spectacular variations of Be stars. Evidence for a temporarily tilted circumstellar disk}",
      journal = {\aap},
         year = 1998,
        month = feb,
       volume = {330},
        pages = {243-252},
       adsurl = {https://ui.adsabs.harvard.edu/abs/1998A&A...330..243H}
}

@ARTICLE{ignace25,
       author = {{Ignace}, Richard and {Fullard}, Andrew G. and {Panopoulou}, Georgia V. and {Hillier}, D. John and {Erba}, Christiana and {Scowen}, Paul A.},
        title = "{Analyzing stellar and interstellar contributions to polarization: modeling approaches for hot stars}",
      journal = {\apss},
         year = 2025,
        month = jun,
       volume = {370},
       number = {6},
          eid = {57},
        pages = {57},
          doi = {10.1007/s10509-025-04445-4},
archivePrefix = {arXiv},
       eprint = {2505.15028},
 primaryClass = {astro-ph.SR},
       adsurl = {https://ui.adsabs.harvard.edu/abs/2025Ap&SS.370...57I}
}

@ARTICLE{intzand01,
   author = {{in't Zand}, J.~J.~M. and {Swank}, J. and {Corbet}, R.~H.~D. and 
	{Markwardt}, C.~B.},
    title = "{Discovery of a 1247 s pulsar in the Be X-ray binary SAX J2239.3+6116}",
  journal = {\aap},
   eprint = {astro-ph/0110695},
     year = 2001,
    month = dec,
   volume = 380,
    pages = {L26-L29},
      doi = {10.1051/0004-6361:20011512},
   adsurl = {http://cdsads.u-strasbg.fr/abs/2001A%26A...380L..26I}
}

@ARTICLE{jones89,
   author = {{Jones}, T.~J.},
    title = "{Infrared polarimetry and the interstellar magnetic field}",
  journal = {\apj},
     year = 1989,
    month = nov,
   volume = 346,
    pages = {728-734},
      doi = {10.1086/168054},
   adsurl = {http://cdsads.u-strasbg.fr/abs/1989ApJ...346..728J}
}

@ARTICLE{jones08,
   author = {{Jones}, C.~E. and {Sigut}, T.~A.~A. and {Porter}, J.~M.},
    title = "{The circumstellar envelopes of Be stars: viscous disc dynamics}",
  journal = {\mnras},
     year = 2008,
    month = jun,
   volume = 386,
    pages = {1922-1930},
      doi = {10.1111/j.1365-2966.2008.13206.x},
   adsurl = {http://cdsads.u-strasbg.fr/abs/2008MNRAS.386.1922J}
}

@ARTICLE{kaur08,
       author = {{Kaur}, Ramanpreet and {Paul}, Biswajit and {Kumar}, Brijesh and {Sagar}, Ram},
        title = "{Multiwavelength study of the transient X-ray binary IGR J01583+6713}",
      journal = {\mnras},
         year = 2008,
        month = jun,
       volume = {386},
       number = {4},
        pages = {2253-2261},
          doi = {10.1111/j.1365-2966.2008.13233.x},
archivePrefix = {arXiv},
       eprint = {0803.1113},
 primaryClass = {astro-ph},
       adsurl = {https://ui.adsabs.harvard.edu/abs/2008MNRAS.386.2253K}
}

@ARTICLE{king14,
   author = {{King}, O.~G. and {Blinov}, D. and {Ramaprakash}, A.~N. and 
	{Myserlis}, I. and {Angelakis}, E. and {Balokovi{\'c}}, M. and 
	{Feiler}, R. and {Fuhrmann}, L. and {Hovatta}, T. and {Khodade}, P. and 
	{Kougentakis}, A. and {Kylafis}, N. and {Kus}, A. and {Modi}, D. and 
	{Paleologou}, E. and {Panopoulou}, G. and {Papadakis}, I. and 
	{Papamastorakis}, I. and {Paterakis}, G. and {Pavlidou}, V. and 
	{Pazderska}, B. and {Pazderski}, E. and {Pearson}, T.~J. and 
	{Rajarshi}, C. and {Readhead}, A.~C.~S. and {Reig}, P. and {Steiakaki}, A. and 
	{Tassis}, K. and {Zensus}, J.~A.},
    title = "{The RoboPol pipeline and control system}",
  journal = {\mnras},
archivePrefix = "arXiv",
   eprint = {1310.7555},
 primaryClass = "astro-ph.IM",
     year = 2014,
    month = aug,
   volume = 442,
    pages = {1706-1717},
      doi = {10.1093/mnras/stu176},
   adsurl = {http://cdsads.u-strasbg.fr/abs/2014MNRAS.442.1706K}
}

@ARTICLE{kiziloglu07b,
   author = {{K{\i}z{\i}lo{\v g}lu}, U. and {K{\i}z{\i}lo{\v g}lu}, N. and 
	{Baykal}, A. and {Yerli}, S.~K. and {{\"O}zbey}, M.},
    title = "{Optical variabilities in the Be/X-ray binary system. GRO J2058+42 (CXOU J205847.5+414637)}",
  journal = {\aap},
archivePrefix = "arXiv",
   eprint = {0705.1873},
     year = 2007,
    month = aug,
   volume = 470,
    pages = {1023-1029},
      doi = {10.1051/0004-6361:20077365},
   adsurl = {http://adsabs.harvard.edu/abs/2007A%26A...470.1023K}
}

@ARTICLE{kravtsov20,
       author = {{Kravtsov}, Vadim and {Berdyugin}, Andrei V. and {Piirola}, Vilppu and {Kosenkov}, Ilia A. and {Tsygankov}, Sergey S. and {Chernyakova}, Maria and {Malyshev}, Denys and {Sakanoi}, Takeshi and {Kagitani}, Masato and {Berdyugina}, Svetlana V. and {Poutanen}, Juri},
        title = "{Orbital variability of the optical linear polarization of the {\ensuremath{\gamma}}-ray binary LS I +61{\textdegree} 303 and new constraints on the orbital parameters}",
      journal = {\aap},
         year = 2020,
        month = nov,
       volume = {643},
          eid = {A170},
        pages = {A170},
          doi = {10.1051/0004-6361/202038745},
archivePrefix = {arXiv},
       eprint = {2010.00999},
 primaryClass = {astro-ph.SR},
       adsurl = {https://ui.adsabs.harvard.edu/abs/2020A&A...643A.170K}
}

@INCOLLECTION{lazarian15,
       author = {{Lazarian}, Alexander and {Andersson}, B.-G. and {Hoang}, Thiem},
        title = "{Grain alignment: Role of radiative torques and paramagnetic relaxation}",
    booktitle = {Polarimetry of Stars and Planetary Systems},
         year = 2015,
       editor = {{Kolokolova}, Ludmilla and {Hough}, James and {Levasseur-Regourd}, Anny-Chantal},
        pages = {81},
          doi = {10.48550/arXiv.1511.03696},
       adsurl = {https://ui.adsabs.harvard.edu/abs/2015psps.book...81L}
}

@ARTICLE{liu25,
       author = {{Liu}, Wei and {Reig}, Pablo and {Yan}, Jingzhi and {Zhang}, Peng and {Li}, Xiukun and {Gao}, Bo and {Xiao}, Guangcheng and {Li}, Zhongmu and {Liu}, Qingzhong},
        title = "{Unraveling the Mysteries of KS 1947+300: Multiwavelength Observations in a Low-eccentricity Be/X-Ray Binary}",
      journal = {\apj},
         year = 2025,
        month = jun,
       volume = {985},
       number = {2},
          eid = {162},
        pages = {162},
          doi = {10.3847/1538-4357/adcac3},
       adsurl = {https://ui.adsabs.harvard.edu/abs/2025ApJ...985..162L}
}

@ARTICLE{mandarakas25,
       author = {{Mandarakas}, N. and {Tassis}, K. and {Skalidis}, R.},
        title = "{3D interstellar medium structure challenges the Serkowski relation}",
      journal = {\aap},
         year = 2025,
        month = jun,
       volume = {698},
          eid = {A168},
        pages = {A168},
          doi = {10.1051/0004-6361/202452263},
archivePrefix = {arXiv},
       eprint = {2409.10317},
 primaryClass = {astro-ph.GA},
       adsurl = {https://ui.adsabs.harvard.edu/abs/2025A&A...698A.168M}
}

@ARTICLE{marcu15,
   author = {{Marcu-Cheatham}, D.~M. and {Pottschmidt}, K. and {K{\"u}hnel}, M. and 
	{M{\"u}ller}, S. and {Falkner}, S. and {Caballero}, I. and {Finger}, M.~H. and 
	{Jenke}, P.~J. and {Wilson-Hodge}, C.~A. and {F{\"u}rst}, F. and 
	{Grinberg}, V. and {Hemphill}, P.~B. and {Kreykenbohm}, I. and 
	{Klochkov}, D. and {Rothschild}, R.~E. and {Terada}, Y. and 
	{Enoto}, T. and {Iwakiri}, W. and {Wolff}, M.~T. and {Becker}, P.~A. and 
	{Wood}, K.~S. and {Wilms}, J.},
    title = "{The Transient Accreting X-Ray Pulsar XTE J1946+274: Stability of X-Ray Properties at Low Flux and Updated Orbital Solution}",
  journal = {\apj},
archivePrefix = "arXiv",
   eprint = {1510.05032},
 primaryClass = "astro-ph.HE",
     year = 2015,
    month = dec,
   volume = 815,
      eid = {44},
    pages = {44},
      doi = {10.1088/0004-637X/815/1/44},
   adsurl = {http://cdsads.u-strasbg.fr/abs/2015ApJ...815...44M}
}

@ARTICLE{marr18,
       author = {{Marr}, K.~C. and {Jones}, C.~E. and {Halonen}, R.~J.},
        title = "{Computing the Polarimetric and Photometric Variability of Be Stars}",
      journal = {\apj},
         year = 2018,
        month = jan,
       volume = {852},
       number = {2},
          eid = {103},
        pages = {103},
          doi = {10.3847/1538-4357/aaa0d0},
archivePrefix = {arXiv},
       eprint = {1712.02858},
 primaryClass = {astro-ph.SR},
       adsurl = {https://ui.adsabs.harvard.edu/abs/2018ApJ...852..103M}
}

@ARTICLE{martin11,
   author = {{Martin}, R.~G. and {Pringle}, J.~E. and {Tout}, C.~A. and {Lubow}, S.~H.
	},
    title = "{Tidal warping and precession of Be star decretion discs}",
  journal = {\mnras},
archivePrefix = "arXiv",
   eprint = {1106.2591},
 primaryClass = "astro-ph.SR",
     year = 2011,
    month = oct,
   volume = 416,
    pages = {2827-2839},
      doi = {10.1111/j.1365-2966.2011.19231.x},
   adsurl = {http://cdsads.u-strasbg.fr/abs/2011MNRAS.416.2827M}
}

@ARTICLE{martin14a,
   author = {{Martin}, R.~G. and {Nixon}, C. and {Armitage}, P.~J. and {Lubow}, S.~H. and 
	{Price}, D.~J.},
    title = "{Giant Outbursts in Be/X-Ray Binaries}",
  journal = {\apjl},
archivePrefix = "arXiv",
   eprint = {1407.5676},
 primaryClass = "astro-ph.HE",
     year = 2014,
    month = aug,
   volume = 790,
      eid = {L34},
    pages = {L34},
      doi = {10.1088/2041-8205/790/2/L34},
   adsurl = {http://cdsads.u-strasbg.fr/abs/2014ApJ...790L..34M}
}

@ARTICLE{martin21,
       author = {{Martin}, Rebecca G. and {Franchini}, Alessia},
        title = "{Nonperiodic Type I Be/X-Ray Binary Outbursts}",
      journal = {\apjl},
         year = 2021,
        month = dec,
       volume = {922},
       number = {2},
          eid = {L37},
        pages = {L37},
          doi = {10.3847/2041-8213/ac3a05},
archivePrefix = {arXiv},
       eprint = {2111.08642},
 primaryClass = {astro-ph.HE},
       adsurl = {https://ui.adsabs.harvard.edu/abs/2021ApJ...922L..37M}
}

@ARTICLE{martin25,
       author = {{Martin}, Rebecca G. and {Lubow}, Stephen H. and {Vallet}, David and {Overton}, Madeline and {Lepp}, Stephen and {Zhu}, Zhaohuan},
        title = "{Formation of Be star decretion discs through boundary layer effects}",
      journal = {\mnras},
         year = 2025,
        month = may,
       volume = {539},
       number = {1},
        pages = {L31-L37},
          doi = {10.1093/mnrasl/slaf019},
archivePrefix = {arXiv},
       eprint = {2503.02014},
 primaryClass = {astro-ph.SR},
       adsurl = {https://ui.adsabs.harvard.edu/abs/2025MNRAS.539L..31M}
}

@ARTICLE{mcdavid01,
   author = {{McDavid}, D.},
    title = "{A Useful Approximation for Computing the Continuum Polarization of Be Stars}",
  journal = {\apj},
   eprint = {astro-ph/0101510},
     year = 2001,
    month = jun,
   volume = 553,
    pages = {1027-1035},
      doi = {10.1086/320945},
   adsurl = {http://cdsads.u-strasbg.fr/abs/2001ApJ...553.1027M}
}

@ARTICLE{moritani11,
   author = {{Moritani}, Y. and {Nogami}, D. and {Okazaki}, A.~T. and {Imada}, A. and 
	{Kambe}, E. and {Honda}, S. and {Hashimoto}, O. and {Ichikawa}, K.
	},
    title = "{Drastic Spectroscopic Variability of the Be/X-Ray Binary Ariel 0535+262/V725 Tau during and after the 2009 Giant Outburst}",
  journal = {\pasj},
archivePrefix = "arXiv",
   eprint = {1105.4721},
 primaryClass = "astro-ph.SR",
     year = 2011,
    month = aug,
   volume = 63,
    pages = {25-29},
      doi = {10.1093/pasj/63.4.L25},
   adsurl = {http://cdsads.u-strasbg.fr/abs/2011PASJ...63L..25M}
}

@ARTICLE{moritani13,
   author = {{Moritani}, Y. and {Nogami}, D. and {Okazaki}, A.~T. and {Imada}, A. and 
	{Kambe}, E. and {Honda}, S. and {Hashimoto}, O. and {Mizoguchi}, S. and 
	{Kanda}, Y. and {Sadakane}, K. and {Ichikawa}, K.},
    title = "{Precessing Warped Be Disk Triggering the Giant Outbursts in 2009 and 2011 in A0535+262/V725Tau}",
  journal = {\pasj},
archivePrefix = "arXiv",
   eprint = {1304.4649},
 primaryClass = "astro-ph.SR",
     year = 2013,
    month = aug,
   volume = 65,
    pages = {83},
      doi = {10.1093/pasj/65.4.83},
   adsurl = {http://cdsads.u-strasbg.fr/abs/2013PASJ...65...83M}
}

@ARTICLE{naghizadeh-khouei93,
       author = {{Naghizadeh-Khouei}, J. and {Clarke}, D.},
        title = "{On the statistical behaviour of the position angle of linear polarization}",
      journal = {\aap},
         year = 1993,
        month = jul,
       volume = {274},
        pages = {968},
       adsurl = {https://ui.adsabs.harvard.edu/abs/1993A&A...274..968N}
}

@ARTICLE{negueruela99,
   author = {{Negueruela}, I. and {Roche}, P. and {Fabregat}, J. and {Coe}, M.~J.
	},
    title = "{The Be/X-ray transient V0332+53: evidence for a tilt between the orbit and the equatorial plane?}",
  journal = {\mnras},
   eprint = {arXiv:astro-ph/9903228},
     year = 1999,
    month = aug,
   volume = 307,
    pages = {695-702},
      doi = {10.1046/j.1365-8711.1999.02682.x},
   adsurl = {http://cdsads.u-strasbg.fr/abs/1999MNRAS.307..695N}
}

@ARTICLE{negueruela01a,
   author = {{Negueruela}, I. and {Okazaki}, A.~T.},
    title = "{The Be/X-ray transient 4U 0115+63/V635 Cassiopeiae. I. A consistent model}",
  journal = {\aap},
   eprint = {arXiv:astro-ph/0011407},
     year = 2001,
    month = apr,
   volume = 369,
    pages = {108-116},
      doi = {10.1051/0004-6361:20010146},
   adsurl = {http://adsabs.harvard.edu/abs/2001A%26A...369..108N}
}

@ARTICLE{negueruela01b,
   author = {{Negueruela}, I. and {Okazaki}, A.~T. and {Fabregat}, J. and 
	{Coe}, M.~J. and {Munari}, U. and {Tomov}, T.},
    title = "{The Be/X-ray transient 4U 0115+63/V635 Cassiopeiae. II. Outburst mechanisms}",
  journal = {\aap},
   eprint = {astro-ph/0101208},
     year = 2001,
    month = apr,
   volume = 369,
    pages = {117-131},
      doi = {10.1051/0004-6361:20010077},
   adsurl = {http://cdsads.u-strasbg.fr/abs/2001A%26A...369..117N}
}

@ARTICLE{nespoli12,
   author = {{Nespoli}, E. and {Reig}, P. and {Zezas}, A.},
    title = "{New insights into the Be/X-ray binary system MXB 0656-072}",
  journal = {\aap},
archivePrefix = "arXiv",
   eprint = {1205.2845},
 primaryClass = "astro-ph.HE",
     year = 2012,
    month = nov,
   volume = 547,
      eid = {A103},
    pages = {A103},
      doi = {10.1051/0004-6361/201219586},
   adsurl = {http://cdsads.u-strasbg.fr/abs/2012A%26A...547A.103N}
}

@ARTICLE{okazaki01,
   author = {{Okazaki}, A.~T. and {Negueruela}, I.},
    title = "{A natural explanation for periodic X-ray outbursts in Be/X-ray binaries}",
  journal = {\aap},
   eprint = {arXiv:astro-ph/0108037},
     year = 2001,
    month = oct,
   volume = 377,
    pages = {161-174},
      doi = {10.1051/0004-6361:20011083},
   adsurl = {http://adsabs.harvard.edu/abs/2001A%26A...377..161O}
}

@ARTICLE{okazaki02,
   author = {{Okazaki}, A.~T. and {Bate}, M.~R. and {Ogilvie}, G.~I. and 
	{Pringle}, J.~E.},
    title = "{Viscous effects on the interaction between the coplanar decretion disc and the neutron star in Be/X-ray binaries}",
  journal = {\mnras},
   eprint = {arXiv:astro-ph/0208288},
     year = 2002,
    month = dec,
   volume = 337,
    pages = {967-980},
      doi = {10.1046/j.1365-8711.2002.05960.x},
   adsurl = {http://adsabs.harvard.edu/abs/2002MNRAS.337..967O}
}

@ARTICLE{okazaki13,
   author = {{Okazaki}, A.~T. and {Hayasaki}, K. and {Moritani}, Y.},
    title = "{Origin of Two Types of X-Ray Outbursts in Be/X-Ray Binaries. I. Accretion Scenarios}",
  journal = {\pasj},
archivePrefix = "arXiv",
   eprint = {1211.5225},
 primaryClass = "astro-ph.HE",
     year = 2013,
    month = apr,
   volume = 65,
    pages = {41},
      doi = {10.1093/pasj/65.2.41},
   adsurl = {http://cdsads.u-strasbg.fr/abs/2013PASJ...65...41O}
}

@ARTICLE{panoglou16,
       author = {{Panoglou}, Despina and {Carciofi}, Alex C. and {Vieira}, Rodrigo G. and {Cyr}, Isabelle H. and {Jones}, Carol E. and {Okazaki}, Atsuo T. and {Rivinius}, Thomas},
        title = "{Be discs in binary systems - I. Coplanar orbits}",
      journal = {\mnras},
         year = 2016,
        month = sep,
       volume = {461},
       number = {3},
        pages = {2616-2629},
          doi = {10.1093/mnras/stw1508},
archivePrefix = {arXiv},
       eprint = {1605.06674},
 primaryClass = {astro-ph.SR},
       adsurl = {https://ui.adsabs.harvard.edu/abs/2016MNRAS.461.2616P}
}

@ARTICLE{panopoulou19,
       author = {{Panopoulou}, Georgia V. and {Hensley}, Brandon S. and {Skalidis}, Raphael and {Blinov}, Dmitry and {Tassis}, Konstantinos},
        title = "{Extreme starlight polarization in a region with highly polarized dust emission}",
      journal = {\aap},
         year = 2019,
        month = apr,
       volume = {624},
          eid = {L8},
        pages = {L8},
          doi = {10.1051/0004-6361/201935266},
archivePrefix = {arXiv},
       eprint = {1903.09684},
 primaryClass = {astro-ph.GA},
       adsurl = {https://ui.adsabs.harvard.edu/abs/2019A&A...624L...8P}
}

@ARTICLE{panopoulou25,
       author = {{Panopoulou}, G.~V. and {Markopoulioti}, L. and {Bouzelou}, F. and {Millar-Blanchaer}, M.~A. and {Tinyanont}, S. and {Blinov}, D. and {Pelgrims}, V. and {Johnson}, S. and {Skalidis}, R. and {Soam}, A.},
        title = "{A Compilation of Optical Starlight Polarization Catalogs}",
      journal = {\apjs},
         year = 2025,
        month = jan,
       volume = {276},
       number = {1},
          eid = {15},
        pages = {15},
          doi = {10.3847/1538-4365/ad8b21},
archivePrefix = {arXiv},
       eprint = {2307.05752},
 primaryClass = {astro-ph.GA},
       adsurl = {https://ui.adsabs.harvard.edu/abs/2025ApJS..276...15P}
}

@ARTICLE{parmar89b,
   author = {{Parmar}, A.~N. and {White}, N.~E. and {Stella}, L. and {Izzo}, C. and 
	{Ferri}, P.},
    title = "{The transient 42 second X-ray pulsar EXO 2030+375. I - The discovery and the luminosity dependence of the pulse period variations}",
  journal = {\apj},
     year = 1989,
    month = mar,
   volume = 338,
    pages = {359-372},
      doi = {10.1086/167204},
   adsurl = {http://cdsads.u-strasbg.fr/abs/1989ApJ...338..359P}
}

@ARTICLE{plaszczynski14,
       author = {{Plaszczynski}, S. and {Montier}, L. and {Levrier}, F. and {Tristram}, M.},
        title = "{A novel estimator of the polarization amplitude from normally distributed Stokes parameters}",
      journal = {\mnras},
         year = 2014,
        month = apr,
       volume = {439},
       number = {4},
        pages = {4048-4056},
          doi = {10.1093/mnras/stu270},
archivePrefix = {arXiv},
       eprint = {1312.0437},
 primaryClass = {astro-ph.CO},
       adsurl = {https://ui.adsabs.harvard.edu/abs/2014MNRAS.439.4048P}
}

@ARTICLE{poeckert79,
   author = {{Poeckert}, R. and {Bastien}, P. and {Landstreet}, J.~D.},
    title = "{Intrinsic polarization of Be stars}",
  journal = {\aj},
     year = 1979,
    month = jun,
   volume = 84,
    pages = {812-830},
      doi = {10.1086/112484},
   adsurl = {http://cdsads.u-strasbg.fr/abs/1979AJ.....84..812P}
}

@ARTICLE{quirrenbach97,
   author = {{Quirrenbach}, A. and {Bjorkman}, K.~S. and {Bjorkman}, J.~E. and 
	{Hummel}, C.~A. and {Buscher}, D.~F. and {Armstrong}, J.~T. and 
	{Mozurkewich}, D. and {Elias}, II, N.~M. and {Babler}, B.~L.
	},
    title = "{Constraints on the Geometry of Circumstellar Envelopes: Optical Interferometric and Spectropolarimetric Observations of Seven Be Stars}",
  journal = {\apj},
     year = 1997,
    month = apr,
   volume = 479,
    pages = {477},
      doi = {10.1086/303854},
   adsurl = {http://cdsads.u-strasbg.fr/abs/1997ApJ...479..477Q}
}

@ARTICLE{raichur10,
   author = {{Raichur}, H. and {Paul}, B.},
    title = "{Apsidal motion in 4U0115+63 and orbital parameters of 2S1417-624 and V0332+53}",
  journal = {\mnras},
     year = 2010,
    month = aug,
   volume = 406,
    pages = {2663-2670},
      doi = {10.1111/j.1365-2966.2010.16862.x},
   adsurl = {http://cdsads.u-strasbg.fr/abs/2010MNRAS.406.2663R}
}

@ARTICLE{ramaprakash19,
       author = {{Ramaprakash}, A.~N. and {Rajarshi}, C.~V. and {Das}, H.~K. and
         {Khodade}, P. and {Modi}, D. and {Panopoulou}, G. and {Maharana}, S. and
         {Blinov}, D. and {Angelakis}, E. and {Casadio}, C. and {Fuhrmann}, L. and
         {Hovatta}, T. and {Kiehlmann}, S. and {King}, O.~G. and {Kylafis}, N. and
         {Kougentakis}, A. and {Kus}, A. and {Mahabal}, A. and {Marecki}, A. and
         {Myserlis}, I. and {Paterakis}, G. and {Paleologou}, E. and
         {Liodakis}, I. and {Papadakis}, I. and {Papamastorakis}, I. and
         {Pavlidou}, V. and {Pazderski}, E. and {Pearson}, T.~J. and
         {Readhead}, A.~C.~S. and {Reig}, P. and {S{\l}owikowska}, A. and
         {Tassis}, K. and {Zensus}, J.~A.},
        title = "{RoboPol: a four-channel optical imaging polarimeter}",
      journal = {\mnras},
         year = "2019",
        month = "May",
       volume = {485},
       number = {2},
        pages = {2355-2366},
          doi = {10.1093/mnras/stz557},
archivePrefix = {arXiv},
       eprint = {1902.08367},
 primaryClass = {astro-ph.IM},
       adsurl = {https://ui.adsabs.harvard.edu/abs/2019MNRAS.485.2355R}
}

@ARTICLE{reig97a,
   author = {{Reig}, P. and {Fabregat}, J. and {Coe}, M.~J.},
    title = "{A new correlation for Be/X-ray binaries: the orbital period-H{$\alpha$} equivalent width diagram.}",
  journal = {\aap},
     year = 1997,
    month = jun,
   volume = 322,
    pages = {193-196},
   adsurl = {http://cdsads.u-strasbg.fr/abs/1997A%26A...322..193R}
}

@ARTICLE{reig97b,
   author = {{Reig}, P. and {Fabregat}, J. and {Coe}, M.~J. and {Roche}, P. and 
	{Chakrabarty}, D. and {Negueruela}, I. and {Steele}, I.},
    title = "{The Be/X-ray binary LS I +61 235/RX J0146.9+6121: physical parameters and V/R variability}",
  journal = {\aap},
     year = 1997,
    month = jun,
   volume = 322,
    pages = {183-192},
   adsurl = {http://cdsads.u-strasbg.fr/abs/1997A%26A...322..183R}
}

@ARTICLE{reig00,
   author = {{Reig}, P. and {Negueruela}, I. and {Coe}, M.~J. and {Fabregat}, J. and 
	{Tarasov}, A.~E. and {Zamanov}, R.~K.},
    title = "{Correlated V/R and infrared photometric variations in the Be/X-ray binary LS I +61{\deg} 235/RX J0146.9+6121}",
  journal = {\mnras},
   eprint = {arXiv:astro-ph/0005033},
     year = 2000,
    month = sep,
   volume = 317,
    pages = {205-210},
      doi = {10.1046/j.1365-8711.2000.03643.x},
   adsurl = {http://cdsads.u-strasbg.fr/abs/2000MNRAS.317..205R}
}

@ARTICLE{reig04,
   author = {{Reig}, P. and {Negueruela}, I. and {Fabregat}, J. and {Chato}, R. and 
	{Blay}, P. and {Mavromatakis}, F.},
    title = "{Discovery of the optical counterpart to the X-ray pulsar  SAX J2103.5+4545}",
  journal = {\aap},
   eprint = {arXiv:astro-ph/0404121},
     year = 2004,
    month = jul,
   volume = 421,
    pages = {673-680},
      doi = {10.1051/0004-6361:20035786},
   adsurl = {http://adsabs.harvard.edu/abs/2004A%26A...421..673R}
}

@ARTICLE{reig05a,
   author = {{Reig}, P. and {Negueruela}, I. and {Papamastorakis}, G. and 
	{Manousakis}, A. and {Kougentakis}, T.},
    title = "{Identification of the optical counterparts of high-mass X-ray binaries through optical photometry and spectroscopy}",
  journal = {\aap},
   eprint = {arXiv:astro-ph/0505319},
     year = 2005,
    month = sep,
   volume = 440,
    pages = {637-646},
      doi = {10.1051/0004-6361:20052684},
   adsurl = {http://cdsads.u-strasbg.fr/abs/2005A%26A...440..637R}
}

@ARTICLE{reig05b,
   author = {{Reig}, P. and {Negueruela}, I. and {Fabregat}, J. and {Chato}, R. and 
	{Coe}, M.~J.},
    title = "{Long-term optical/IR variability of the Be/X-ray binary <ASTROBJ>LS V +44 17</ASTROBJ>/<ASTROBJ>RX J0440.9+4431</ASTROBJ>}",
  journal = {\aap},
   eprint = {arXiv:astro-ph/0506230},
     year = 2005,
    month = sep,
   volume = 440,
    pages = {1079-1086},
      doi = {10.1051/0004-6361:20053124},
   adsurl = {http://cdsads.u-strasbg.fr/abs/2005A%26A...440.1079R}
}

@ARTICLE{reig07a,
   author = {{Reig}, P.},
    title = "{On the neutron star-disc interaction in Be/X-ray binaries}",
  journal = {\mnras},
   eprint = {arXiv:astro-ph/0703073},
     year = 2007,
    month = may,
   volume = 377,
    pages = {867-873},
      doi = {10.1111/j.1365-2966.2007.11657.x},
   adsurl = {http://adsabs.harvard.edu/abs/2007MNRAS.377..867R}
}

@ARTICLE{reig07b,
   author = {{Reig}, P. and {Larionov}, V. and {Negueruela}, I. and {Arkharov}, A.~A. and 
	{Kudryavtseva}, N.~A.},
    title = "{The Be/X-ray transient 4U 0115+63/V635 Cassiopeiae. III. Quasi-cyclic variability}",
  journal = {\aap},
   eprint = {arXiv:astro-ph/0611516},
     year = 2007,
    month = feb,
   volume = 462,
    pages = {1081-1089},
      doi = {10.1051/0004-6361:20066217},
   adsurl = {http://adsabs.harvard.edu/abs/2007A%26A...462.1081R}
}

@ARTICLE{reig10b,
   author = {{Reig}, P. and {Zezas}, A. and {Gkouvelis}, L.},
    title = "{The optical counterpart to IGR J06074+2205: a Be/X-ray binary showing disc loss and V/R variability}",
  journal = {\aap},
archivePrefix = "arXiv",
   eprint = {1006.4935},
 primaryClass = "astro-ph.GA",
     year = 2010,
    month = nov,
   volume = 522,
      eid = {A107},
    pages = {A107},
      doi = {10.1051/0004-6361/201014788},
   adsurl = {http://adsabs.harvard.edu/abs/2010A%26A...522A.107R}
}

@ARTICLE{reig11,
   author = {{Reig}, P.},
    title = "{Be/X-ray binaries}",
  journal = {\apss},
archivePrefix = "arXiv",
   eprint = {1101.5036},
 primaryClass = "astro-ph.HE",
     year = 2011,
    month = mar,
   volume = 332,
    pages = {1-29},
      doi = {10.1007/s10509-010-0575-8},
   adsurl = {http://cdsads.u-strasbg.fr/abs/2011Ap%26SS.332....1R}
}

@ARTICLE{reig14a,
   author = {{Reig}, P. and {Zezas}, A.},
    title = "{Disc-loss episode in the Be shell optical counterpart to the high-mass X-ray binary IGR J21343+4738}",
  journal = {\aap},
archivePrefix = "arXiv",
   eprint = {1311.3093},
 primaryClass = "astro-ph.HE",
     year = 2014,
    month = jan,
   volume = 561,
      eid = {A137},
    pages = {A137},
      doi = {10.1051/0004-6361/201321408},
   adsurl = {http://cdsads.u-strasbg.fr/abs/2014A%26A...561A.137R}
}

@ARTICLE{reig14d,
       author = {{Reig}, P. and {Blinov}, D. and {Papadakis}, I. and {Kylafis}, N. and {Tassis}, K.},
        title = "{The high optical polarization in the Be/X-ray binary EXO 2030+375}",
      journal = {\mnras},
         year = 2014,
        month = dec,
       volume = {445},
       number = {4},
        pages = {4235-4240},
          doi = {10.1093/mnras/stu2322},
archivePrefix = {arXiv},
       eprint = {1409.8411},
 primaryClass = {astro-ph.SR},
       adsurl = {https://ui.adsabs.harvard.edu/abs/2014MNRAS.445.4235R}
}

@ARTICLE{reig15,
   author = {{Reig}, P. and {Fabregat}, J.},
    title = "{Long-term variability of high-mass X-ray binaries. I. Photometry}",
  journal = {\aap},
archivePrefix = "arXiv",
   eprint = {1411.7163},
 primaryClass = "astro-ph.HE",
     year = 2015,
    month = feb,
   volume = 574,
      eid = {A33},
    pages = {A33},
      doi = {10.1051/0004-6361/201425008},
   adsurl = {http://cdsads.u-strasbg.fr/abs/2015A%26A...574A..33R}
}

@ARTICLE{reig16,
   author = {{Reig}, P. and {Nersesian}, A. and {Zezas}, A. and {Gkouvelis}, L. and 
	{Coe}, M.~J.},
    title = "{Long-term optical variability of high-mass X-ray binaries. II. Spectroscopy$\lt$xref ref-type=''fn'' rid=''FN1''$\gt$}",
  journal = {\aap},
archivePrefix = "arXiv",
   eprint = {1603.08327},
 primaryClass = "astro-ph.HE",
     year = 2016,
    month = may,
   volume = 590,
      eid = {A122},
    pages = {A122},
      doi = {10.1051/0004-6361/201628271},
   adsurl = {http://cdsads.u-strasbg.fr/abs/2016A%26A...590A.122R}
}

@ARTICLE{reig17,
       author = {{Reig}, P. and {Blay}, P. and {Blinov}, D.},
        title = "{The optical counterpart to the Be/X-ray binary SAX J2239.3+6116}",
      journal = {\aap},
         year = "2017",
        month = "Feb",
       volume = {598},
          eid = {A16},
        pages = {A16},
          doi = {10.1051/0004-6361/201629603},
archivePrefix = {arXiv},
       eprint = {1609.09215},
 primaryClass = {astro-ph.SR},
       adsurl = {https://ui.adsabs.harvard.edu/abs/2017A&A...598A..16R}
}

@ARTICLE{reig18b,
       author = {{Reig}, P. and {Blinov}, D.},
        title = "{Warped disks during type II outbursts in Be/X-ray binaries: evidence from optical polarimetry}",
      journal = {\aap},
         year = "2018",
        month = "Oct",
       volume = {619},
          eid = {A19},
        pages = {A19},
          doi = {10.1051/0004-6361/201833649},
archivePrefix = {arXiv},
       eprint = {1808.08726},
 primaryClass = {astro-ph.HE},
       adsurl = {https://ui.adsabs.harvard.edu/abs/2018A&A...619A..19R}
}

@ARTICLE{reig20,
       author = {{Reig}, P. and {Fabregat}, J. and {Alfonso-Garz{\'o}n}, J.},
        title = "{Optical counterpart to Swift J0243.6+6124}",
      journal = {\aap},
         year = 2020,
        month = aug,
       volume = {640},
          eid = {A35},
        pages = {A35},
          doi = {10.1051/0004-6361/202038333},
archivePrefix = {arXiv},
       eprint = {2006.10303},
 primaryClass = {astro-ph.HE},
       adsurl = {https://ui.adsabs.harvard.edu/abs/2020A&A...640A..35R}
}

@ARTICLE{reiz98,
       author = {{Reiz}, A. and {Franco}, G.~A.~P.},
        title = "{UBV polarimetry of 361 A- and F-type stars in selected areas}",
      journal = {\aaps},
         year = 1998,
        month = may,
       volume = {130},
        pages = {133-140},
          doi = {10.1051/aas:1998216},
       adsurl = {https://ui.adsabs.harvard.edu/abs/1998A&AS..130..133R}
}

@ARTICLE{rivinius13a,
   author = {{Rivinius}, T. and {Carciofi}, A.~C. and {Martayan}, C.},
    title = "{Classical Be stars. Rapidly rotating B stars with viscous Keplerian decretion disks}",
  journal = {\aapr},
archivePrefix = "arXiv",
   eprint = {1310.3962},
 primaryClass = "astro-ph.SR",
     year = 2013,
    month = oct,
   volume = 21,
    pages = {69},
      doi = {10.1007/s00159-013-0069-0},
   adsurl = {http://cdsads.u-strasbg.fr/abs/2013A%26ARv..21...69R}
}

@ARTICLE{sarty09,
   author = {{Sarty}, G.~E. and {Kiss}, L.~L. and {Huziak}, R. and {Catalan}, L.~J.~J. and 
	{Luciuk}, D. and {Crawford}, T.~R. and {Lane}, D.~J. and {Pickard}, R.~D. and 
	{Grzybowski}, T.~A. and {Closas}, P. and {Johnston}, H. and 
	{Balam}, D. and {Wu}, K.},
    title = "{Periodicities in the high-mass X-ray binary system RXJ0146.9+6121/LSI+61{\deg}235}",
  journal = {\mnras},
archivePrefix = "arXiv",
   eprint = {0810.4577},
     year = 2009,
    month = jan,
   volume = 392,
    pages = {1242-1252},
      doi = {10.1111/j.1365-2966.2008.14138.x},
   adsurl = {http://adsabs.harvard.edu/abs/2009MNRAS.392.1242S}
}

@ARTICLE{serkowski70,
       author = {{Serkowski}, K.},
        title = "{Intrinsic Polarization of Early-Type Stars with Extended Atmospheres}",
      journal = {\apj},
         year = 1970,
        month = jun,
       volume = {160},
        pages = {1083},
          doi = {10.1086/150496},
       adsurl = {https://ui.adsabs.harvard.edu/abs/1970ApJ...160.1083S}
}

@ARTICLE{serkowski75,
   author = {{Serkowski}, K. and {Mathewson}, D.~S. and {Ford}, V.~L.},
    title = "{Wavelength dependence of interstellar polarization and ratio of total to selective extinction}",
  journal = {\apj},
     year = 1975,
    month = feb,
   volume = 196,
    pages = {261-290},
      doi = {10.1086/153410},
   adsurl = {http://cdsads.u-strasbg.fr/abs/1975ApJ...196..261S}
}

@ARTICLE{simmons85,
       author = {{Simmons}, J.~F.~L. and {Stewart}, B.~G.},
        title = "{Point and interval estimation of the true unbiased degree of linear polarization in the presence of low signal-to-noise ratios}",
      journal = {\aap},
         year = 1985,
        month = jan,
       volume = {142},
       number = {1},
        pages = {100-106},
       adsurl = {https://ui.adsabs.harvard.edu/abs/1985A&A...142..100S}
}

@ARTICLE{tomsick11,
   author = {{Tomsick}, J.~A. and {Heinke}, C. and {Halpern}, J. and {Kaaret}, P. and 
	{Chaty}, S. and {Rodriguez}, J. and {Bodaghee}, A.},
    title = "{Confirmation of IGR J01363+6610 as a Be X-ray Binary with Very Low Quiescent X-ray Luminosity}",
  journal = {\apj},
archivePrefix = "arXiv",
   eprint = {1012.2817},
 primaryClass = "astro-ph.HE",
     year = 2011,
    month = feb,
   volume = 728,
      eid = {86},
    pages = {86},
      doi = {10.1088/0004-637X/728/2/86},
   adsurl = {http://cdsads.u-strasbg.fr/abs/2011ApJ...728...86T}
}

@ARTICLE{tram22,
       author = {{Tram}, Le Ngoc and {Hoang}, Thiem},
        title = "{Recent progress in theory and observational study of dust grain alignment and rotational disruption in star-forming regions}",
      journal = {Frontiers in Astronomy and Space Sciences},
         year = 2022,
        month = oct,
       volume = {9},
          eid = {923927},
        pages = {923927},
          doi = {10.3389/fspas.2022.923927},
archivePrefix = {arXiv},
       eprint = {2208.13195},
 primaryClass = {astro-ph.GA},
       adsurl = {https://ui.adsabs.harvard.edu/abs/2022FrASS...9.3927T}
}

@ARTICLE{treiber25,
       author = {{Treiber}, H. and {Vasilopoulos}, G. and {Bailyn}, C.~D. and {Haberl}, F. and {Udalski}, A.},
        title = "{Two decades of optical variability of Small Magellanic Cloud high-mass X-ray binaries}",
      journal = {\aap},
         year = 2025,
        month = feb,
       volume = {694},
          eid = {A43},
        pages = {A43},
          doi = {10.1051/0004-6361/202451132},
archivePrefix = {arXiv},
       eprint = {2501.13147},
 primaryClass = {astro-ph.HE},
       adsurl = {https://ui.adsabs.harvard.edu/abs/2025A&A...694A..43T}
}

@ARTICLE{tycner05,
   author = {{Tycner}, C. and {Lester}, J.~B. and {Hajian}, A.~R. and {Armstrong}, J.~T. and 
	{Benson}, J.~A. and {Gilbreath}, G.~C. and {Hutter}, D.~J. and 
	{Pauls}, T.~A. and {White}, N.~M.},
    title = "{Properties of the H{$\alpha$}-emitting Circumstellar Regions of Be Stars}",
  journal = {\apj},
   eprint = {arXiv:astro-ph/0501552},
     year = 2005,
    month = may,
   volume = 624,
    pages = {359-371},
      doi = {10.1086/429126},
   adsurl = {http://cdsads.u-strasbg.fr/abs/2005ApJ...624..359T}
}

@ARTICLE{vaillancourt06,
       author = {{Vaillancourt}, John E.},
        title = "{Placing Confidence Limits on Polarization Measurements}",
      journal = {\pasp},
         year = 2006,
        month = sep,
       volume = {118},
       number = {847},
        pages = {1340-1343},
          doi = {10.1086/507472},
archivePrefix = {arXiv},
       eprint = {astro-ph/0603110},
 primaryClass = {astro-ph},
       adsurl = {https://ui.adsabs.harvard.edu/abs/2006PASP..118.1340V}
}

@ARTICLE{verrecchia02,
       author = {{Verrecchia}, F. and {Israel}, G.~L. and {Negueruela}, I. and {Covino}, S. and {Polcaro}, V.~F. and {Clark}, J.~S. and {Steele}, I.~A. and {Gualandi}, R. and {Speziali}, R. and {Stella}, L.},
        title = "{The identification of the optical/IR counterpart of the 15.8-s transient X--ray pulsar XTE J1946+274}",
      journal = {\aap},
         year = 2002,
        month = oct,
       volume = {393},
        pages = {983-989},
          doi = {10.1051/0004-6361:20021087},
archivePrefix = {arXiv},
       eprint = {astro-ph/0207587},
 primaryClass = {astro-ph},
       adsurl = {https://ui.adsabs.harvard.edu/abs/2002A&A...393..983V}
}

@ARTICLE{vince95,
       author = {{Vince}, I. and {Arsenijevi{\'c}}, J. and {Markovi{\'c}-Kr{\v{s}}ljanin}, S. and {Jankov}, S. and {Skuljan}, Lj.},
        title = "{Polarization Measurements of Some Be Stars}",
      journal = {International Amateur-Professional Photoelectric Photometry Communications},
         year = 1995,
        month = mar,
       volume = {59},
        pages = {32},
       adsurl = {https://ui.adsabs.harvard.edu/abs/1995IAPPP..59...32V}
}

@ARTICLE{wang10,
       author = {{Wang}, W.},
        title = "{Discovery of a magnetic neutron star in X-ray transient IGR J01583+6713}",
      journal = {\aap},
         year = 2010,
        month = jun,
       volume = {516},
          eid = {A15},
        pages = {A15},
          doi = {10.1051/0004-6361/200913196},
archivePrefix = {arXiv},
       eprint = {0912.0337},
 primaryClass = {astro-ph.HE},
       adsurl = {https://ui.adsabs.harvard.edu/abs/2010A&A...516A..15W}
}

@ARTICLE{whittet78,
   author = {{Whittet}, D.~C.~B. and {van Breda}, I.~G.},
    title = "{The correlation of the interstellar extinction law with the wavelength of maximum polarization}",
  journal = {\aap},
     year = 1978,
    month = may,
   volume = 66,
    pages = {57-63},
   adsurl = {http://cdsads.u-strasbg.fr/abs/1978A%26A....66...57W}
}

@ARTICLE{wilson98,
       author = {{Wilson}, Colleen A. and {Finger}, Mark H. and {Harmon}, B. Alan and {Chakrabarty}, Deepto and {Strohmayer}, Tod},
        title = "{Discovery of the 198 Second X-Ray Pulsar GRO J2058+42}",
      journal = {\apj},
         year = 1998,
        month = may,
       volume = {499},
       number = {2},
        pages = {820-827},
          doi = {10.1086/305677},
archivePrefix = {arXiv},
       eprint = {astro-ph/9802324},
 primaryClass = {astro-ph},
       adsurl = {https://ui.adsabs.harvard.edu/abs/1998ApJ...499..820W}
}

@ARTICLE{wilson08,
   author = {{Wilson}, C.~A. and {Finger}, M.~H. and {Camero-Arranz}, A.},
    title = "{Outbursts Large and Small from EXO 2030+375}",
  journal = {\apj},
archivePrefix = "arXiv",
   eprint = {0804.1375},
     year = 2008,
    month = may,
   volume = 678,
    pages = {1263-1272},
      doi = {10.1086/587134},
   adsurl = {http://cdsads.u-strasbg.fr/abs/2008ApJ...678.1263W}
}

@ARTICLE{wilson18,
       author = {{Wilson-Hodge}, Colleen A. and {Malacaria}, Christian and
         {Jenke}, Peter A. and {Jaisawal}, Gaurava K. and {Kerr}, Matthew and
         {Wolff}, Michael T. and {Arzoumanian}, Zaven and {Chakrabarty}, Deepto and
         {Doty}, John P. and {Gendreau}, Keith C. and {Guillot}, Sebastien and
         {Ho}, Wynn C.~G. and {LaMarr}, Beverly and {Markwardt}, Craig B. and
         {{\"O}zel}, Feryal and {Prigozhin}, Gregory Y. and {Ray}, Paul S. and
         {Ramos-Lerate}, Mercedes and {Remillard}, Ronald A. and
         {Strohmayer}, Tod E. and {Vezie}, Michael L. and {Wood}, Kent S. and
         {NICER Science Team}},
        title = "{NICER and Fermi GBM Observations of the First Galactic Ultraluminous X-Ray Pulsar Swift J0243.6+6124}",
      journal = {\apj},
         year = "2018",
        month = "Aug",
       volume = {863},
       number = {1},
          eid = {9},
        pages = {9},
          doi = {10.3847/1538-4357/aace60},
archivePrefix = {arXiv},
       eprint = {1806.10094},
 primaryClass = {astro-ph.HE},
       adsurl = {https://ui.adsabs.harvard.edu/abs/2018ApJ...863....9W}
}

@ARTICLE{wisniewski10,
       author = {{Wisniewski}, John P. and {Draper}, Zachary H. and {Bjorkman}, Karen S. and {Meade}, Marilyn R. and {Bjorkman}, Jon E. and {Kowalski}, Adam F.},
        title = "{Disk-Loss and Disk-Renewal Phases in Classical Be Stars. I. Analysis of Long-Term Spectropolarimetric Data}",
      journal = {\apj},
         year = 2010,
        month = feb,
       volume = {709},
       number = {2},
        pages = {1306-1320},
          doi = {10.1088/0004-637X/709/2/1306},
archivePrefix = {arXiv},
       eprint = {0912.1504},
 primaryClass = {astro-ph.SR},
       adsurl = {https://ui.adsabs.harvard.edu/abs/2010ApJ...709.1306W}
}

@ARTICLE{wood96,
   author = {{Wood}, K. and {Bjorkman}, J.~E. and {Whitney}, B.~A. and {Code}, A.~D.
	},
    title = "{The Effect of Multiple Scattering on the Polarization from Axisymmetric Circumstellar Envelopes. I. Pure Thomson Scattering Envelopes}",
  journal = {\apj},
     year = 1996,
    month = apr,
   volume = 461,
    pages = {828},
      doi = {10.1086/177105},
   adsurl = {http://cdsads.u-strasbg.fr/abs/1996ApJ...461..828W}
}

@ARTICLE{yan12a,
       author = {{Yan}, Jingzhi and {Zurita Heras}, Juan Antonio and {Chaty}, Sylvain and {Li}, Hui and {Liu}, Qingzhong},
        title = "{Multi-wavelength Study of the Be/X-Ray Binary MXB 0656-072}",
      journal = {\apj},
         year = 2012,
        month = jul,
       volume = {753},
       number = {1},
          eid = {73},
        pages = {73},
          doi = {10.1088/0004-637X/753/1/73},
archivePrefix = {arXiv},
       eprint = {1205.0063},
 primaryClass = {astro-ph.HE},
       adsurl = {https://ui.adsabs.harvard.edu/abs/2012ApJ...753...73Y}
}

@ARTICLE{yudin01,
   author = {{Yudin}, R.~V.},
    title = "{Statistical analysis of intrinsic polarization, IR excess and projected rotational velocity distributions of classical Be stars}",
  journal = {\aap},
     year = 2001,
    month = mar,
   volume = 368,
    pages = {912-931},
      doi = {10.1051/0004-6361:20000577},
   adsurl = {http://cdsads.u-strasbg.fr/abs/2001A%26A...368..912Y}
}

@ARTICLE{zamanov01,
   author = {{Zamanov}, R.~K. and {Reig}, P. and {Mart{\'{\i}}}, J. and {Coe}, M.~J. and 
	{Fabregat}, J. and {Tomov}, N.~A. and {Valchev}, T.},
    title = "{Comparison of the H{$\alpha$} circumstellar disks in Be/X-ray binaries and Be stars}",
  journal = {\aap},
   eprint = {arXiv:astro-ph/0012371},
     year = 2001,
    month = mar,
   volume = 367,
    pages = {884-890},
      doi = {10.1051/0004-6361:20000533},
   adsurl = {http://adsabs.harvard.edu/abs/2001A%26A...367..884Z}
}

@ARTICLE{zamanov13,
   author = {{Zamanov}, R. and {Stoyanov}, K. and {Mart{\'{\i}}}, J. and 
	{Tomov}, N.~A. and {Belcheva}, G. and {Luque-Escamilla}, P.~L. and 
	{Latev}, G.},
    title = "{H{$\alpha$} observations of the {$\gamma$}-ray-emitting Be/X-ray binary LSI+61{\deg}303: orbital modulation, disk truncation, and long-term variability}",
  journal = {\aap},
archivePrefix = "arXiv",
   eprint = {1309.3947},
 primaryClass = "astro-ph.SR",
     year = 2013,
    month = nov,
   volume = 559,
      eid = {A87},
    pages = {A87},
      doi = {10.1051/0004-6361/201321991},
   adsurl = {http://cdsads.u-strasbg.fr/abs/2013A%26A...559A..87Z}
}

\onecolumn

\begin{appendix}

\section{Long-term polarimetric variability of targets }
\label{app:longterm}

This appendix gives the evolution of the polarization observations of the
targets. 

\FloatBarrier
\begin{figure*}[h]
    \centering

    \includegraphics[width=0.43\textwidth]{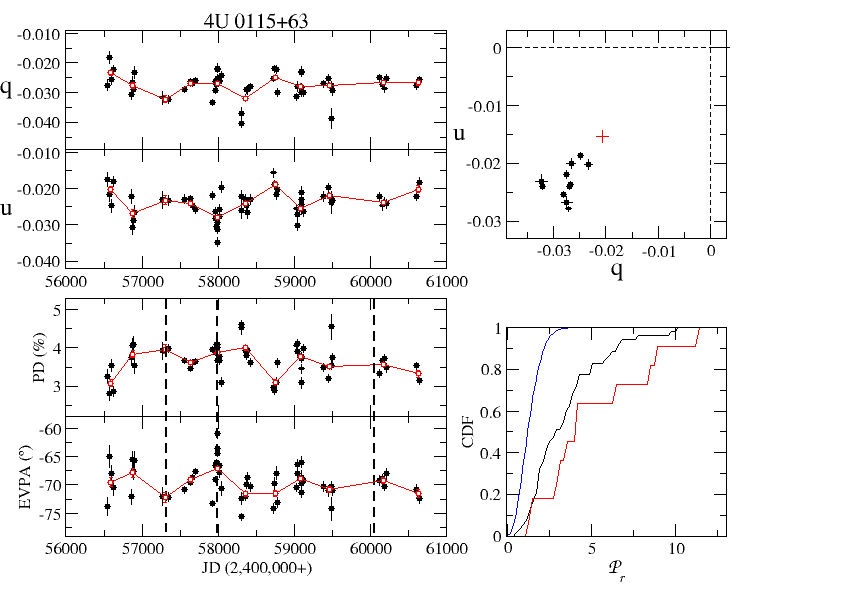}     
    \includegraphics[width=0.43\textwidth]{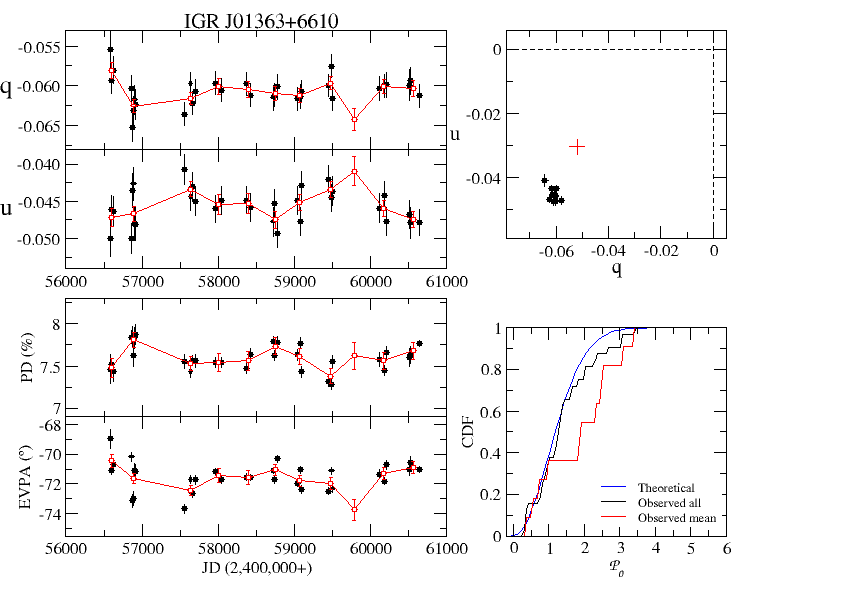}\\[-12pt]
    \includegraphics[width=0.43\textwidth]{./RXJ0146.9+6121_VAR.png}
    \includegraphics[width=0.43\textwidth]{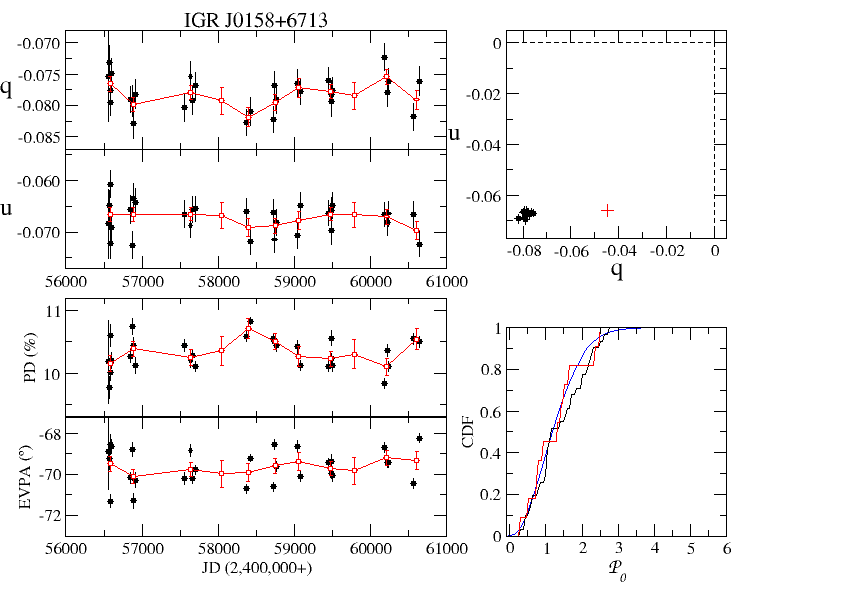}\\[-12pt] 
    \includegraphics[width=0.43\textwidth]{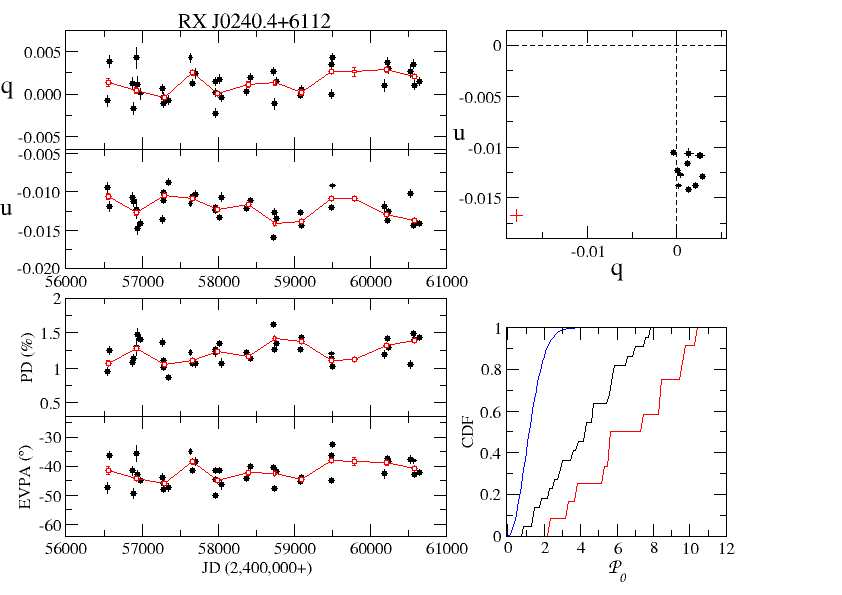} 
    \includegraphics[width=0.43\textwidth]{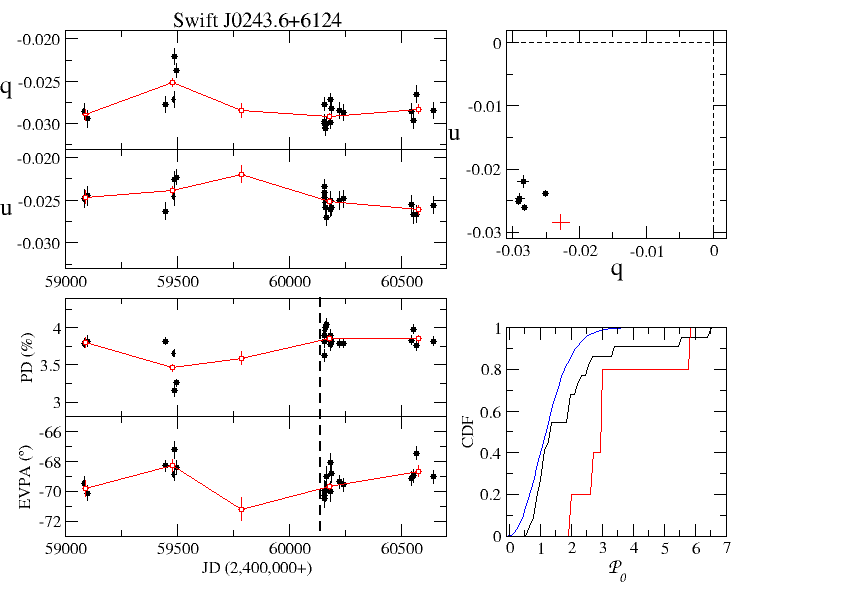} \\[-12pt]
    \includegraphics[width=0.43\textwidth]{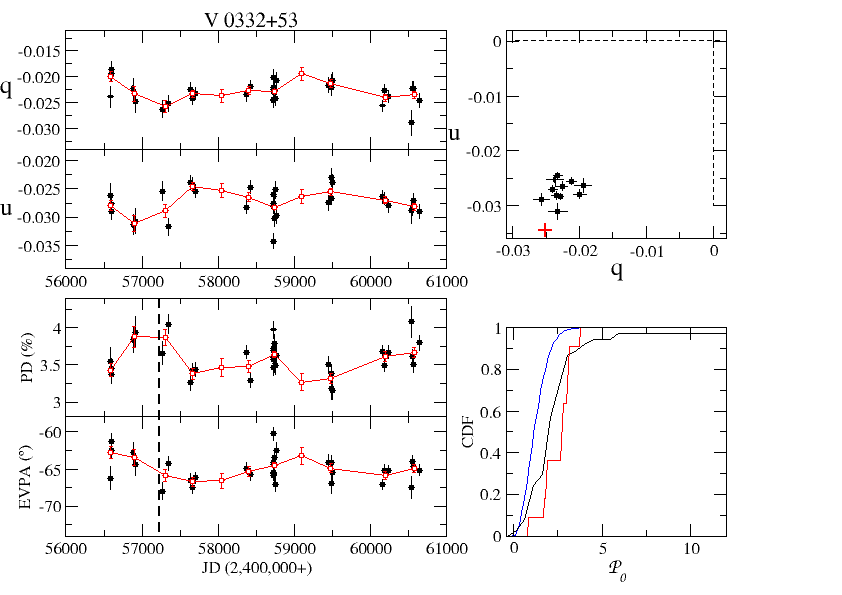}      
    \includegraphics[width=0.43\textwidth]{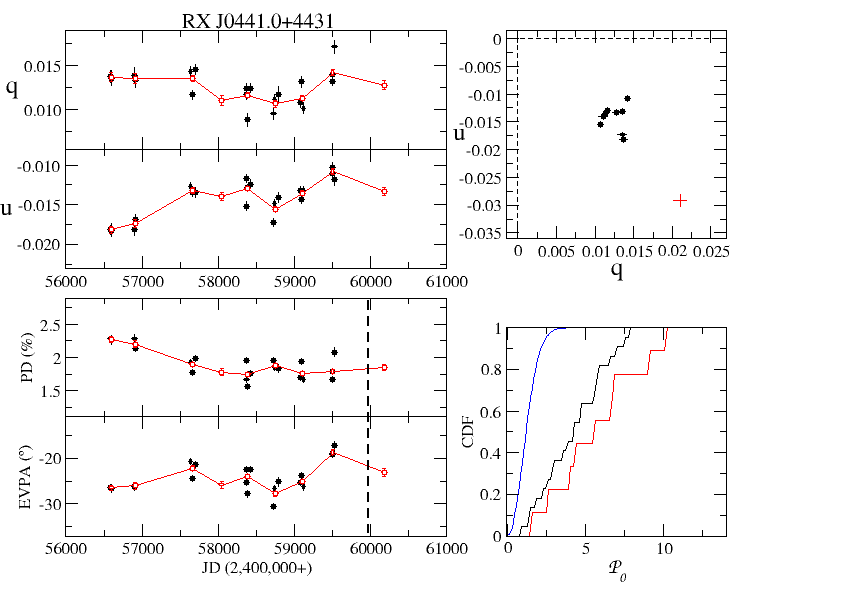}

    \caption{{\em Left panels}: Evolution of the Stokes parameters, polarization degree
and angle; {\em Top-right panel}: $q-u$ plane. Weighted mean of the source observations
(black circles) calculated yearly and of the field stars (red cross); {\em Bottom-right panel}:  
EDF of measured polarization using all data points (black line) or the weighted
averaged points (red line) compared with expected (theoretical) CDF of polarization
measurements (blue line).}
\end{figure*}

\begin{figure*}

\ContinuedFloat
    \centering

    \includegraphics[width=0.49\textwidth]{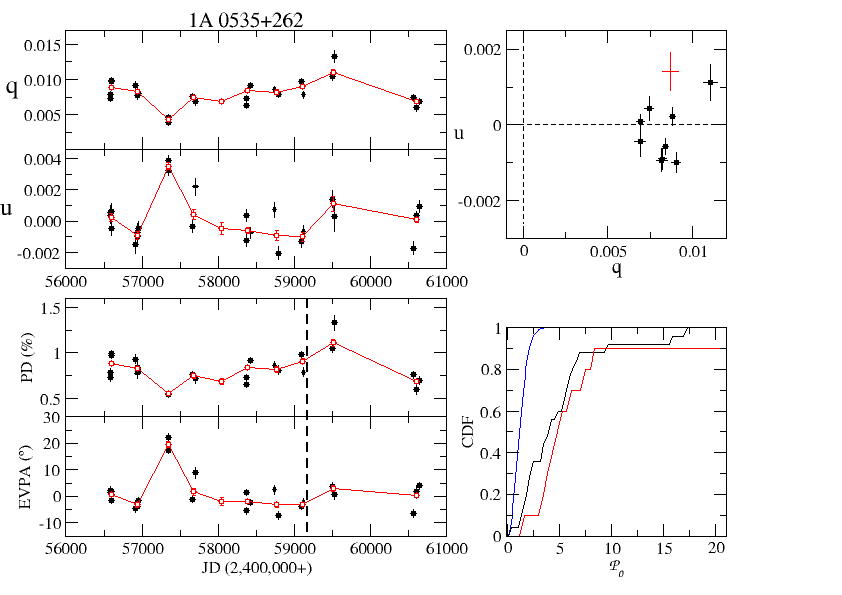}      
    \includegraphics[width=0.49\textwidth]{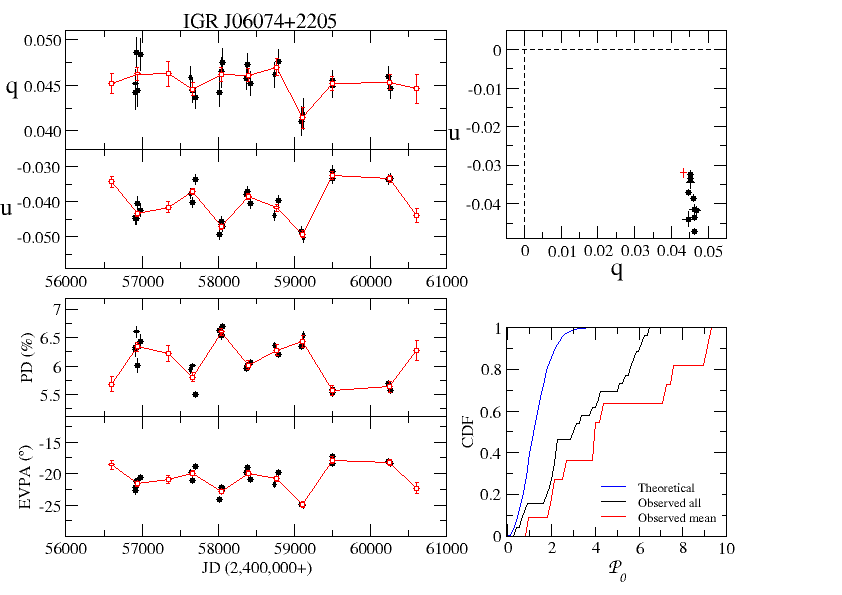}\\[-12pt]

    \includegraphics[width=0.49\textwidth]{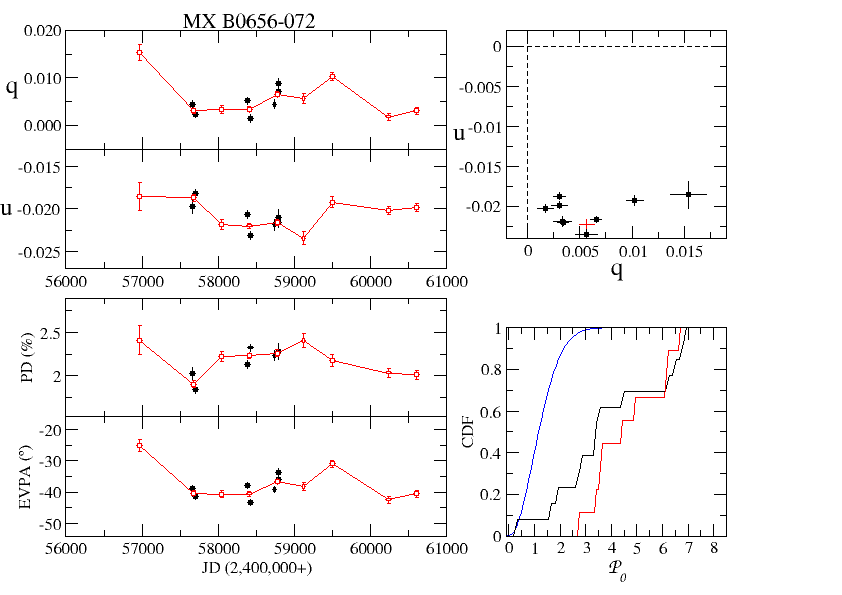}   
    \includegraphics[width=0.49\textwidth]{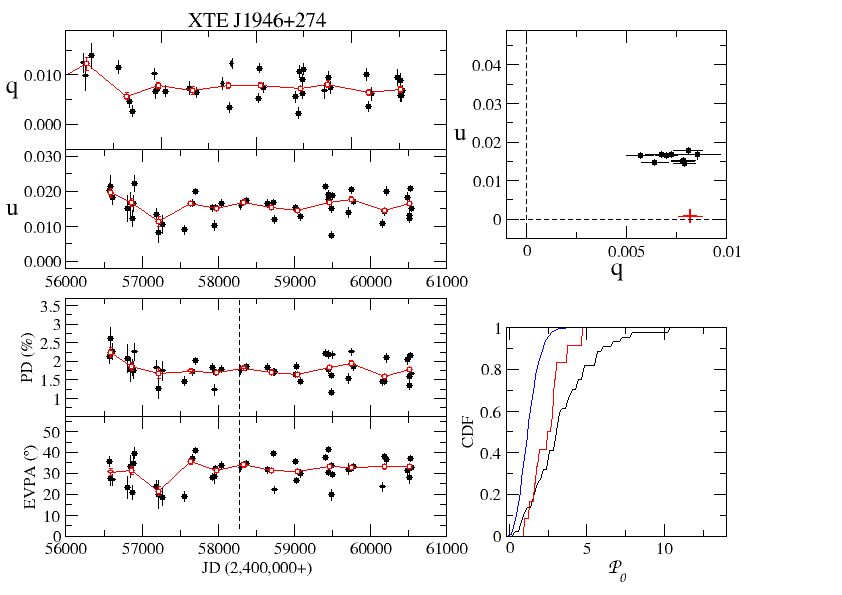} \\[-12pt] 
    
    \includegraphics[width=0.49\textwidth]{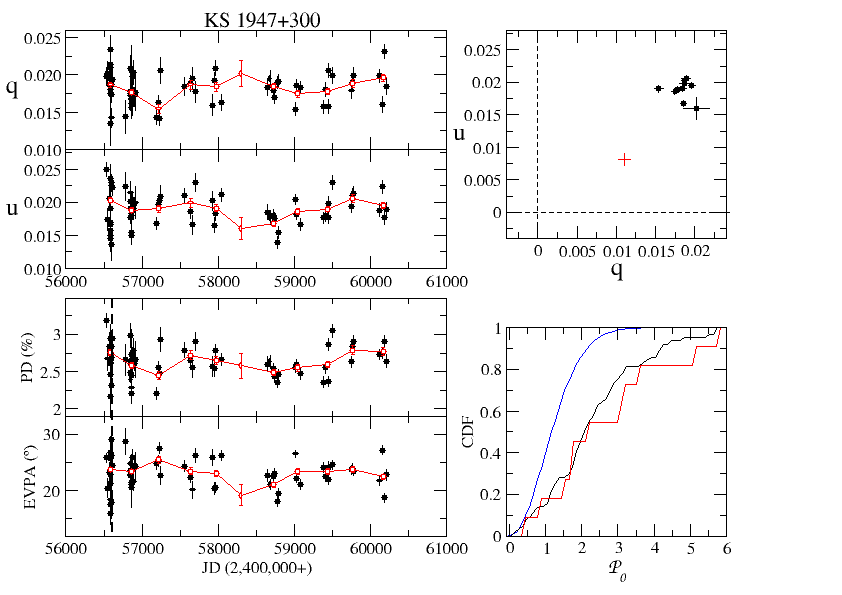}    
    \includegraphics[width=0.49\textwidth]{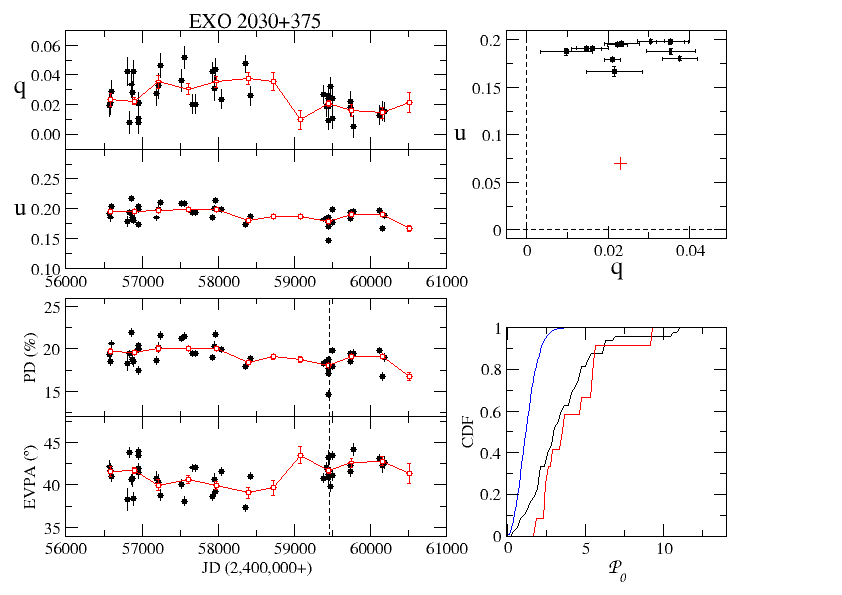} \\[-12pt]  
    
    \includegraphics[width=0.49\textwidth]{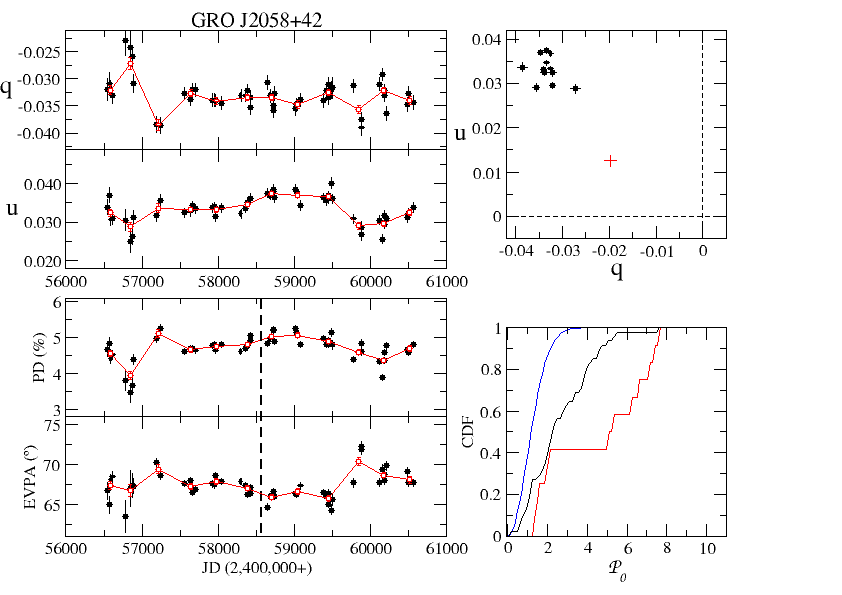}   
    \includegraphics[width=0.49\textwidth]{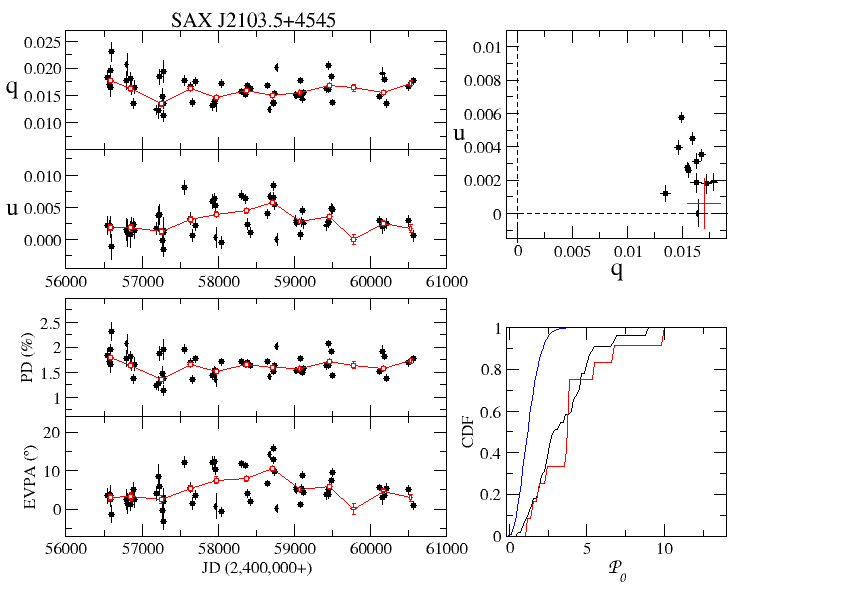}

    \caption[]{ Continued.}
\end{figure*}

\begin{figure*}
\ContinuedFloat
    \centering

    
    \includegraphics[width=0.49\textwidth]{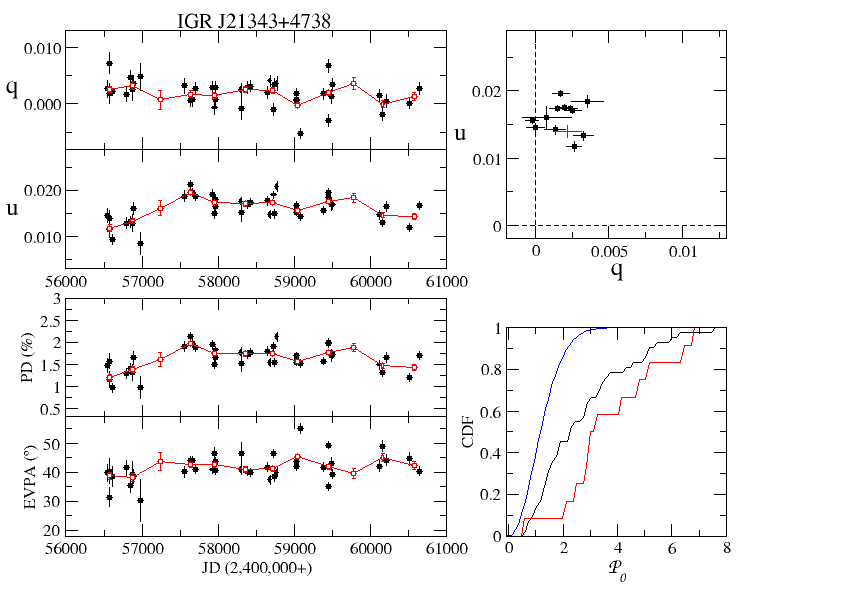}
    \includegraphics[width=0.49\textwidth]{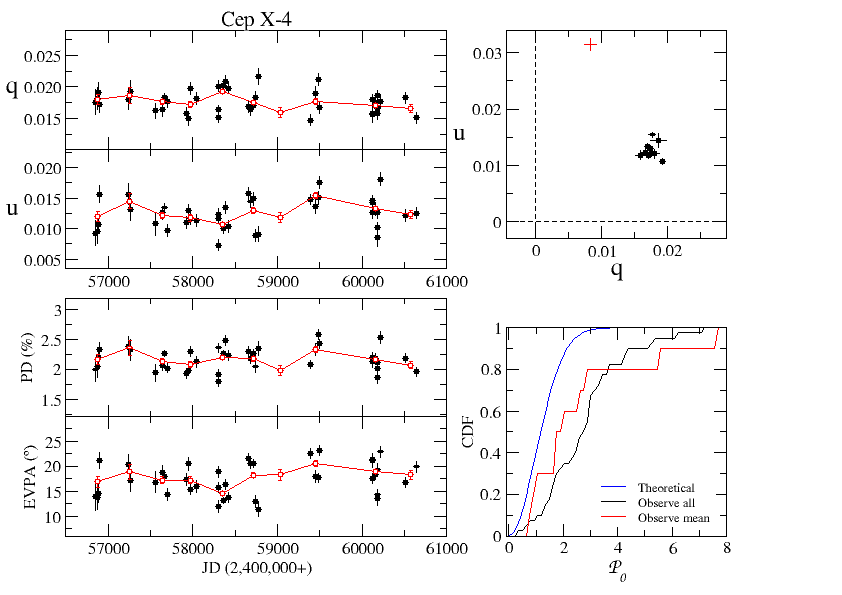}\\[-12pt]

    \includegraphics[width=0.49\textwidth]{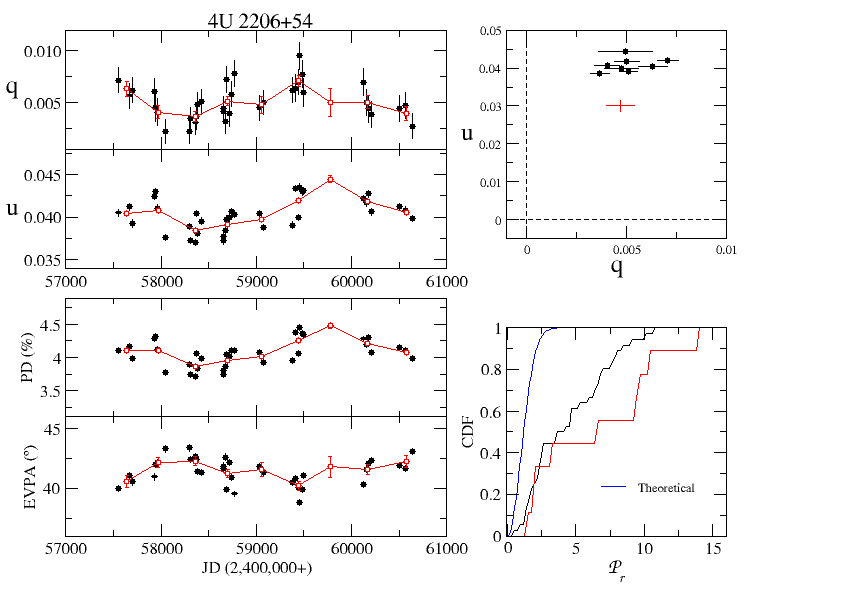}
    \includegraphics[width=0.49\textwidth]{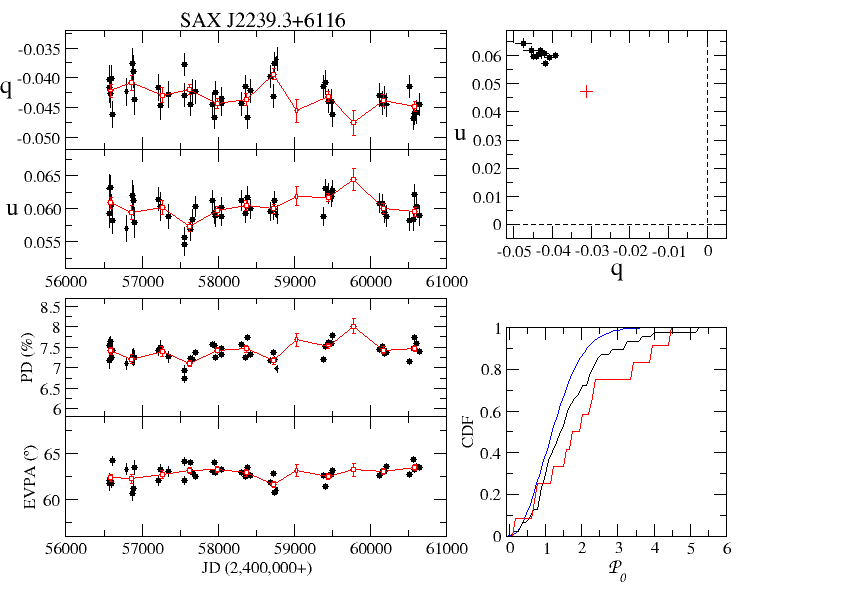}\\[-12pt]
    
    \includegraphics[width=0.49\textwidth]{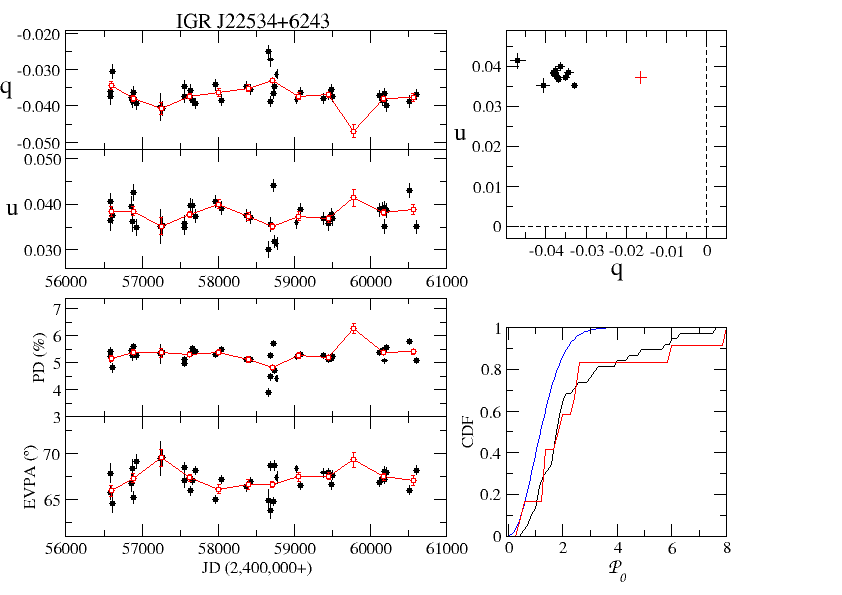} 

   \caption[]{ Continued.}
\end{figure*}
\FloatBarrier 


\section{Extinction curves}
\label{app:extinction}

This appendix shows the extinction curves for each individual target.

\FloatBarrier
\begin{figure*}[h]
    \centering

    \includegraphics[width=0.45\textwidth]{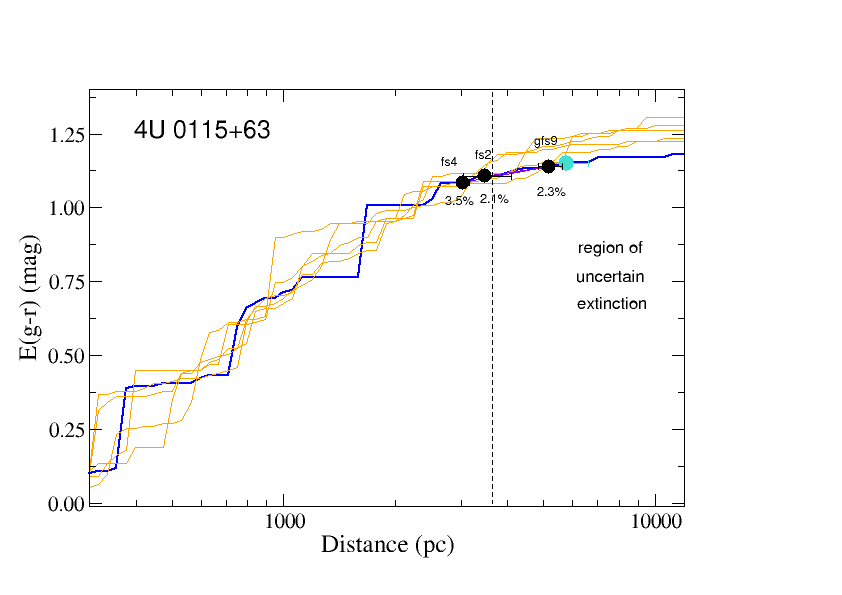}     
    \includegraphics[width=0.45\textwidth]{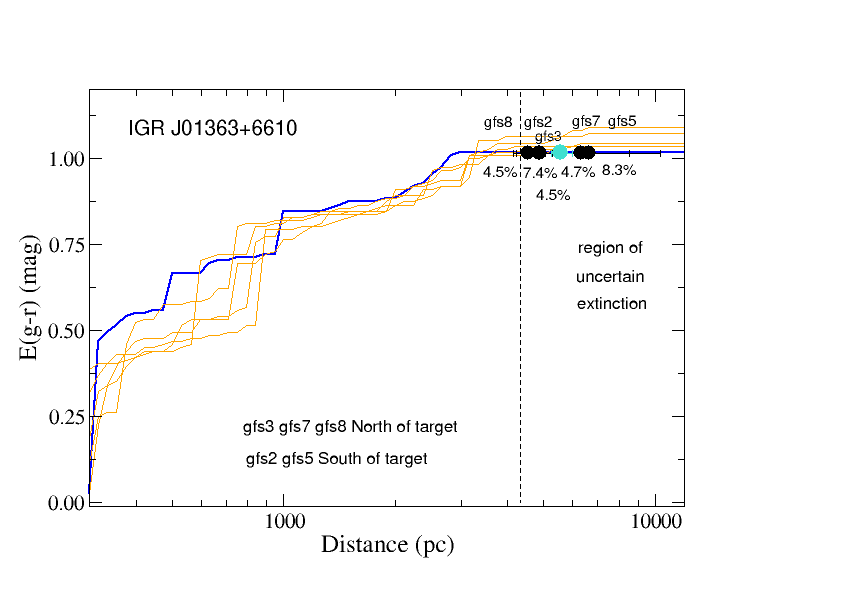}\\[-13pt]
    \includegraphics[width=0.45\textwidth]{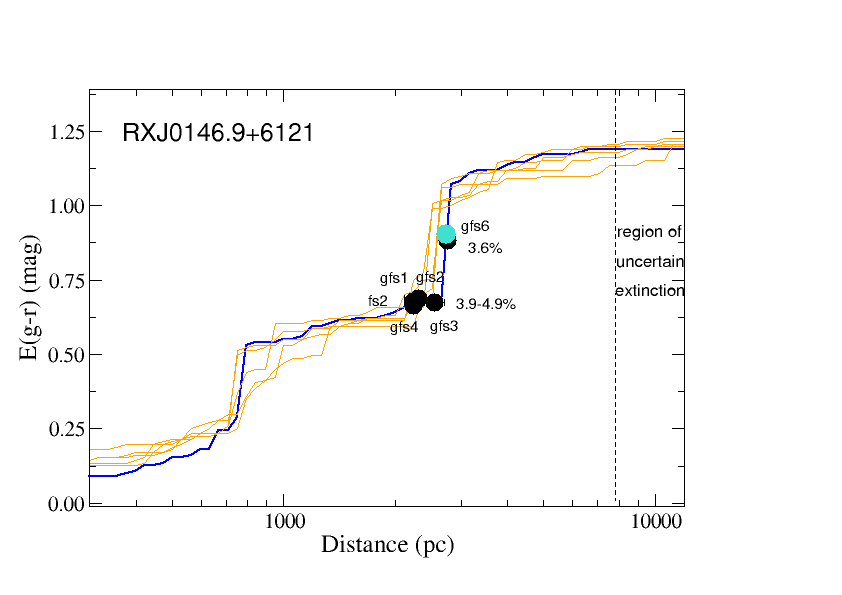}
    \includegraphics[width=0.45\textwidth]{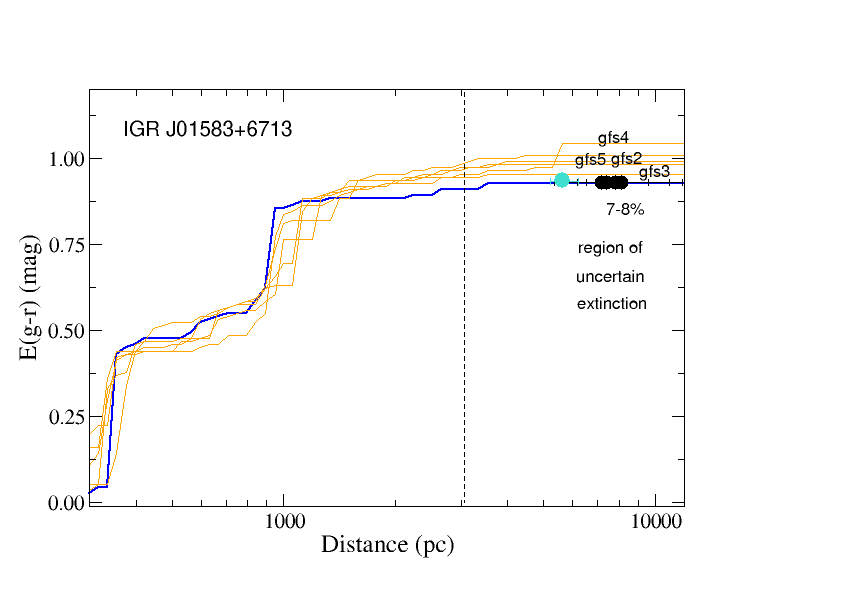}\\[-13pt] 
    \includegraphics[width=0.45\textwidth]{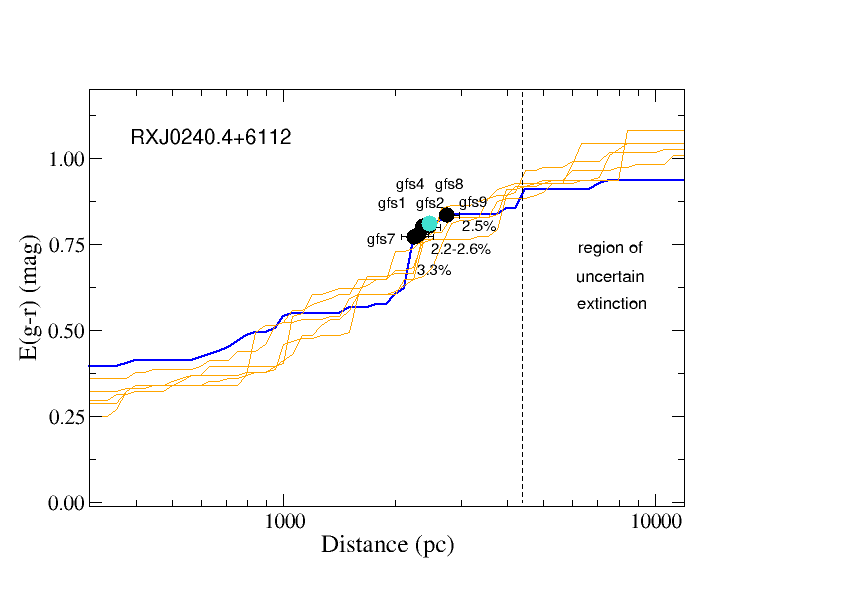} 
    \includegraphics[width=0.45\textwidth]{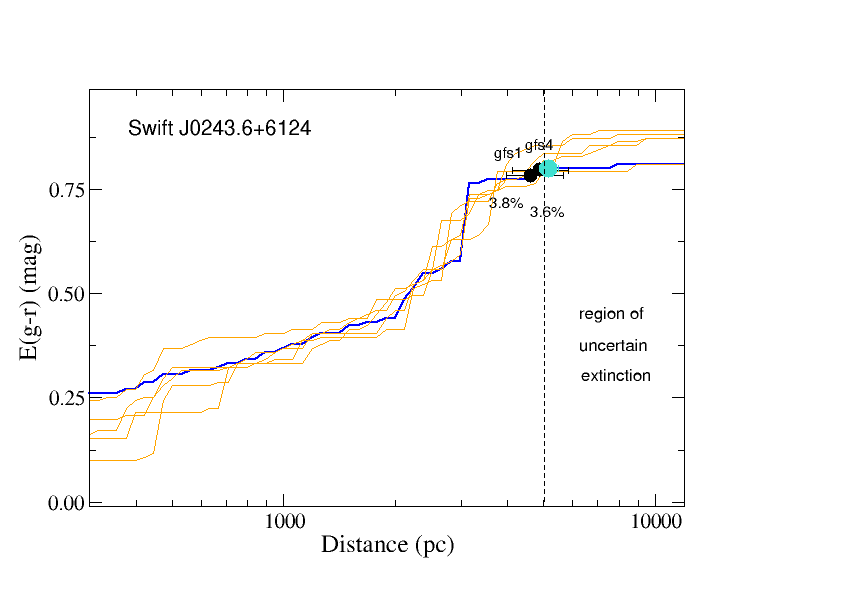} \\[-13pt]
    \includegraphics[width=0.45\textwidth]{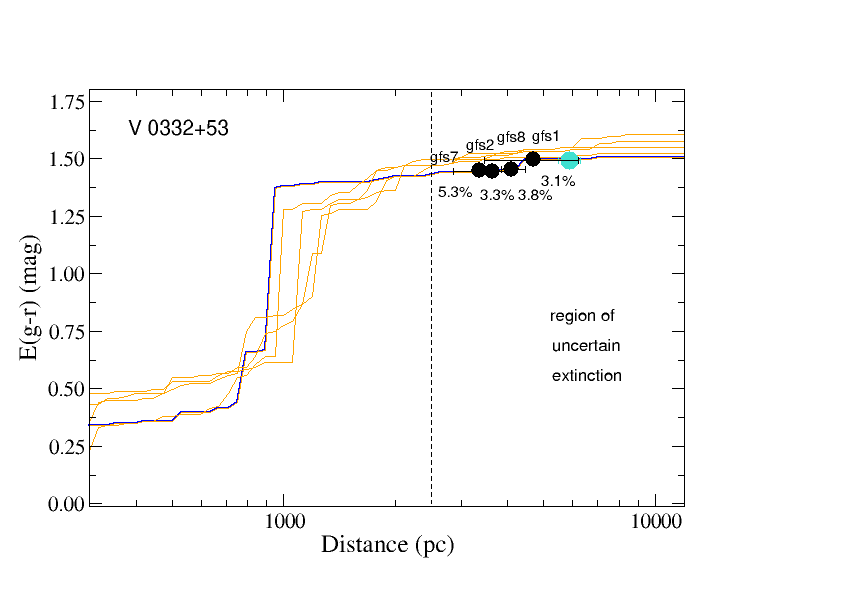}      
    \includegraphics[width=0.45\textwidth]{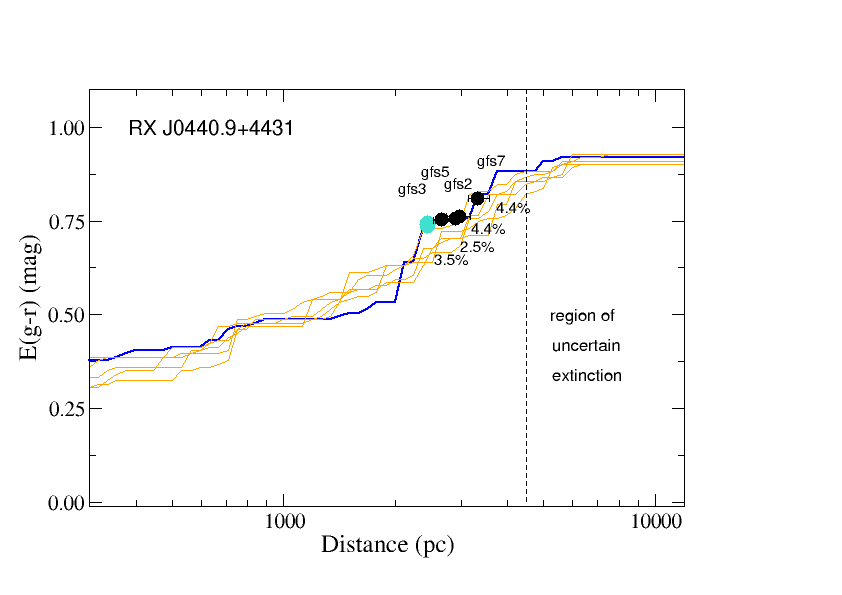}\\[-13pt]

    \caption{Extinction as a function of distance in the direction of the targets. The turquoise circle 
 represents the target, while the black circles correspond to the field stars that were used to correct 
 for ISM polarization.}
\end{figure*}

\begin{figure*}

    \centering

    \includegraphics[width=0.49\textwidth]{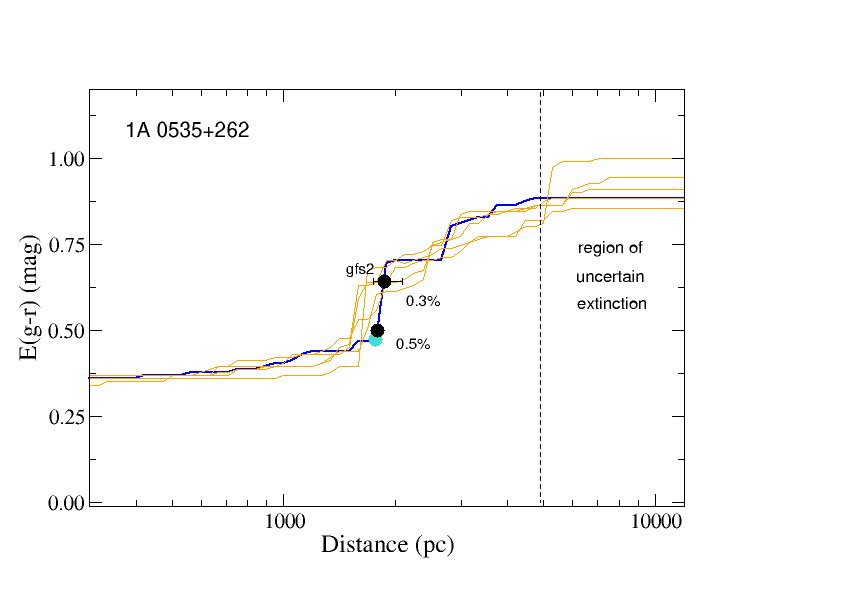}      
    \includegraphics[width=0.49\textwidth]{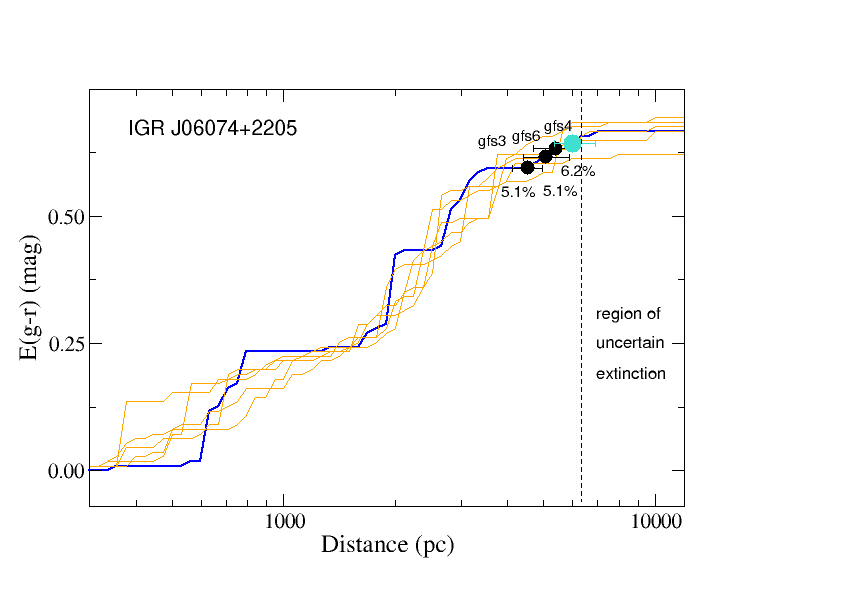}\\[-12pt] 

    \includegraphics[width=0.49\textwidth]{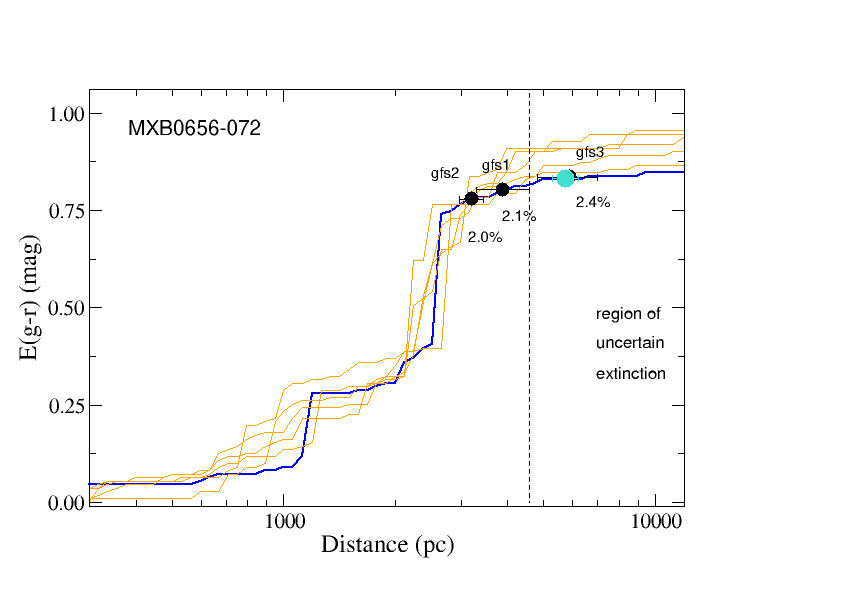}   
    \includegraphics[width=0.49\textwidth]{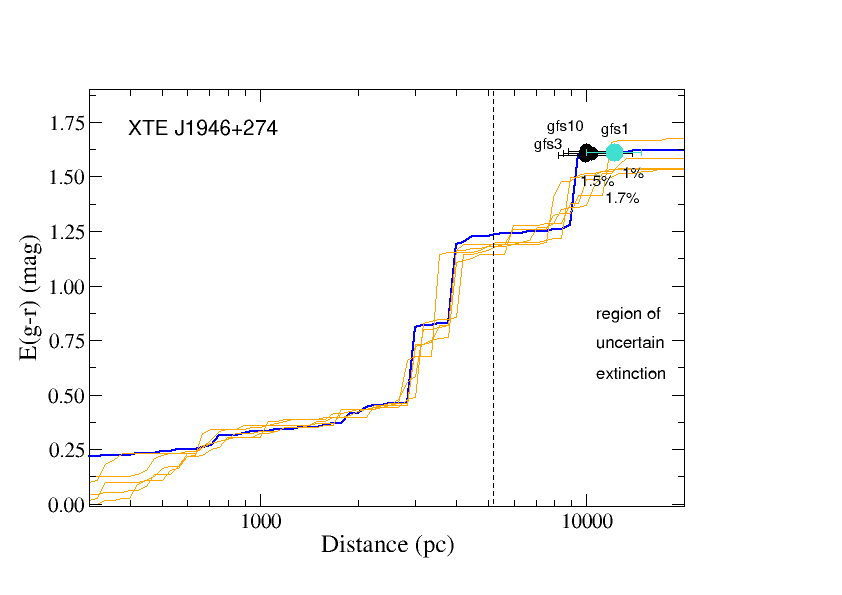}\\[-12pt] 
    
    \includegraphics[width=0.49\textwidth]{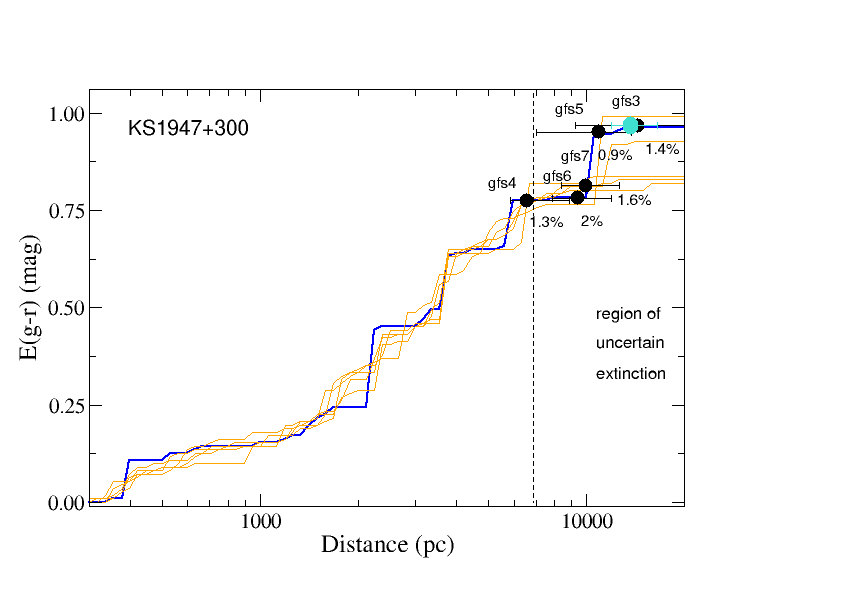}    
    \includegraphics[width=0.49\textwidth]{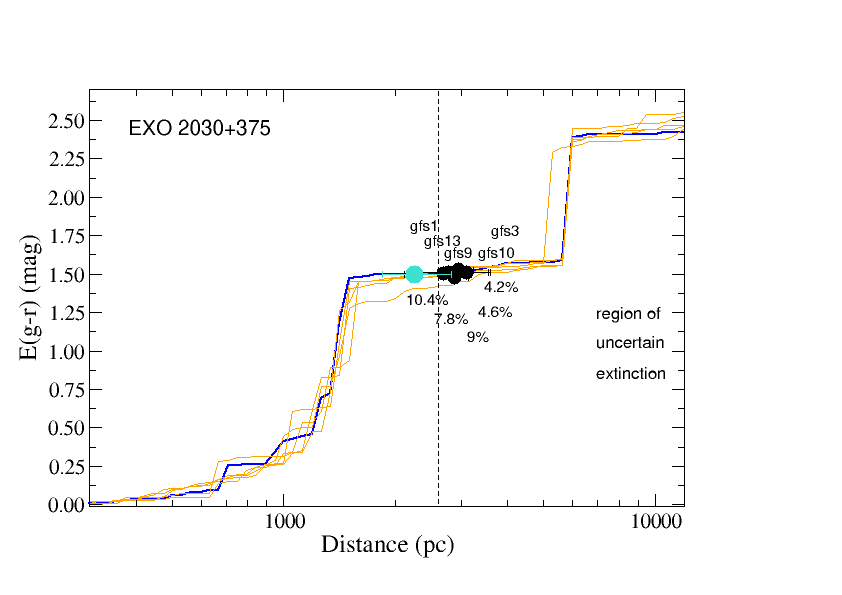}\\[-12pt] 
    
    \includegraphics[width=0.49\textwidth]{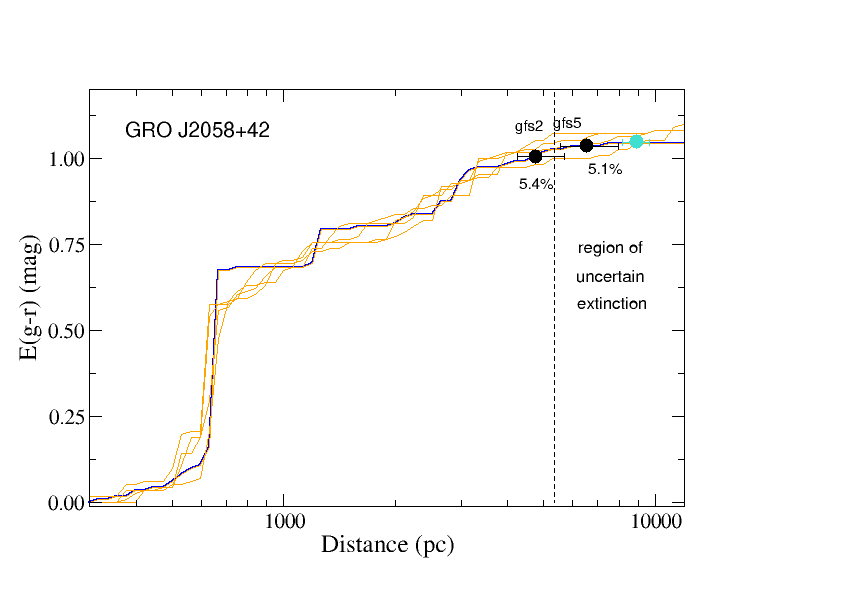}   
    \includegraphics[width=0.49\textwidth]{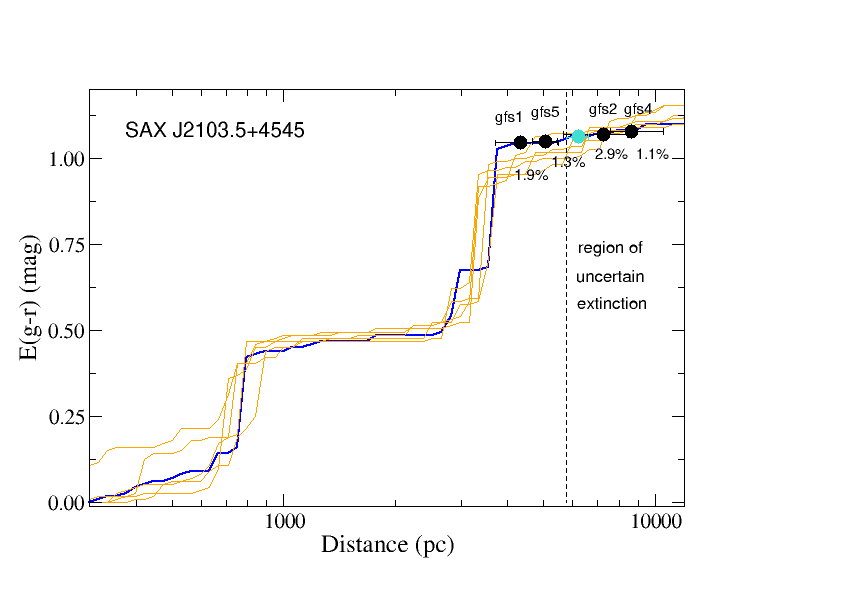}\\[-12pt]

    \caption[]{ Continued.}
\end{figure*}

\begin{figure*}
    \centering

    
    \includegraphics[width=0.49\textwidth]{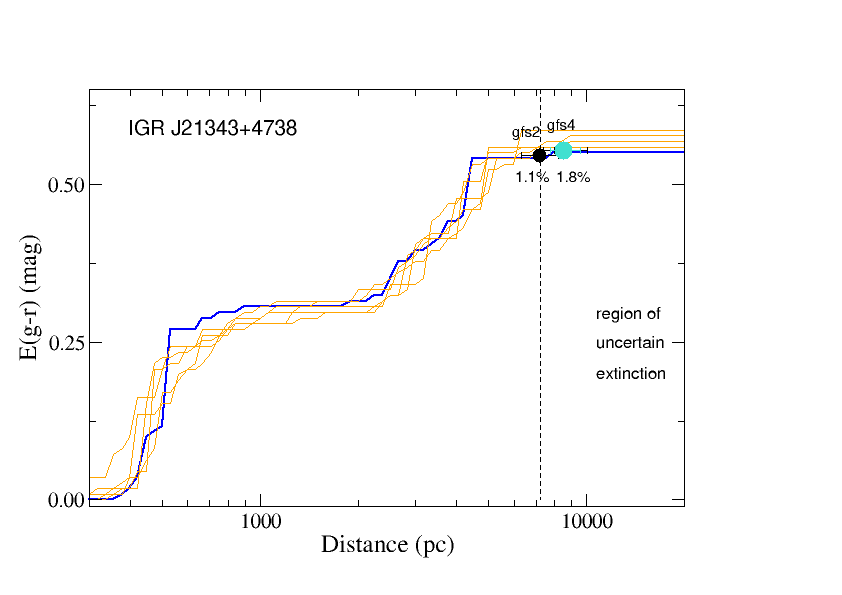}
    \includegraphics[width=0.49\textwidth]{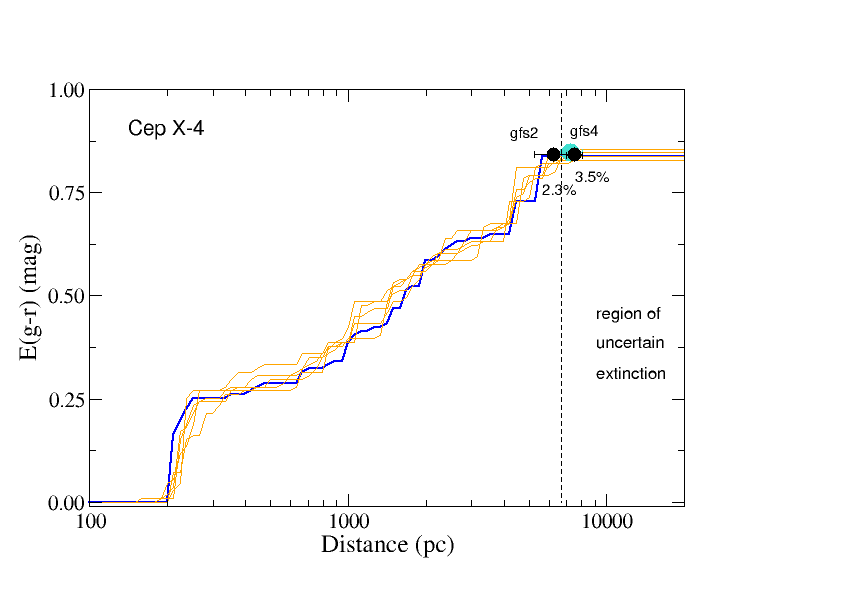}\\[-12pt]

    \includegraphics[width=0.49\textwidth]{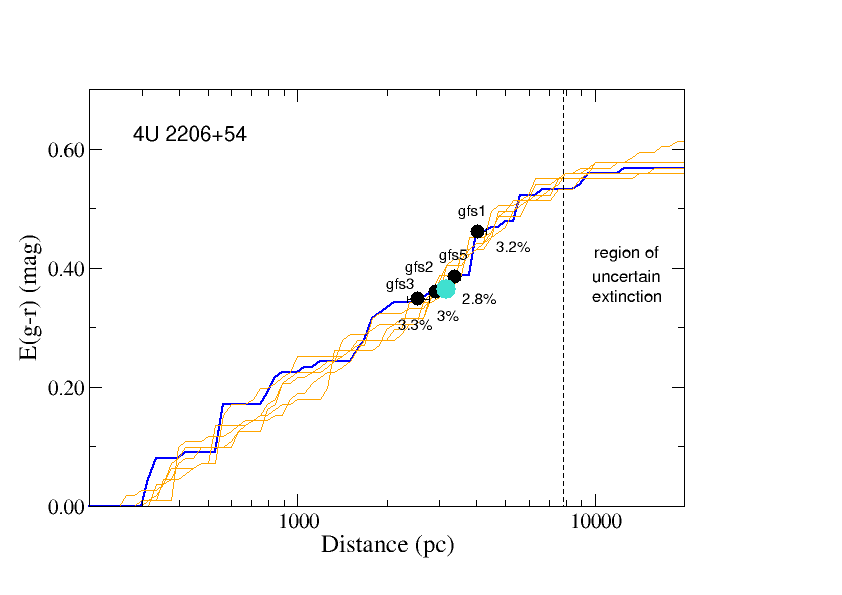}
    \includegraphics[width=0.49\textwidth]{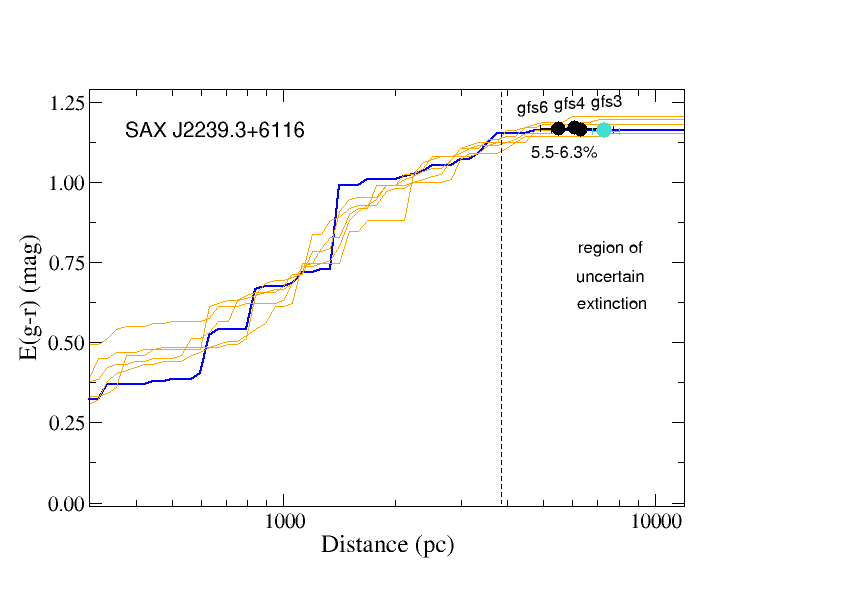}\\[-12pt]
    
    \includegraphics[width=0.49\textwidth]{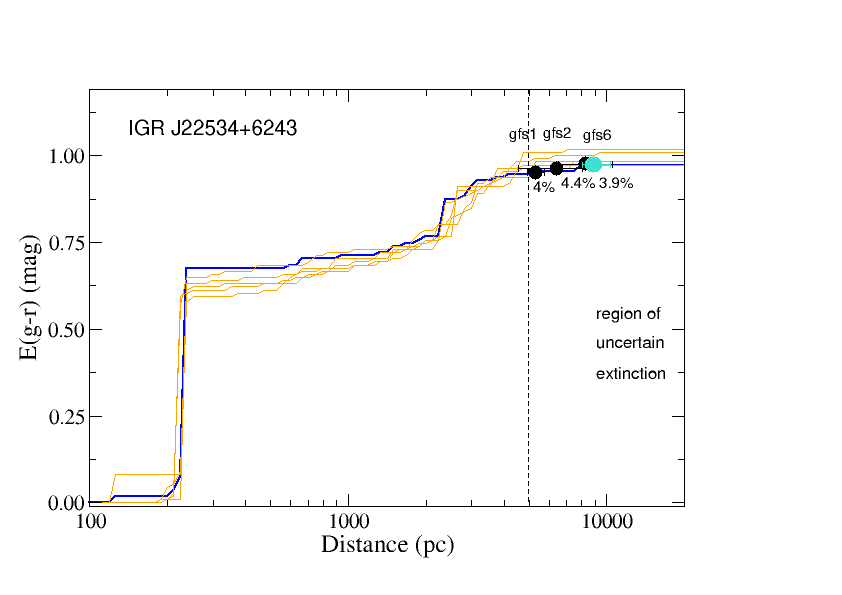} 

   \caption[]{ Continued.}
\end{figure*}
\FloatBarrier

\clearpage

\section{Sky charts of field stars}
\label{app:charts}

Sky charts of the targets. The field of view  is $12.5 \times 12.5$
arcmin. North is up, East is left. The target is marked in red, while the field
stars are marked with blue circles.

\FloatBarrier

\begin{figure*}[h]
    \centering

    \includegraphics[width=0.33\textwidth]{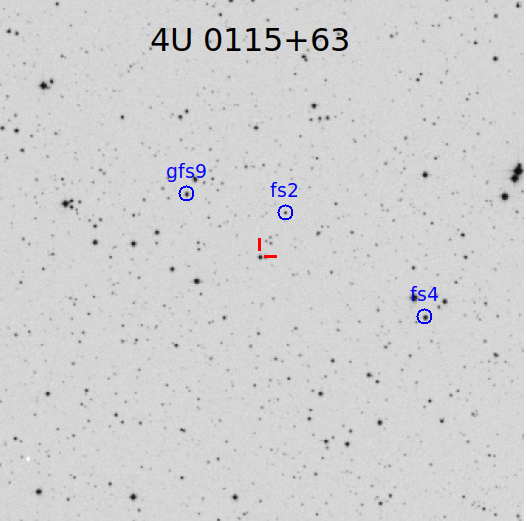}\hspace{1pt}%
    \includegraphics[width=0.33\textwidth]{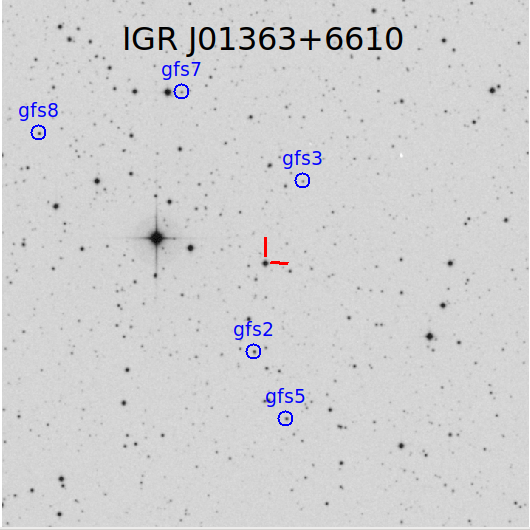}\hspace{1pt}%
    \includegraphics[width=0.33\textwidth]{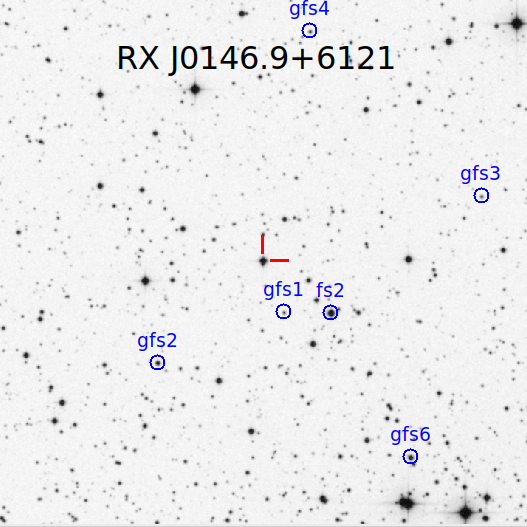}\\[2pt] 
    \includegraphics[width=0.33\textwidth]{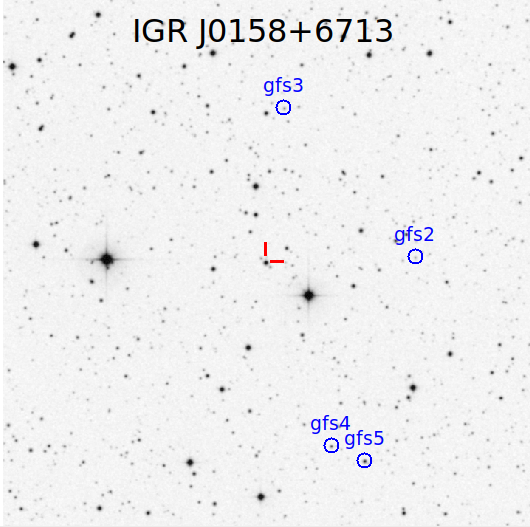}\hspace{1pt}%
    \includegraphics[width=0.33\textwidth]{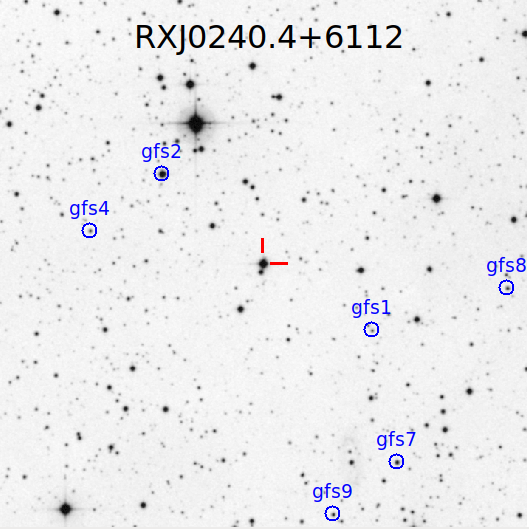}\hspace{1pt}%
    \includegraphics[width=0.33\textwidth]{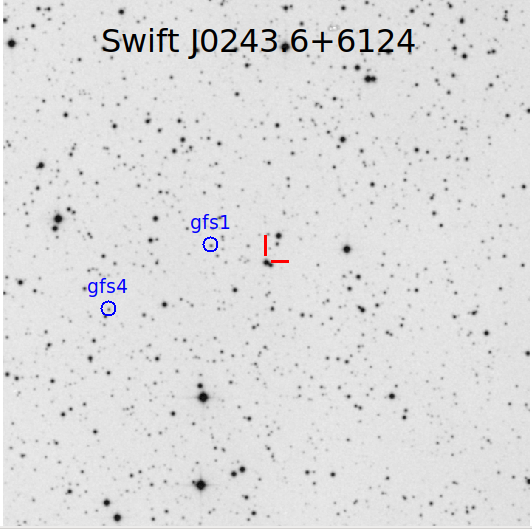}\\[2pt]
    \includegraphics[width=0.33\textwidth]{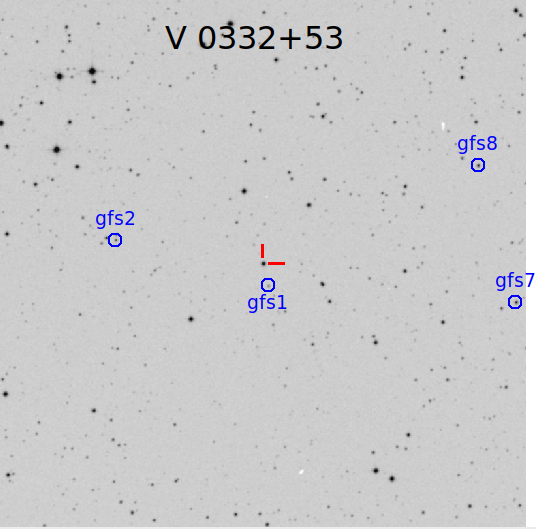}\hspace{1pt}%
    \includegraphics[width=0.33\textwidth]{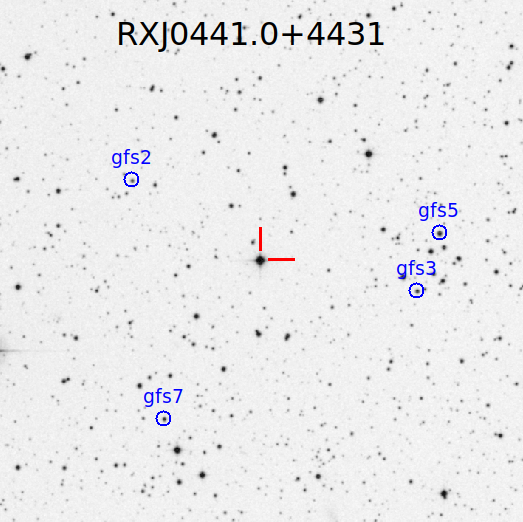}\hspace{1pt}%
    \includegraphics[width=0.33\textwidth]{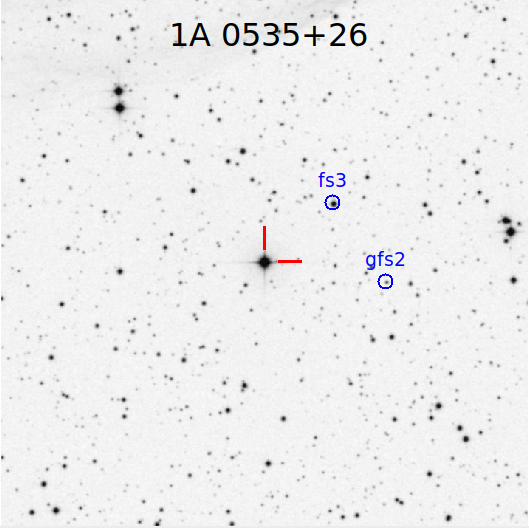}\\[2pt]

    \caption{Sky charts with the identification of the target (red lines) and
    the field stars (blue circles).}
    \label{charts}
\end{figure*}

\begin{figure*}

    \centering

    \includegraphics[width=0.33\textwidth]{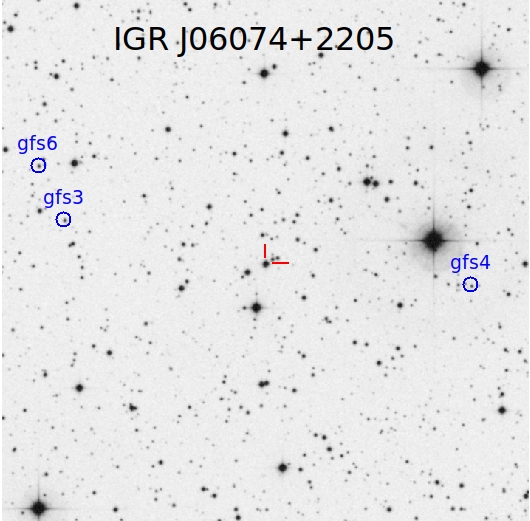}\hspace{1pt}%
    \includegraphics[width=0.33\textwidth]{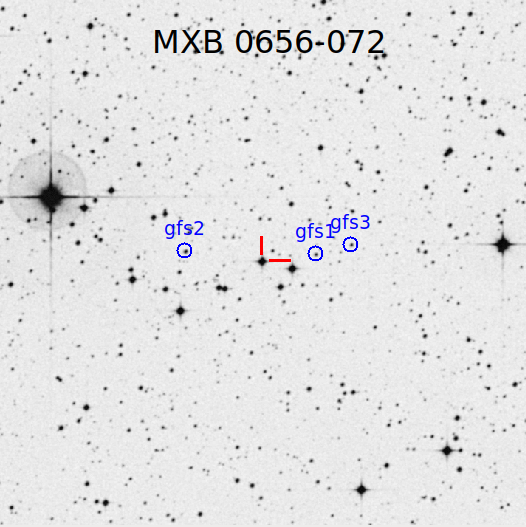}\hspace{1pt}%
    \includegraphics[width=0.33\textwidth]{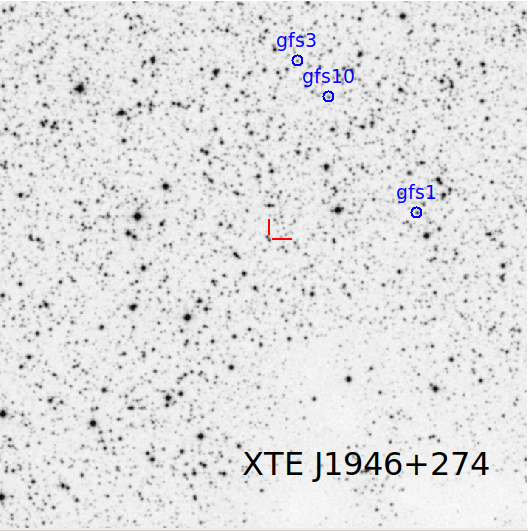}\\[2pt] 
    \includegraphics[width=0.33\textwidth]{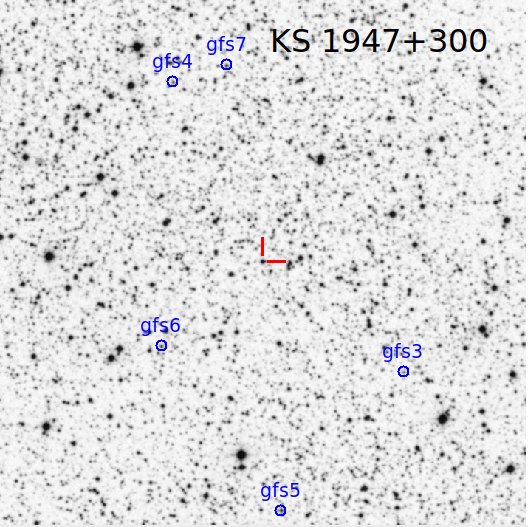}\hspace{1pt}%
    \includegraphics[width=0.33\textwidth]{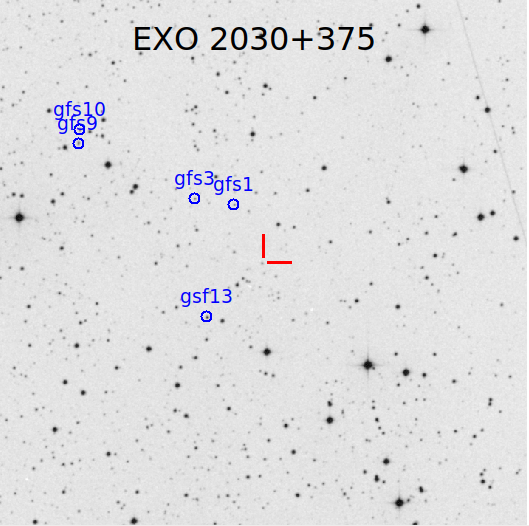}\hspace{1pt}%
    \includegraphics[width=0.33\textwidth]{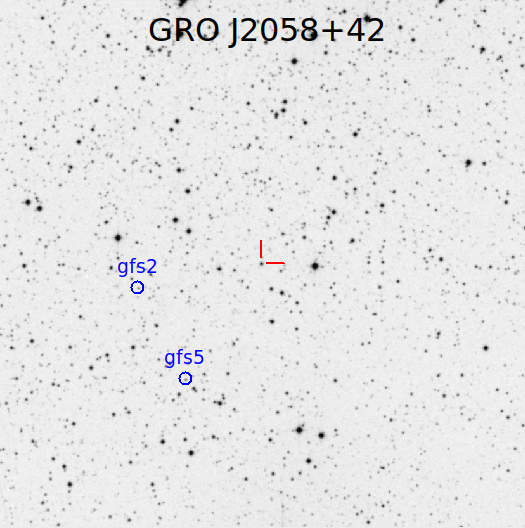}\\[2pt]
    \includegraphics[width=0.33\textwidth]{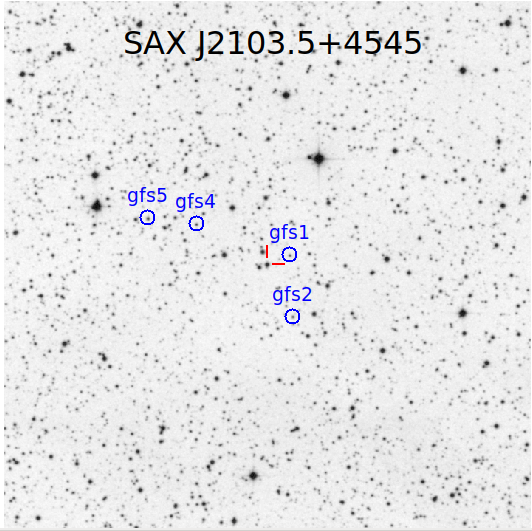}\hspace{1pt}%
    \includegraphics[width=0.33\textwidth]{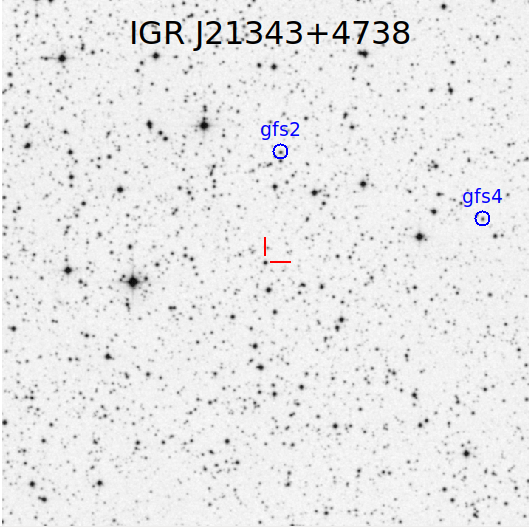}\hspace{1pt}%
    \includegraphics[width=0.33\textwidth]{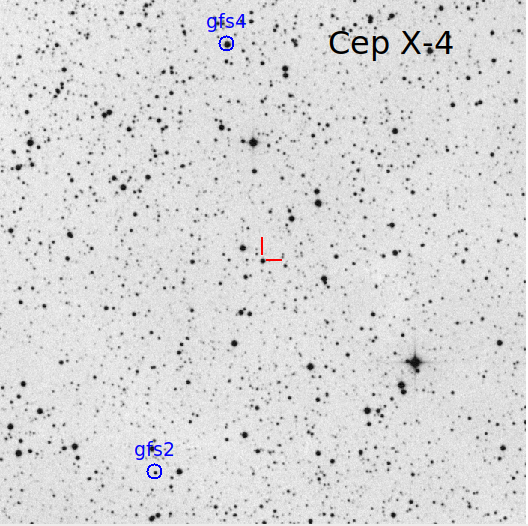}\\[2pt]
 
    \includegraphics[width=0.33\textwidth]{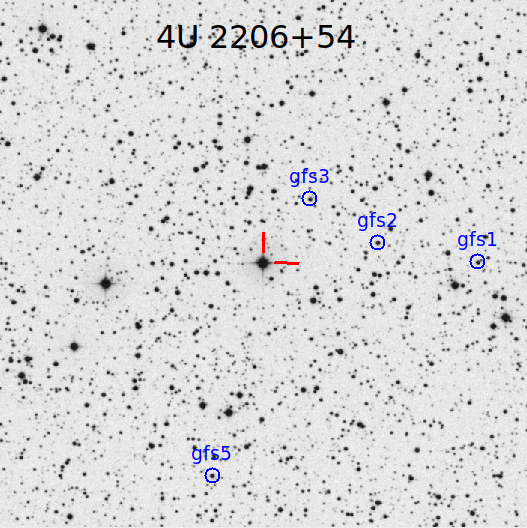}\hspace{1pt}%
    \includegraphics[width=0.33\textwidth]{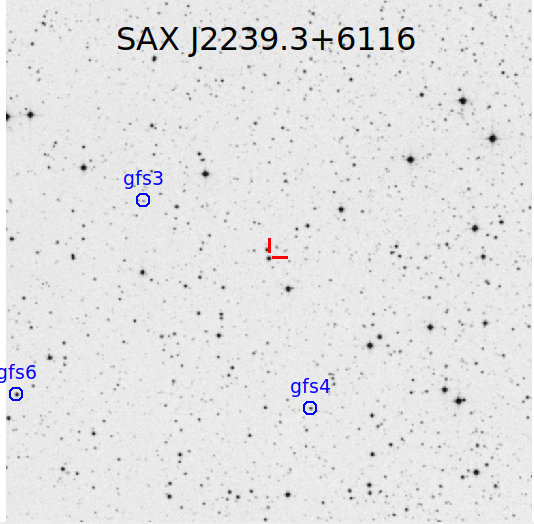}\hspace{1pt}%
    \includegraphics[width=0.33\textwidth]{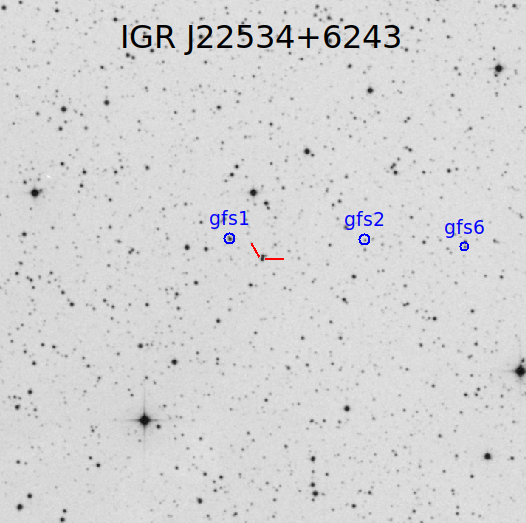}\\[2pt]

    \caption[]{ Continued.}
\end{figure*}

\FloatBarrier

\end{appendix}
\end{document}